\documentclass[a4paper,11pt]{article}
\pdfoutput=1 
\usepackage[english]{babel}
\usepackage{microtype,xspace}
\usepackage[utf8]{inputenc}	

\usepackage{jheppub} 
\usepackage{bm,amssymb,amsmath,slashed,graphicx,multirow,soul,mathtools,xspace,array}
\usepackage{siunitx}
\usepackage[shortlabels]{enumitem}
\SetEnumerateShortLabel{i}{\textit{\roman*}}

\usepackage{float}
\allowdisplaybreaks
\usepackage{ bbold }
\usepackage{subfigure}

\newcommand{\todo}[1]{{\color{red} \ifmmode\else[todo]\fi #1}}
\usepackage[usenames,dvipsnames]{xcolor}
     \definecolor{hgreen}{rgb}{0,.3,0}
      \definecolor{darkgreen}{rgb}{0.3,.8,0.2}
     \definecolor{hred}{rgb}{.3,0,0}
     \definecolor{hblue}{rgb}{0,0,.3}
     \definecolor{LightGray}{gray}{0.95}

\newcommand{\rep}[1]{\mathbf{#1}}
\newcommand{\repbar}[1]{\overline{\mathbf{#1}}}

\newcommand{\GeV}{\text{GeV}}

\newcommand{\beq}{\begin{equation} }
\newcommand{\eeq}{\end{equation}}
\newcommand{\bi}{\begin{itemize} }
\newcommand{\ei}{\end{itemize} }

\newcommand{\hc}{\; + \; \mathrm{H.c.} \;}
\newcommand{\andeq}{\quad \mathrm{and} \quad}

\renewcommand{\L}{\mathcal{L}}
\newcommand{\LL}{\mathrm{L}}
\newcommand{\RR}{\mathrm{R}}
\newcommand{\U}{\mathrm{U}}
\newcommand{\SU}{\mathrm{SU}}
\newcommand{\eminus}{\vcenter{\hbox{\scalebox{0.6}[1]{$ - $}}}}	
\newcommand{\dd}{\mathop{}\!\mathrm{d}}
\newcommand{\sscript}[1]{{\scriptscriptstyle \mathrm{#1}}}
\newcommand{\RK}{$ R_{K^{(\ast)}} $\xspace}
\newcommand{\Umt}{$ \U(1)_{L_\mu\eminus L_\tau} $\xspace}

\newcommand{\gmu}{$ (g-2)_\mu $\xspace}

\definecolor{Red}{rgb}{1.,0.,0.}
\definecolor{Grn}{rgb}{0.,0.75,0.}
\definecolor{Blu}{rgb}{0.,0.,1.}

\definecolor{red}{rgb}{0.6,.0706,.1373}
\definecolor{blue}{rgb}{0,0.396,0.741}

\DeclareMathOperator{\diag}{diag}

\DeclareMathOperator{\Tr}{Tr}

\usepackage[T1]{fontenc} 

\usepackage{amsmath,amssymb,epsfig,color,slashed}

\allowdisplaybreaks

\title{\boldmath Muonic Force Behind Flavor Anomalies}

\author[1]{Admir Greljo,}
\author[2]{Yotam Soreq,}
\author[1]{Peter Stangl,}
\author[1]{Anders Eller Thomsen,}
\author[3]{Jure Zupan}

\affiliation[1]{Albert Einstein Center for Fundamental Physics, Institute for Theoretical Physics, University of Bern, CH-3012 Bern, Switzerland}
\affiliation[2]{Physics Department, Technion – Israel Institute of Technology, Haifa 3200003, Israel}
\affiliation[3]{Department of Physics, University of Cincinnati, Cincinnati, Ohio 45221,USA}

\emailAdd{admir.greljo@unibe.ch}
\emailAdd{soreqy@physics.technion.ac.il}
\emailAdd{stangl@itp.unibe.ch}
\emailAdd{thomsen@itp.unibe.ch}
\emailAdd{zupanje@ucmail.uc.edu}

\abstract{
We develop an economical theoretical framework for combined explanations of the flavor physics anomalies involving muons: $(g-2)_\mu$, $R_{K^{(*)}}$, and $b \to s \mu^+ \mu^-$ angular distributions and branching ratios, that was first initiated by some of us in Ref.~\cite{Greljo:2021xmg}.
The Standard Model~(SM) is supplemented with a lepton-flavored $\mathrm{U}(1)_X$ gauge group.
The $\mathrm{U}(1)_X$ gauge boson with the mass of $\mathcal{O}(0.1)$\,GeV resolves the $(g-2)_\mu$ tension.
A TeV-scale leptoquark, charged under the $\mathrm{U}(1)_X$, carries a muon number and mediates $B$-decays without prompting charged lepton flavor violation or inducing proton decay.
We explore the theory space of the chiral, anomaly-free $\mathrm{U}(1)_X$ gauge extensions featuring the above scenario, and identify many suitable charge assignments for the SM$+3\nu_R$ fermion content with the integer charges in the range $X_{F_i} \in [-10,10]$.
We then carry out a comprehensive phenomenological study of the muonic force in  representative benchmark models.
Interestingly, we found models which can resolve the tension without conflicting the complementary constraints, and all of the viable parameter space will be tested in future muonic resonance searches.
Finally, the catalog of the anomaly-free lepton-non-universal charge assignments motivated us to explore different directions in model building.
We present a model in which the muon mass and the $(g-2)_\mu$ are generated radiatively from a common short-distance dynamics after the $\mathrm{U}(1)_X$ breaking.
We also show how to charge a vector leptoquark under $\mathrm{U}(1)_{\mu - \tau}$ in a complete gauge model.}

\begin{document}

\maketitle

\flushbottom

\section{Introduction}
\label{sec:intro}

Two sets of observables may be pointing to a new muonic force:
 $i)$  that the measurements of the anomalous magnetic moment of the muon, $(g-2)_\mu$,~\cite{Bennett:2006fi,Aoyama:2020ynm} are disagreeing with the $R$-ratio data-driven theory prediction~\cite{Aoyama:2020ynm,Colangelo:2020lcg,aoyama:2012wk,Aoyama:2019ryr,czarnecki:2002nt,gnendiger:2013pva,davier:2017zfy,keshavarzi:2018mgv,colangelo:2018mtw,hoferichter:2019gzf,davier:2019can,keshavarzi:2019abf,kurz:2014wya,melnikov:2003xd,masjuan:2017tvw,Colangelo:2017fiz,hoferichter:2018kwz,gerardin:2019vio,bijnens:2019ghy,colangelo:2019uex,Blum:2019ugy,colangelo:2014qya} (however, see also~\cite{Borsanyi:2020mff})
 and $ii)$ the deviations from predictions in rare $B$ meson decay observables, in particular $b \to s \mu^+ \mu^-$ angular distributions~\cite{LHCb:2020lmf,LHCb:2020gog} and branching ratios~\cite{LHCb:2020zud,LHCb:2021awg,LHCb:2021vsc,LHCb:2014cxe,LHCb:2015wdu,LHCb:2016ykl,LHCb:2021zwz} (we will collectively refer to them as ``$b\to s\mu\mu$ anomalies'') and the lepton-flavor university (LFU) ratios \RK~\cite{LHCb:2017avl,LHCb:2021trn}.
Among these, the observables \RK stand out in particular, because their predictions are extremely clean in the Standard Model~(SM)~\cite{Hiller:2003js,Bordone:2016gaq,Isidori:2020acz}.
The most recent update of $R_K$ increased the significance of the anomaly and, for the first time, LHCb declared evidence for lepton flavor universality violation~(LFUV)~\cite{LHCb:2021trn}.
While \RK could be explained by New Physics~(NP) coupling to electrons or muons, an explanation in terms of muonic NP provides a consistent combined explanation of all anomalies in rare $B$ decays, i.e.\ the $b\to s\mu\mu$ anomalies and \RK (for recent global fits see e.g.~\cite{Altmannshofer:2021qrr,Geng:2021nhg,Alguero:2021anc,Hurth:2021nsi,Ciuchini:2020gvn}).

In contrast to the possible new flavor-diagonal couplings of NP to muons there is striking absence of any such NP hints in lepton-flavor-violating~(LFV) transitions, such as $\mu\to e\gamma$ and $\mu \to 3 e$.
Numerically, taking as an example NP that induces the dimension-5 dipole moment operators after the electroweak symmetry breaking,
$\mathcal{L}_{\rm eff} \supset - { e \,v}\, \bar \ell^i_{\LL} \sigma^{\mu \nu} \ell^j_{\RR } F_{\mu \nu}/{(4 \pi \Lambda_{ij})^2} + {\rm h.c.}$, with $v  =\SI{246}{GeV} $ the electroweak vacuum expectation value~(VEV), the NP scale required to explain the $(g-2)_\mu$ anomaly  is $\Lambda_{22} \simeq \SI{15}{TeV}$.
This should be compared with the much more stringent bound $\Lambda_{12 (21)}  \gtrsim \SI{3600}{TeV}$ on the effective NP scale implied by the absence of the $\mu \to e \gamma$ transition~\cite{TheMEG:2016wtm}.
The hierarchy between the two effective scales persists even if the couplings to electrons are Yukawa suppressed: $\Lambda_{12 (21)} \sqrt{m_e / m_\mu}  \gtrsim \SI{250}{TeV}$.
Qualitatively similar
 implications follow from the absence of $\tau \to \mu \gamma$ transitions~\cite{BaBar:2009hkt}.

The high suppression of flavor-violating effects strongly hints at NP with an almost exact  muon-number symmetry, $\U(1)_{L_\mu}$.
The $\U(1)_{L_\mu}$ symmetry forbids flavor-violating transitions with muons, but still allows for deviations from flavor universality, i.e., different flavor-diagonal couplings to electrons, muons, and taus.
A number of proposed solutions to the experimental anomalies are based on an anomaly-free lepton-flavored symmetry group \Umt~\cite{Baek:2001kca,Ma:2001md,Harigaya:2013twa,Altmannshofer:2014pba,Altmannshofer:2019zhy,Crivellin:2016ejn,Crivellin:2015mga,Crivellin:2018qmi,Altmannshofer:2014cfa,Altmannshofer:2015mqa}, one of three such anomaly-free symmetries of the SM in the absence of right-handed neutrinos which allow a SM-like Yukawa sector for the quarks and diagonal charged lepton Yukawa couplings (up to an overall hypercharge shift)~\cite{He:1990pn,He:1991qd,Altmannshofer:2019xda}.
For instance, a light \Umt gauge boson $X_\mu$ can generate a 1-loop contribution to $(g-2)_\mu$ of the right size in order to explain the deviations in the experimental measurements, while at the same time not being excluded by other complementary searches~\cite{Altmannshofer:2014pba,Altmannshofer:2019zhy}.
For the \RK anomalies, on the other hand, a heavy gauge field $X_\mu$ was used along with a set of vector-like quarks~\cite{Altmannshofer:2014cfa,Altmannshofer:2015mqa}.
The other gauging choices used to explain \RK are $\U(1)_{B_3-L_\mu}$~\cite{Alonso:2017uky,Bonilla:2017lsq,Allanach:2020kss}, third family hypercharge~\cite{Allanach:2018lvl,Allanach:2019iiy,Bhatia:2017tgo}, and other alternatives~\cite{Altmannshofer:2019xda,AristizabalSierra:2015vqb,Celis:2015ara,Falkowski:2015zwa,Chiang:2016qov,Boucenna:2016wpr,Boucenna:2016qad,Ko:2017lzd,Alonso:2017bff,Tang:2017gkz,Bhatia:2017tgo,Fuyuto:2017sys,Bian:2017xzg,King:2018fcg,Duan:2018akc}.
Note, however, that
it is not possible to explain both $(g-2)_\mu$ and \RK in an effective field theory (EFT) with a single light mediator $X_\mu$ that provides the dominant contribution. We show this in complete generality in Section~\ref{sec:single_mediator}: the size of the $X_\mu$ contribution required to explain the deviations in $(g-2)_\mu$ and \RK is ruled out by a combination of constraints from $B \to K \nu \bar \nu$, $\tau \to 3 \mu$ and neutrino trident production.

 A different class of mediators that can successfully explain the flavor anomalies are the leptoquarks~\cite{Dorsner:2016wpm}. The lepton-flavored $\U(1)_X$ gauge symmetry would constrain the leptoquark couplings, which can be crucial for having a viable phenomenology~\cite{Hambye:2017qix,Davighi:2020qqa,Greljo:2021xmg}.
 For instance, the TeV-scale \textit{muoquarks}, i.e., the leptoquarks that interact only with muons and not with electrons or taus, are motivated by both
 $i)$ the already mentioned muon anomalies in $(g-2)_\mu$ and rare $B$ decays;
 and  $ii)$ by the stringent constraints on the charged LFV.
 A subset of chiral anomaly-free $\U(1)_X$ extensions of the SM, under which leptoquarks are charged, provide natural theories for muoquarks while also addressing the absence of proton decay.
 A pragmatic proposal for the common explanation of the muon anomalies utilizes a light $X_\mu$ for $(g-2)_\mu$ and a heavy muoquark $S_{3} = (\repbar{3},\, \rep{3},\, 1/3)_{X_{S_3}}$ connected by the underlying $\U(1)_{X}$ gauge symmetry~\cite{Greljo:2021xmg}.

In this paper, we systematically explore the theory space of anomaly-free $\U(1)_X$ gauge extensions of the SM, extending the scenario in Ref.~\cite{Greljo:2021xmg}.
In Section~\ref{sec:models} we study the anomaly-cancellation conditions and identify a complete set of quark-flavor-universal and third-family-quark $\U(1)_X$ models with appropriate rational charge assignments for the SM$+3\nu_R$ chiral fermions with the maximal charge ratios $\leq 10$.
The solutions are classified as vector-like (Section~\ref{sec:category-I}) or chiral (Section~\ref{sec:not:all}) depending on the charges of the left- and the right-handed $e$, $\mu$ and $\tau$.

We then address whether {\Umt is a unique} gauge group that can lead to a successful phenomenology: an explanation of $(g-2)_\mu$ with the light $X_\mu$ while avoiding all other constraints.
As we will show, there are {very few} other anomaly-free gauge group choices that are not excluded by the present experimental constraints and can simultaneously explain the $(g-2)_\mu$ anomaly.
We carry out a detailed phenomenological study to confront the preferred region (Section~\ref{sec:g-2}) with the complementary constraints from neutrino trident production (Section~\ref{sec:trident}), {non-standard neutrino interactions,} Borexino and light resonance searches (Section~\ref{sec:Borexino_lightX}).
The constraints are applied to several carefully chosen benchmark models in Section~\ref{sec:benchmarks} {to illustrate possible scenarios.}

The lepton-flavored $\U(1)_X$ systematics outlined in Section~\ref{sec:models} opens up new directions in model building beyond the scenario of Ref.~\cite{Greljo:2021xmg}.
We illustrate this with two  examples.
In Section~\ref{sec:radiative} we present a model in which the muon mass and the $(g-2)_\mu$ are both radiatively generated at one-loop level by the TeV-scale muoquarks after the $\U(1)_X$-breaking scalar obtains a VEV.
This is made possible by the chiral solutions of the anomaly cancellation conditions, which forbid the dimension-4 muon Yukawa in the unbroken phase.
Remarkably, the $(g-2)_\mu$ tension and the muon mass sharply predict the leptoquark mass in the range that is of interest for direct searches at colliders. A different type of a model building example is presented in Section~\ref{sec:4321}, where we show how to construct an ultraviolet~(UV) completion of the vector muoquark model.
This example also gives a possible unification scenario of the $\U(1)_X$ into a simple Lie group.

Finally, Section~\ref{sec:conc} contains our conclusions, while Appendices
contain further details on the equivalence of charge assignments for the products of $\U(1)$ subgroups (Appendix~\ref{app:reparam}), the mass basis of the gauge sector (Appendix~\ref{app:gauge_mass_basis}), the RG running of the kinetic mixing (Appendix~\ref{app:epsilon_running}),
the $X$ boson decay channels (Appendix~\ref{app:branchingratios}), the contributions to \RK from a light $X$ vector boson (Appendix~\ref{app:RKX}) and the generators of $\U(1)_X$ embeddings in $\SU(5)$ (Appendix~\ref{app:su5}).

\section{Model classification}
\label{sec:models}
We start by classifying the anomaly free models that, in addition to the SM, contain a new gauge group $\U(1)_X$ and a muoquark, that is, a leptoquark that only couples to muon flavored fermions (muons and muon neutrinos). We assume that all the couplings allowed by the gauge symmetry are nonzero. As such the fact that muoquark only couples to muons is imposed by the choice of charge assignments under $\U(1)_X$, Eq.~\eqref{eq:S3quark1}. Similarly, the charge assignments, Eq.~\eqref{eq:S3quark2}, forbid the proton decay, while quark Yukawas are fully allowed in Eq.~\eqref{eq:B:requirement} or partially in Eq.~\eqref{eq:3rdcondt}. In the rest of the section we discuss these requirements in detail.

\subsection{General gauged flavor $\U(1)_X$}

Throughout the manuscript we assume that the SM is extended by three right-handed neutrinos. The
chiral fermions of the theory thus carry the following charges under the   $\SU(3)_C \times \SU(2)_L \times \U(1)_Y\times \U(1)_X$ gauge group,
\beq
\begin{matrix*}[l]
    Q_i \sim ( \rep{3},\rep{2},\tfrac{1}{6},X_{Q_i}), &
   \qquad  U_i \sim (\rep{3},\rep{1},\tfrac{2}{3},X_{U_i}), &
    \qquad D_i \sim (\rep{3},\rep{1},-\tfrac{1}{3},X_{D_i}),
    \\
    L_i \sim (\rep{1},\rep{2},-\tfrac{1}{2},X_{L_i}), &
    \qquad  E_i \sim (\rep{1},\rep{1},-1,X_{E_i}), &
     \qquad  N_i \sim (\rep{1},\rep{1},0,X_{N_i}), \label{eq:chiralF}
    \end{matrix*}
\eeq
with $i = 1,2,3$ the flavor index. The $\SU(2)_L$ doublets (singlets) are left (right) Weyl spinors under Lorentz symmetry.

A consistent ultraviolet (UV) gauge theory has to be free of chiral anomalies. In this work we require that the $\U(1)_X$ charge assignments for the field content in Eq.~\eqref{eq:chiralF} are already anomaly free.\footnote{Our construction could be viewed as a low-energy effective theory in which anomalies could alternatively be canceled by a higher-dimension Wess-Zumino-Witten  operator~\cite{Wess:1971yu}. The WZW operator is generated by integrating out heavy chiral fermions in the UV. In general, it is not always clear how to make these fermions heavy enough to satisfy the self-consistency of the effective theory assumptions. For an example see, e.g., Ref.~\cite{Davighi:2021oel}.} This results in six conditions corresponding to the cancellation of (mixed) triangle anomalies between $ \U(1)_X $, SM gauge groups, and gravity~\cite{Allanach:2018vjg},
\begin{align}
    \SU(3)_C^2 \times \U(1)_X &: \; \; \sum_{i=1}^3 (2 X_{Q_i} - X_{U_i} - X_{D_i})= 0~,\label{eq:1}\\
    \SU(2)_L^2 \times \U(1)_X &: \; \; \sum_{i=1}^3 (3 X_{Q_i} + X_{L_i})= 0~,\\
    \U(1)_Y^2 \times \U(1)_X &: \; \; \sum_{i=1}^3 (X_{Q_i} + 3 X_{L_i} - 8 X_{U_i} - 2 X_{D_i} - 6 X_{E_i} )= 0~,\\
    \text{Gravity}^2 \times \U(1)_X &: \; \; \sum_{i=1}^3 (6 X_{Q_i} + 2 X_{L_i} - 3 X_{U_i} - 3 X_{D_i} - X_{E_i} - X_{N_i} )= 0~,\\
    \U(1)_Y \times \U(1)_X^2 &: \; \; \sum_{i=1}^3 (X^2_{Q_i} - X^2_{L_i} - 2 X^2_{U_i} + X^2_{D_i} + X^2_{E_i} )= 0~,\\
    \U(1)_X^3 &: \; \; \sum_{i=1}^3 (6 X^3_{Q_i} + 2 X^3_{L_i} - 3 X^3_{U_i} - 3 X^3_{D_i} - X^3_{E_i} - X^3_{N_i} )= 0~.\label{eq:6}
\end{align}
We consider only rational solutions motivated by the unification scenario, i.e., embedding the $\U(1)_X$ into a simple Lie group at high-energies.
We can work with integer charges without loss of generality, since for any set of rational charges $\{ p_{F_i} / q_{F_i} \}$, there is an equivalent set of integer charges obtained by rescaling the gauge coupling $g_X$
with the least common denominator.
Any set of integer charges $\{ X_{F_i} \}$ satisfying the anomaly conditions~\eqref{eq:1}--\eqref{eq:6} can be used to generate up to $ (3!)^6 $ inequivalent solutions (and a correspondingly smaller set, if some of the charges for different families coincide), by permuting the flavor specific charges within each species. Below, we list the solutions to the Diophantine equations~\eqref{eq:1}--\eqref{eq:6} up to this freedom of family permutations.

Still, this leaves us with infinitely many integer solutions of the anomaly cancellation conditions.
For concreteness, we limit the maximal ratio of the largest and the smallest nonzero charge magnitudes to be $\leq 10$.\footnote{As a point of reference, this ratio is 6 for the SM hypercharge.}
In the following we then give an exhaustive set of inequivalent integer solutions of Eqs.~\eqref{eq:1}--\eqref{eq:6} with
\begin{equation}
    - 10 \leq X_{F_i} \leq 10 ~\text{ for every $F_i$ in Eq.~\eqref{eq:chiralF}},
\end{equation}
building on the work of Ref.~\cite{Allanach:2018vjg}, while imposing further constraints to produce viable muoquark models.

\subsection{Quark flavor universal $\U(1)_X$}
\label{sec:qfu}

The symmetry-breaking scalar fields are
\begin{equation}
    H = (\rep{1},\rep{2},\frac{1}{2},X_H)~,
    \qquad \phi = (\rep{1},\rep{1},0,X_{\phi})~,
\end{equation}
where $ H $ is the SM Higgs (with $U(1)_X $ charge $X_H$) and $ \phi $ is the SM singlet responsible for the breaking of $ \U(1)_X$.
Shifting the $\U(1)_X$ charge assignments for all fields $f$ by a universal multiple of the hypercharge, $X_f \to X_f - a Y_f$, gives a physically equivalent theory, cf. Appendix~\ref{app:reparam}. In particular, after a linear invertible field transformation $q_f = (Y_f, X_f)^\intercal$ becomes
\beq
\tilde{q}_f = L^\intercal q_f \quad \text{where} \quad
L = \begin{pmatrix}
1 & -a \\
0 & 1
\end{pmatrix}~.
\eeq
The ambiguity in charge assignments is a direct consequence of the freedom in defining the $ \U(1) $ subgroups for a symmetry group with several
Abelian factors.
A familiar example is the QCD, which, ignoring the anomalies, has a global $ \U(1)_V \times \U(1)_A $ or, equivalently, a $ \U(1)_\LL \times \U(1)_\RR $ symmetry.

In what follows, we use the above reparameterization invariance to make $H$ a $\U(1)_X$ singlet,
\beq\label{eq:HiggsCharge}
    X_H = 0~,
\eeq
and thus $H$ is the usual SM Higgs.
To simplify the discussion further, we require all quarks to have the same $\U(1)_X$ charge,
\beq
    \label{eq:B:requirement}
    X_{Q_i} = X_{U_j} = X_{D_k} \equiv X_q,\quad \text{for all}\quad i,j,k=1,2,3,
\eeq
such that their masses and the CKM mixing matrix are allowed by the gauge symmetry, i.e. $Y_u^{ij} \, \bar Q^i \tilde H u^j$ and $Y_d^{ij}\, \bar Q^i H d^j$ where $\tilde H = \epsilon H^*$.
The conditions \eqref{eq:B:requirement}
reduce the number of inequivalent sets of $X_{F_i}$ charges in the range $[-10, 10]$ from the original $ 21\,546\,920 $ of Ref.~\cite{Allanach:2018vjg} to
276.

The theory will also have a leptoquark field charged under $\U(1)_X$ coupled exclusively to muons~\cite{Hambye:2017qix,Davighi:2020qqa,Greljo:2021xmg}. To realize the \textit{muoquark}~\cite{Greljo:2021xmg}, we further impose:
\begin{itemize}
    \item The leptoquark coupling ($q\ell$-LQ) is allowed for $\mu$ but not for $e$ and $\tau$,
    \begin{equation}
        \label{eq:S3quark1}
        \quad X_{\ell_2} \ne X_{\ell_{1,3}}~,
    \end{equation}
    where $\ell$ is $L$ or $E$, i.e., the three  flavors of $L_i$ or $E_i$ ($i=1,2,3$) are not all charged the same.
    Out of 276 sets, 273 satisfy this criteria.

    \item
    The diquark couplings ($q q$-LQ or $q q H$-LQ) are forbidden and thus proton decay is suppressed, postponing the potential $ \U(1)_B $ violating effects to dimension-6 operators, the same as in the SM EFT. Given that the color contraction in a diquark ($q q$-LQ) coupling is $\rep{3} \times \rep{3} \times \rep{3}$ while in the leptoquark coupling ($q\ell$-LQ) it is $\rep{3} \times \repbar{3}$, this implies
    \begin{subequations}
     \label{eq:S3quark2}
    \begin{align}
     \label{eq:S3quark2:first}
    X_{\ell_2} ~&\ne ~+3 X_q~\text{for $\SU(2)_L$  doublet scalars, and vectors $\rep{1(3)}_{2/3}$}~,
    \\
     \label{eq:S3quark2:second}
    X_{\ell_2} ~&\ne ~-3 X_q~\text{for $\SU(2)_L$  singlet and triplet scalars,  and vector doublets}.
    \end{align}
    \end{subequations}
    This is satisfied for 272 and 273 sets, respectively.

\item The $\phi$ charge should be chosen such that $qq\phi$-LQ dimension-5 operators are forbidden, i.e. the $\U(1)_B$ stays an accidental symmetry up to dimension-6 Lagrangian.

\end{itemize}

The observed structure of neutrino masses and mixings may impose further nontrivial constraints on the setup. We make no attempt to impose these constraints when listing the anomaly-free $U(1)_X$ models below, since they depend on whether or not there are additional scalars in the theory.
For instance, the $\U(1)_{B-3 L_\mu}$ model proposed in Ref.~\cite{Greljo:2021xmg} requires no additional scalars, since it already admits  the minimal realization of the neutrino masses. The nontrivial change from the requirements listed in the bullets above is that  now the charge $X_\phi$ is determined from the structure of the neutrino mass matrix, and in general  the proton decay inducing dimension-5 operator $qq\phi-$LQ could be allowed. This does not happen in the $\U(1)_{B-3 L_\mu}$ model and no dimension-5 proton decay operator is allowed in the case of a realistic type-I seesaw model that gives the neutrino masses and mixings in agreement with the neutrino oscillations data.

 The situation is expected to be different for a generic $\U(1)_X$ gauge model. Assuming only the minimal breaking sector will normally impose a texture of the Majorana mass matrix that is too restrictive and will not be able to accommodate the observed neutrino mixing and mass patterns. For example, the minimal type-I seesaw realization of the neutrino mass in \Umt introduces dimension-5 proton  decay plus shows some tension in fitting $\theta_{23}$ and $\sum_i m_{\nu_i}$~\cite{Asai:2019ciz}, calling for additional structure to be added~\cite{Araki:2019rmw}. In general, it is always possible to introduce additional $\U(1)_X$ symmetry-breaking scalars whose VEVs then populate the missing entries in the mass matrix. For example, the mass matrix of the right-handed neutrinos can be populated by $\phi_{i j} \bar N_i^c N_j$ where $X_{\phi_{i j}} = - X_{N_i} - X_{N_j}$. In such extensions some care needs to be taken to remove the potential Goldstone bosons, as well as to avoid baryon number violating operators at dimension-5.  While the catalog of the models derived in this manuscript provides a good starting point, a detailed discussion of the neutrino sector is beyond the scope of the present work and is left for future studies.

With the above caveat about neutrino masses in mind let us now move to the classification of different anomaly free $\U(1)_X$ models. It is remarkable that almost all anomaly-free charge assignments $X_{F_i} \in [-10,10]$ in the quark flavor universal $\U(1)_X$ automatically satisfy the muoquark conditions.
The list of charge assignments can be classified into two categories:
\begin{align}
\text{{\bf vector category}} &: \;\; X_{L_i}=X_{E_i} \quad \text{for all}\quad i=1,2,3~, \label{eq:charged:lep:req}\\
\text{{\bf chiral category}} &: \;\; \text{the rest}.
\end{align}
In the vector category models the charged lepton Yukawas for all three generations are allowed by the $U(1)_X$ symmetry, while in the chiral category models at least some of the charged lepton Yukawas are forbidden and thus all the lepton masses are generated only after the $U(1)_X$ symmetry is spontaneously broken.

Before discussing each of the two categories in more detail, let us consider several examples of muoquarks adopting the nomenclature from Ref.~\cite{Dorsner:2016wpm}:
\begin{itemize}

    \item The scalar leptoquark $S_3 \equiv (\repbar{3},\rep{3},1/3, X_{S_3})$, where $X_{S_3} = -X_q - X_{L_2}$, gives $V-A$ contribution to $b \to s \mu^+ \mu^-$ transitions, see e.g.~\cite{Greljo:2021xmg,Davighi:2020qqa,Hiller:2014yaa,Dorsner:2016wpm,Buttazzo:2017ixm,Crivellin:2017zlb,Hiller:2017bzc, Gherardi:2020qhc,Angelescu:2021lln,Marzocca:2018wcf,Dorsner:2017ufx,Babu:2020hun}. The condition in Eq.~\eqref{eq:S3quark2:second} implies $ X_{L_2} \neq - 3 X_q$ such that the dimension-4 operator $Q Q S_3$ is forbidden.

    \item The scalar leptoquark $S_1 \equiv (\repbar{3},\rep{1},1/3, X_{S_1})$, where $X_{S_1} = -X_q - X_{L_2}$ or $X_{S_1} = -X_q - X_{E_2}$, implemented in ``vector category'' models, couples to both $L_2$ and $E_2$ to give the $m_t$-enhanced contribution to $(g-2)_\mu$, see e.g.~\cite{Greljo:2021xmg,Dorsner:2016wpm,Bauer:2015knc,Dorsner:2019itg,Babu:2020hun,Gherardi:2020qhc,Brdar:2020quo,Queiroz:2014zfa}. The condition in Eq.~\eqref{eq:S3quark2:second} is $X_{\ell_2} \neq - 3 X_q$.

    \item The scalar leptoquark $R_2 \equiv (\rep{3}, \rep{2}, 7/6, X_{R_2})$, where $X_{R_2} = X_q - X_{L_2}$ or $X_{R_2} = X_q - X_{E_2}$, and the condition in Eq.~\eqref{eq:S3quark2:first} is $X_{\ell_2} \neq 3 X_q$ such that dimension-5 operator $d d H^\dagger R_2$ is forbidden. Note that otherwise such operators would lead to excessive proton decay even when suppressed by the Planck scale~\cite{Arnold:2013cva,Assad:2017iib,Dorsner:2016wpm}.
    This scalar leptoquark representation is also used to address the $(g-2)_\mu$, see e.g.~\cite{Queiroz:2014zfa,Dorsner:2016wpm,Dorsner:2019itg,Babu:2020hun}. We will employ it in Section~\ref{sec:radiative} to build a model for radiative muon mass and $(g-2)_\mu$.

    \item The vector leptoquark $U_1 \equiv (\rep{3}, \rep{1}, 2/3, X_{U_1})$, where $X_{U_1} = X_q - X_{L_2}$ or $X_{U_1} = X_q - X_{E_2}$. The baryon number violating dimension-5 operator $Q d H^\dagger U_1$ is forbidden when $X_{\ell_2} \neq 3 X_q$, Eq. \eqref{eq:S3quark2:first}. Possible UV completions for the $U_1$ vector muoquark will be presented in Section~\ref{sec:4321}. This leptoquark representation was extensively discussed in the literature to address the $B$-decay anomalies, see e.g.~\cite{Barbieri:2015yvd,DiLuzio:2017vat,Greljo:2018tuh,Bordone:2017bld,Bordone:2018nbg,Cornella:2019hct,Fornal:2018dqn,Blanke:2018sro,Fuentes-Martin:2019ign,Guadagnoli:2020tlx,Heeck:2018ntp,Fuentes-Martin:2020bnh,Fuentes-Martin:2019bue,Fuentes-Martin:2020luw,Fuentes-Martin:2020hvc}.

\end{itemize}

\subsubsection{Vector category $U(1)_X$ charge assignments}
\label{sec:category-I}

The vector category is defined such that the left-handed and the right-handed $e$, $\mu$ and $\tau$ leptons carry the same $X$ charge.  Solutions to the anomaly conditions~\eqref{eq:1}--\eqref{eq:6} that further satisfy Eqs.~\eqref{eq:B:requirement} and~\eqref{eq:charged:lep:req} are parameterized by~\cite{Altmannshofer:2019xda}
\begin{equation}
    X_{F}
=   X_{e} T_{L_{e}}+X_{\mu} T_{L_{\mu}}+X_{\tau}
    T_{L_{\tau}}-\left(\frac{X_{e}+X_{\mu}+X_{\tau}}{3}\right)
    T_{B} + a Y_{F} + \sum_i X_{N_i} T_{N_i},
    \label{eq:Tx}
\end{equation}
where $\{T_B,T_{L_{e}},T_{L_{\mu}},T_{L_{\tau}}\}$ are the usual baryon and lepton numbers for the SM fermions, while $T_{N_i}$ are the right-handed neutrino numbers (also $ T_{B , L_{e,\mu,\tau}} N_i = 0 $). The reparameterization invariance ($X_F \to X_F - a Y_F$) allows to restrict the discussion to the case $a = 0$ in agreement with Eqs.~\eqref{eq:HiggsCharge}-\eqref{eq:B:requirement}, that is $-9 X_q = X_{e}+X_{\mu}+X_{\tau}$. The coefficients $\{X_{e},X_{\mu},X_{\tau}, X_{N_1},X_{N_2},X_{N_3}  \}$ in Eq.~\eqref{eq:Tx} need to satisfy the Diophantine equations~\cite{Dobrescu:2020evn,Allanach:2018vjg}
\beq
\label{eq:nu:dioph}X_e+X_\mu+X_\tau=\sum_i X_{N_i} ~, \qquad X_e^3+X_\mu^3+X_\tau^3=\sum_i X_{N_i}^3 ~.
\eeq

We group the solutions to the above equations in three non-exclusive classes, up to arbitrary permutations in flavor indices $\{e,\mu, \tau\}$ and $\{N_1, N_2, N_3\}$,
\begin{align}
\text{Class 1}:& \qquad X_e = X_{N_1}, \quad X_\mu = X_{N_2} , \quad X_\tau = X_{N_3},
 \label{eq:class_1}\\
\text{Class 2}: & \qquad X_e=X_{N_1}, \quad X_\mu=-X_\tau, \quad X_{N_2}=- X_{N_3},
\label{eq:class_2} \\
\begin{split}
\text{Class 3}: & \qquad  \text{the rest.}
\end{split}
\end{align}
This generalizes the results from Ref.~\cite{Altmannshofer:2019xda}, which mainly considers Class 1 solutions (but also explores the use of a Class 3 solution to introduce LFV in the neutrino sector).
The Class 1 and Class 2 solutions are three-parameter family of solutions, with $\{X_e, X_\mu, X_\tau\}$ and $\{X_e, X_\mu, X_{N_2}\}$ taken as free parameters in Eqs. \eqref{eq:class_1} and \eqref{eq:class_2}, respectively. The solutions that have $X_\mu=X_{N_2}=-X_\tau=-X_{N_3}$ are both of Class 1 and Class 2.

Scanning over the general results from Ref.~\cite{Allanach:2018vjg} (see also~\cite{Costa:2019zzy}) for the anomaly-free $\U(1)_X$ extensions of the SM, we find the charge assignments that satisfy the anomaly conditions~\eqref{eq:1}--\eqref{eq:6}, the charged lepton condition \eqref{eq:charged:lep:req}, the muoquark conditions \eqref{eq:S3quark1}--\eqref{eq:S3quark2},
and have the ratio between the largest and the smallest nonzero charge magnitudes 10 or less. There are in total 252 for Eq.~\eqref{eq:S3quark2:second} and 251 for Eq.~\eqref{eq:S3quark2:first}. Out of these 77 (or 76) belong to Class 1, 185 to Class 2
(with 12 both in Class 1 and Class 2),
while there are two exceptional (Class 3) charge assignments. These are (up to flavor permutations),
\begin{subequations}
\begin{align}
\{X_e, X_\mu, X_\tau\}&=\{-5,-1,6\}, & \{X_{N_1}, X_{N_2}, X_{N_3}\}&=\{-3,-2,5\},
\\
\{X_e, X_\mu, X_\tau\}&=\{-3,-2,5\}, & \{X_{N_1}, X_{N_2}, X_{N_3}\}&=\{-5,-1,6\}.
\end{align}
\end{subequations}
Class~1 and~2 models can be considered to be vector-like solutions to the Diophantine Eq.~\eqref{eq:nu:dioph}, in the usual physics nomenclature from anomaly cancellations. Indeed, any numbers satisfying Eqs.~\eqref{eq:class_1}--\eqref{eq:class_2} automatically satisfy the Diophantine equations \eqref{eq:1}--\eqref{eq:6}. Upon relaxing our search requirement $ |X_{F_i}| \leq 10 $, all solutions of Class~1 and~2 are parameterized by three arbitrary integers $ (X_e,\, X_\mu,\, X_\tau)  $ and $ (X_e,\, X_\mu,\, X_{N_2}) $, respectively.
Class~3 models, corresponding to chiral solutions of the Diophantine equation, are not easily parameterized beyond $ |X_{F_i}| \leq 10 $. However, given some effort this problem has been solved~\cite{Allanach:2020zna} (see also~\cite{Costa:2019zzy}).

\begin{table}[t]
\vspace{0.2cm}
\begin{center}{
\begin{tabular}{ccccccc}
\hline\hline
$X_{L_1}$ & $X_{L_2}$ & $X_{L_3}$ & $b_{E_1}$ & $X_{N_1}$       & $X_{N_2}$ & $X_{N_3}$
\\
\hline
$-1$ & $-1$ & $2$ &  $-1$ & $-2$ & $-1$ & $3$
\\
$-6$ & $-1$ & $7$ &  $-1$ & $-7$ & $-2$ & $9$
\\
$-5$ & $-2$ & $7$ &  $-2$ & $-7$ & $-3$ & $10$
\\
$-3$ & $-2$ & $5$ &  $-2$ & $-9$ & $-1$ & $10$
\\
$-5$ & $-1$ & $6$ &  $-1$ & $-9$ & $-1$ & $10$
\\
$-2$ & $5$ & $6$ &  $2$ & $1$ & $3$ & $5$
\\
$-1$ & $5$ & $5$ &  $2$ & $1$ & $3$ & $5$
\\
$-3$ & $5$ & $7$ &  $2$ & $1$ & $3$ & $5$
\\
$0$ & $4$ & $5$ &  $1$ & $1$ & $3$ & $5$
\\
$2$ & $2$ & $5$ &  $-1$ & $1$ & $3$ & $5$
\\
$-4$ & $5$ & $8$ &  $2$ & $1$ & $3$ & $5$
\\
$-3$ & $4$ & $8$ &  $1$ & $-1$ & $3$ & $7$
\\
$-2$ & $4$ & $7$ &  $1$ & $-1$ & $4$ & $6$
\\
$-3$ & $6$ & $6$ &  $3$ & $3$ & $3$ & $3$
\\
$1$ & $1$ & $7$ &  $-2$ & $-3$ & $5$ & $7$
\\
$-2$ & $2$ & $9$ &  $-1$ & $-6$ & $5$ & $10$
\\
$-1$ & $2$ & $8$ &  $-1$ & $-6$ & $7$ & $8$
\\
$3$ & $5$ & $10$ &  $-1$ & $2$ & $6$ & $10$
\\
\hline\hline
\end{tabular}}
\end{center}
\caption{Charge assignments (up to family permutations) for $b_{E_1}=-b_{E_2}/2=b_{E_3}$ case (Category II) with maximal ratio of nonzero charges less or equal to 10. See Section~\ref{sec:not:all} for details.  \label{tab:charge:assign}}
\end{table}

\subsubsection{Chiral category $U(1)_X$ charge assignments}
\label{sec:not:all}

There are additional 21 charge assignments for $X_{F_i} \in [-10,10]$ for which the right-handed $e$, $\mu$ and $\tau$ charges change to
\begin{align}
X_{E_1} &= X_{L_1} + b_{E_1},
\\
X_{E_2} &= X_{L_2} + b_{E_2},
\\
X_{E_3} &= X_{L_3} + b_{E_3}.
\end{align}
The universal quark charge is  $-9 X_q = X_{L_1}+X_{L_2}+X_{L_3}$, while the right-handed neutrino charges are $X_{N_1}$, $X_{N_2}$ and $X_{N_3}$.
 The 18 solutions that have
\beq
b_{E_1}=-b_{E_2}/2=b_{E_3},
\eeq
are listed in Table \ref{tab:charge:assign} (up to flavor permutations).  The remaining three solutions are
\begin{align}
\{X_{L_{1,2,3}}\}=&\{-7,0,7\}, &\{b_{E_{1,2,3}}\}&=\{-1,3,-2\},&\{X_{N_{1,2,3}}\}&=\{-5,-3,8\},
\\
\{X_{L_{1,2,3}}\}=&\{-5,-3,8\}, &\{b_{E_{1,2,3}}\}&=\{-2,3,-1\}, &\{X_{N_{1,2,3}}\}&=\{-6,-4,10\},\\
\{X_{L_{1,2,3}}\}=&\{-5,6,8\}, &\{b_{E_{1,2,3}}\}&=\{1,-3,2\},~&\{X_{N_{1,2,3}}\}&=\{0,3,6\}.
\end{align}
These solutions are particularly interesting as they facilitate models in which the  muon mass and the $(g-2)_\mu$ are both generated at one-loop order (see Section~\ref{sec:radiative}).

\subsection{Third-family-quark $\U(1)_X$}
\label{sec:3rdFam}

One can relax the assumption of universal $\U(1)_X$ charges for quarks,  Eq.~\eqref{eq:B:requirement}, and instead allow for family-dependent quark charges.  The quark Yukawa matrices $Y_u^{i j}$ and $Y_d^{i j}$ are then no longer arbitrary $ 3 \times 3 $ complex matrices but, rather, have a texture restricted by the gauge symmetry.
The ``$2+1$'' quark charge assignment is particularly well-motivated by phenomenology. In this case, the $\U(1)_X$ charge of the third quark family differs from that of the first two families, the latter still taken to be universal:
\beq
\begin{split}\label{eq:3rdcondt}
    X_{Q_i} &= X_{U_j} = X_{D_k} \equiv X_{q_{12}} \quad \text{for all}\quad i,j,k=1,2, \quad \text{and}
    \\
    X_{Q_3} &= X_{U_3} = X_{D_3} \equiv X_{q_{3}}~.
\end{split}
\eeq
The anomaly cancellation conditions~\eqref{eq:1}--\eqref{eq:6} are identical to the quark flavor-universal case (Section~\ref{sec:qfu}) provided that
\beq
2 X_{q_{12}} + X_{q_{3}} = 3 X_{q}~,
\eeq
where $X_{q}$ is defined in Eq.~\eqref{eq:B:requirement}. The quark flavor-universal solutions found in Section~\ref{sec:qfu} can, therefore, immediately be extended to the $2+1$ case. Each flavor-universal solution generates a one-parameter family of $2+1$ charge assignments. $X_{q_{3}}$ can be taken as a free parameter, while $X_{q_{12}}$ is set to $X_{q_{12}} =(3 X_{q} - X_{q_{3}})/2$, with $X_q$ the flavor-universal quark charge assignment for a given solution listed in Section~\ref{sec:qfu}. In the phenomenological studies (Section~\ref{sec:benchmarks}), we will focus on the scenario where $X_{q_{12}} = 0$ and $ X_{q_{3}} = 3 X_{q} \neq 0 $.

The non-Abelian accidental symmetry of the renormalizable Lagrangian without Yukawa interactions is the $\SU(2)_Q \times \SU(2)_U \times \SU(2)_D$ flavor symmetry, under which the first two generations transform as doublets, while the third generation is a singlet~\cite{Barbieri:2011ci,Kagan:2009bn}. As discussed in the literature~\cite{Barbieri:2011ci,Kagan:2009bn,Fuentes-Martin:2019mun}, the minimal set of the symmetry-breaking spurions capable of explaining the observed quark masses and the CKM mixing matrix consists of a doublet $V \sim (\bf{2},\bf{1},\bf{1})$ and two bidoublets $\Delta_{U} \sim (\bf{2},\bf{\bar 2},\bf{1})$ and $\Delta_{D} \sim (\bf{2},\bf{1},\bf{\bar 2})$. For the $2+1$ charge assignments, the bidoublet spurions are allowed in the Yukawa interactions already at the dimension-4 level. The doublet $V$ is generated only at the dimension-5 level,
\beq \label{eq:dim5yuk}
\mathcal{L} \supset \frac{x_i^u}{\Lambda} \overline{Q}_i \tilde H \phi U_3 + \frac{x_i^d}{\Lambda} \overline{Q}_i H \phi D_3 \hc ,
\eeq
after the $\U(1)_X$-symmetry-breaking scalar $\phi$ gets a VEV (here $ X_\phi = -X_{q_3} $ and $i=1,2$). If there is a hierarchy between the $\phi$ VEV and the masses of integrated modes, $ \langle \phi \rangle / \Lambda \ll 1 $, this can explain the smallness of the CKM parameters $V \approx (V_{ts}, V_{td})^T$. As detailed in the next section, the light $X_\mu$ solution of the $(g-2)_\mu$ anomaly predicts $ \langle \phi \rangle $ close to the EW scale and, therefore, $x^{u}_2 / \Lambda \approx V_{ts} / \langle \phi \rangle \sim (10$~TeV$)^{-1}$.

A simple UV completion of the operators in Eq.~\eqref{eq:dim5yuk} is to integrate out at tree-level heavy vector-like quarks in the gauge representations of the right-handed up and down quarks. For instance, integrating out vector-like $\Psi \sim (\rep{3},\rep{1},\tfrac{2}{3},0)$ would generate the first operator in \eqref{eq:dim5yuk} but not the second, leading to a down-alignment. This alignment is favored by the strong constraints on the flavor-changing neutral current (FCNC) processes in the down-quark sector, see e.g. Ref.~\cite{Fuentes-Martin:2019mun}.

The down-alignment is also useful in the presence of a light flavor-violating mediator. The non-universal quark charges, after flavor rotations to the mass basis, lead to flavor-violating $X_\mu$ couplings.  Decays such as $B \to K X$ impose very stringent limits on a sub-GeV vector boson $X_\mu$, see Eq.~\eqref{eq:gbs_bound}. These can be avoided in the down-alignment scenario ($ x_i^d = 0 $ and $x_i^u \neq 0$). In this case, the $3 \times 3$ down quark Yukawa matrix is a sum of a $2 \times 2$ matrix for the light quarks and a single entry for the third generation.
The $X_\mu$ couplings within these two subspaces are proportional to the identity matrix and are not affected by flavor rotations required for down quark mass diagonalization.
In contrast, the $X_\mu$ couplings to the up quarks receive off-diagonal entries after up-quark mass diagonalization. Due to the residual $\U(2)$ protection, the $Xcu$ coupling has the CKM suppression $V_{cb}^* V_{ub}$. The relevant observables are: $D \to \pi X$ (below the dimuon threshold $X$ mostly decays to neutrinos) and $D - \bar D$ mixing. We will derive the bounds in Section~\ref{sec:9B3}.

Finally, the muoquark conditions from Section~\ref{sec:qfu} change with non-universal quark charges. Motivated by the flavor anomalies, we choose the leptoquark charge $X_{\rm{LQ}}$ to allow for the leptoquark couplings with the third-generation quarks and second-generation leptons at the renormalisable level while preventing all other leptoquark and diquark couplings.\footnote{The choice of the third quark generation is indeed advantageous in many ways. For instance, the one-loop leptoquark contribution to the $(g-2)_\mu$ is enhanced when the top quark is running in the loop (Section~\ref{sec:radiative}).}

Let us consider as an illustration a scalar leptoquark $S_3 \equiv (\repbar{3},\rep{3},1/3, X_{S_3})$. Assuming $X_{q_3} \neq 0$ and $X_{q_{12}} = 0$, we further require:
\begin{enumerate}[i)]
\item the interaction $Q_3 L_2 S_3$ is allowed,
\item $Q_3 L_{1,3} S_3$, $Q_{1,2} L_{1,3} S_3$, and $Q_{1,2,3} Q_{1,2,3} S^\dagger_3$ are forbidden.
\end{enumerate}
For this to be true, the following set of conditions needs to be satisfied
\beq\label{eq:MuCondS3}
X_{L_2} \neq \{ X_{L_{1,3}}, X_{L_{1,3}} - X_{q_3}, - X_{q_3}, - 2 X_{q_3}, -3 X_{q_3} \}~.
\eeq
These criteria are met by 171 inequivalent sets of $X_{F_i}$ charges in the range $[-10, 10]$ out of which 158 belong to the vector category (cf. Eq.~\eqref{eq:charged:lep:req}) and 13 are in the chiral category. Consider for example the sub-GeV $X_\mu$ vector boson of the gauged $\U(1)_X$ with:
\begin{align} \label{eq:3rd_family_charges}
X_{q_{12}} = 0~,~X_{q_3} &= -3~,\nonumber\\
X_{L_{1,2,3}} = X_{E_{1,2,3}} = X_{N_{1,2,3}} &= \{0,1,8 \},~\{0,2,7 \},~\{0,4,5 \},\textrm{or}~\{0,-1,10 \}.
\end{align}
These benchmarks satisfy Eq.~\eqref{eq:MuCondS3} and do not couple $X$ to electrons or valence quarks.

The $S_3$ muoquark at tree-level contributes to $b \to s \mu \mu$ decays and can fit the data well, see e.g. Ref.~\cite{Greljo:2021xmg}. The coupling to the strange quark is generated in a way similar to the CKM matrix, i.e., by a dimension-5 operator
\beq \label{eq:dim5yLQ}
\mathcal{L} \supset \frac{z_i^u}{\Lambda} \overline{Q}_i^C L_2 S_3 \phi^\dagger \hc ,
\eeq
where $i=1,2$. This operator is allowed by gauge symmetry despite the $\U(1)_X$ charges already being fixed by Eqs.~\eqref{eq:dim5yuk} and \eqref{eq:MuCondS3}. The simplest way to generate this operator without spoiling the down-alignment of $X_\mu$ interactions is to introduce a vector-like lepton $\chi \sim \left(\rep{1},\rep{2},-\tfrac{1}{2},- X_{S_3}\right)$. More precisely, the interactions $\overline{Q}^C_{1,2} \chi_L S_3$ and $\bar \chi_R \phi^\dagger L_2$ generate the operator in Eq.~\eqref{eq:dim5yLQ} when the $\chi$ field gets integrated out at tree-level.

\section{Light $X_\mu$ phenomenology}
\label{sec:pheno}

When the $\U(1)_X$ gauge boson $X_\mu$ is light,
it can give the dominant new physics contribution to  $(g-2)_\mu$ and potentially resolve the discrepancy between the measurements and the SM prediction, see e.g.~\cite{Jegerlehner:2009ry}.
In this section we show that the $(g-2)_\mu$ anomaly can indeed be
explained, without violating other experimental constraints, by a  sub-GeV vector boson $X_\mu$ in a broad class of $\U(1)_X$ gauge models. The $\U(1)_X$ models that we consider all admit the muoquark solution of the rare $B$ decay anomalies in the $R_{K^{(*)}}$ ratios and $b \to s \mu \mu$ angular distributions and branching ratios.

 The model independent discussion in Sections~\ref{sec:g-2}, \ref{sec:trident}, and \ref{sec:Borexino_lightX} is limited to the flavor-conserving $X_\mu$ couplings applicable for the $\U(1)_X$ gauge models. Section~\ref{sec:single_mediator} contains also a short discussion of challenges facing a light vector boson that would be flavor-violating~\cite{Altmannshofer:2016brv}.
 The main goal of Section~\ref{sec:single_mediator} is to show that a single light $X_\mu$ cannot simultaneously resolve both the $(g-2)_\mu$ and rare $B$ decay anomalies, and therefore another heavy mediator such as $S_3 \equiv (\repbar{3},\rep{3},1/3, X_{S_3})$ is required.
 Section~\ref{sec:benchmarks} then contains several benchmark models that can solve both the $(g-2)_\mu$ and rare $B$ decay anomalies, with most of the phenomenological discussion focused on $X_\mu$ while for the $S_3$ phenomenology we refer to~\cite{Gherardi:2020qhc,Greljo:2021xmg,Davighi:2020qqa,Buttazzo:2017ixm,Angelescu:2021lln}.

\subsection{Muon $(g-2)_\mu$}
\label{sec:g-2}

Recently, the Muon $g-2 $ collaboration at the Fermilab announced its first preliminary measurement of the muon anomalous magnetic moment~\cite{Abi:2021gix,Albahri:2021kmg,Albahri:2021ixb}, consistent with the BNL result~\cite{Bennett:2006fi}.
The combination of the two experimental results ($a_\mu^{\rm avg}$) differs by $4.2\sigma$~\cite{Abi:2021gix} from  the SM prediction ($a^{\rm SM}_\mu$),\footnote{Obtained from the measurements of $R$-ratios, see~\cite{Aoyama:2020ynm} and reference therein. This prediction is supported by the electroweak precision tests, where the same $e^+e^-\to$hadrons data is used to calculate $\alpha(m_Z)$~\cite{Crivellin:2020zul,Keshavarzi:2020bfy}.}
\begin{align}
    \label{eq:anu_exp}
    \Delta a_\mu^{R}
=   a^{\rm avg}_\mu - a^{\rm SM}_\mu
=   \left( 251 \pm 59 \right) \times 10^{-11} \,.
\end{align}
The SM prediction obtained on the lattice by the BMW collaboration~\cite{Borsanyi:2020mff}, on the other hand, differs by only $1.6\,\sigma$ from the experimental average, and is  thus consistent with $\Delta a_\mu =0$.
In the numerical studies we use Eq.~\eqref{eq:anu_exp} with the caveat that the situation calls for further studies of the SM prediction.

The light gauge boson $X_\mu$ that couples to muons,
\beq
    \L \supset
    g_X\, \overline{\mu}\, \slashed X (q_{V} - q_{A} \gamma_5) \mu~,\label{eq:lag-gm2}
\eeq
gives the following 1-loop contribution to the muon anomalous magnetic moment, see, \textit{e.g.}, Refs.~\cite{Jackiw:1972jz,Jegerlehner:2009ry},
\beq
   \begin{split}
    \label{eq:amu_NP}
    \Delta a_\mu
=   &  \dfrac{g_X^2 }{8\pi^2}  r_\mu^2
    \left[q_V^2\, I_V (r_{\mu})+q_A^2\, I_A (r_{\mu})\right] =  \frac{g_X^2 }{8\pi^2}
    \left\{
    \begin{matrix}
    q_V^2 -2\, r_\mu^2\, q_A^2,  & \qquad m_X \ll m_\mu
    \\
    \frac{2}{3} r_\mu^2 \left[  q_V^2 - 5\, q_A^2 \right],
    & \qquad m_X \gg m_\mu
    \end{matrix}
    \right. \, ,
\end{split}
\eeq
where
$r_\ell=m_\ell/m_X$
and
\begin{align}
    I_V(r_\ell) &= \int_0^1 \!\! \dd x \frac{2 x^2 (1-x)}{1-x + r_\ell^2 x^2}, &
    I_A(r_\ell) &= -\int_0^1 \!\! \dd x \frac{2 x (1-x) (4-x) + 4r_\ell^2 x^3}{1-x + r_\ell^2 x^2}
\end{align}

Both $I_V(r_\ell)$ and $I_A(r_\ell)$ are monotonic functions of $m_X$ such that
the vector\,(axial) contributions to $a_\mu$ are always positive\,(negative).
Thus, in order to account for the central value of $\Delta a_\mu$  in Eq.~\eqref{eq:anu_exp},  the vector coupling needs to be nonzero, $q_V\ne0$.
Numerically,
\beq
    \label{eq:gXqV}
    g_X
=   \left( \frac{\Delta a_\mu}{251 \times 10^{-11}} \right)^{1/2}
    \left\{\begin{matrix}
    4.5 \times 10^{-4}
    \big[ q_V^2 -2\,q_A^2\, r_\mu^2\big]^{-1/2},&\qquad
    m_X \ll m_\mu, \\
    5.5 \times 10^{-4} r_\mu^{-1/2} \big[q_V^2-5\,q_A^2\big]^{-1/2}, &\qquad
    m_X \gg m_\mu.
    \end{matrix} \right.
\eeq
For $m_X$ below or comparable with the muon mass, the gauge coupling required to explain the observed $\Delta a_\mu$ is $g_X\sim {\mathcal O}(10^{-4})$. To get the correct sign,
Eq.~\eqref{eq:gXqV} implies that $X_\mu$ needs to predominantly couple vectorially, with the axial-to-vector ratio of couplings required to be below $\big|q_A/q_V\big| <  m_X/(\sqrt{2}m_\mu)\ll1$ when $m_X \ll m_\mu$, and below  $\big|q_A/q_V\big| < 1/\sqrt{5}\,$  when $m_\mu \ll m_X$.

\subsection{Neutrino trident production}
\label{sec:trident}
The neutrino trident production provides important bounds on the muonic force explanations of the $(g-2)_\mu $ anomaly~\cite{Altmannshofer:2014pba}.
The muon neutrino and the left-handed muon form an electroweak doublet and thus share the same coupling to $X_\mu$, proportional to $\propto (q_V + q_A) $.
Since any explanation of the $(g-2)_\mu $ anomaly must be mostly vectorial (see Eq.~\eqref{eq:gXqV}), a flavor diagonal explanation of the $(g-2)_\mu$ anomaly necessarily implies NP contributions to the trident production  $ \nu_\mu N \to \nu_\mu N \mu^+ \mu^- $ induced by the $\nu_\mu$ neutrino  scattering on nucleus $N$.
The strongest bound on the NP contributions to the trident production cross section is from the CCFR experiment that reported $ \sigma_\sscript{CCFR}/ \sigma_\sscript{SM} = 0.82\pm 0.28$ ~\cite{Mishra:1991bv}, where $ \sigma_\sscript{CCFR}$ is the measured cross section and $\sigma_\sscript{SM} $ the SM prediction. For comparison, we also show in Fig.~\ref{fig:qAqV}  a weaker constraint from the NuTeV experiment~\cite{NuTeV:1999wlw}. Note that the NuTeV analysis claims to have identified an additional background that was not included by the CCFR.

The calculation of the trident production cross section in the presence of NP was performed in Ref.~\cite{Altmannshofer:2014pba} (see also~\cite{Altmannshofer:2019zhy,Ballett:2019xoj,Ballett:2018uuc}), by considering the scattering of neutrinos on the  potential photons sourced by the nucleus, $ \nu_\mu \gamma \to \nu_\mu \gamma \mu^+ \mu^- $.
We use the public code of Ref.~\cite{Altmannshofer:2019zhy} to calculate the SM and NP cross sections, including the kinematical cuts~\cite{Mishra:1991bv}.
The public code was modified to include the light spin-1 with both vectorial and axial-vector couplings. Only the dominant flux ($\nu_\mu$) is used.

\subsection{Other constraints}
\label{sec:Borexino_lightX}

\subsection*{Borexino}
\label{sec:Borexio}
The Borexino experiment measured a cross section for the elastic scattering of $^{7}$Be  solar neutrinos on electrons~\cite{Bellini:2011rx,Borexino:2017rsf}. Because of neutrino oscillations the flux on Earth is composed of all three neutrino flavors incoherently scattering on electrons. The tree-level $X_\mu$ exchanges can modify the scattering rate from the SM expectation.
Since no deviation was observed, this implies bounds on $X_\mu$ couplings to fermions that are due to a combination of direct $ \U(1)_X $ charges and induced couplings from kinetic mixing of $X_\mu$ with the photon (in particular to the electron).
For the numerical estimates we closely follow the analysis in Ref.~\cite{Altmannshofer:2019zhy}.
The bound becomes stronger if $X_\mu$ also couples to tau and electron neutrinos in addition to the muon neutrino and becomes weaker if the direct coupling of $X_\mu$ to the electron cancels against the kinetic-mixing-induced one.

\subsection*{Light resonance searches}
\label{sec:lightX}

New light resonances can be probed by a number of intensity frontier experiments, summarized, e.g., in Refs.~\cite{Ilten:2018crw,Bauer:2018onh}.
In the numerical estimates we mostly use \textsc{DarkCast}~\cite{Ilten:2018crw} to recast the existing and future projections of dark photon bounds. The $X\mu\mu$ coupling is currently probed by NA62 ($K\to \mu\nu X$ decays)~\cite{Krnjaic:2019rsv,NA62:2021bji}, and BaBar (searches for $X\to\mu\mu$ decays in the $4\mu$ final state)~\cite{BaBar:2016sci}.
In case of couplings to baryon number and/or to electron via kinetic mixing there are additional bounds from the LHCb dark photon searches~\cite{Ilten:2016tkc,LHCb:2017trq,LHCb:2019vmc}, NA64~\cite{Banerjee:2019pds}, BaBar~\cite{BaBar:2014zli} and NA62 (invisible $\pi^0$ decays)~\cite{NA62:2019meo}.

For future projections on the sensitivity to $X\mu\mu$ coupling we consider  NA64$_\mu$~\cite{Chen:2018vkr,Batell:2016ove,Gninenko:2014pea,NA64:2016oww}, M$^3$~\cite{Kahn:2018cqs}, Belle-II~\cite{Jho:2019cxq}, NA62~\cite{Krnjaic:2019rsv} and ATLAS~\cite{Galon:2019owl}.
For other couplings (e.g. to hadrons) we also consider the projections for LHCb from Ref.~\cite{Ilten:2016tkc}.

\subsection*{Astrophysics and cosmology}
\label{sec:astro_cosmo}

The parameter region of interest easily passes the astrophysical and cosmological constraints. The $ X_\mu $ decays to neutrinos before the onset of BBN for $m_X \gtrsim 6$\,MeV~\cite{Kamada:2018zxi}.
The potential supernova 1987A limits discussed in~\cite{Croon:2020lrf} (see~\cite{Bar:2019ifz} for the robustness of the bound) apply to much smaller couplings in the considered mass range. That is, the values of $g_X$ relevant for the explanation of the $(g-2)_\mu$ anomaly lead to $X_\mu$ trapping in the proto-neutron star and, therefore, to no cooling constraints.

\subsection*{Non-standard neutrino interactions}
\label{sec:NSI}
Neutrino non-standard interactions (NSI) change the oscillation of the neutrinos as they propagate through matter via the MSW mechanism~\cite{Wolfenstein:1977ue,Mikheyev:1985zog}. Such changes, in turn, influence the global fit to oscillation data and constrain the size of the interactions~\cite{Esteban:2018ppq}.
\begin{equation} \label{eq:eff_neutrino_matter_int}
    \mathcal{L}_\sscript{NSI} = -\frac{G_F}{2\sqrt{2}} \sum_{f,\alpha\beta} \varepsilon_{\alpha \beta}^f (\overline{f} \gamma_\mu f) (\overline{\nu}_\alpha P_\LL \nu_\beta)
\end{equation}
The oscillation is decided by the forward scattering limit, where an effective theory description is suited even for small $ X$ masses. The couplings will then depend on the charges of the matter fields, $ f=\{e, p, n\} $, and neutrinos, as well as the $ g_X/ m_X$ setting the overall strength of the interactions.
Since ordinary matter (atoms) are electrically neutral, any contribution to the interactions from kinetic mixing with the photon cancels in the end~\cite{Heeck:2018nzc}. With that in mind, we use the bounds of Ref.~\cite{Coloma:2020gfv} directly to constrain NSI in our benchmark models irrespective of the kinetic mixing.

Other bounds on neutrino interactions stem from the coherent scattering on nuclei~\cite{Freedman:1973yd,Drukier:1984vhf}, which was observed by the COHERENT experiment~\cite{COHERENT:2017ipa}. Since the momentum transfer of the neutrinos in COHERENT is $\sim \SI{50}{MeV} $, right in the middle of the viable $ m_X $ window for a \gmu explanation, the effective interaction~\eqref{eq:eff_neutrino_matter_int}, which was considered e.g. in \cite{Altmannshofer:2018xyo,Heeck:2018nzc,Coloma:2019mbs,Denton:2020hop},
overestimates the NP contribution to the cross section.
Therefore, we determine the COHERENT bounds on a dynamical light vector boson. To this end, we have implemented the relevant contributions into the public Python code \texttt{7stats}~\cite{Denton:2020hop}.
This code includes the COHERENT CsI data~\cite{COHERENT:2017ipa,COHERENT:2018imc} and provides a $\chi^2$-function that we use to set bounds on our benchmark models.

\subsection{A single mediator for both $(g-2)_\mu$ and rare $B$ decay anomalies?}
\label{sec:single_mediator}

An interesting question is whether a single mediator could be responsible for both the $(g-2)_\mu$ and rare $B$ decay anomalies.
Here we explore to what extent this is possible when the mediator is a neutral spin-1 boson, $X_\mu$.
We keep the discussion quite general so that the results also apply to the different $\U(1)_X$ models.

We construct an EFT with the dynamical gauge field $X_\mu$ and the $\U(1)_X$ symmetry-breaking scalar $\phi$, while the rest of the BSM spectrum is integrated out. In particular, the quark flavor-violating $X b s$ interaction, needed to explain the anomalies in rare $B$ decays, is a result of some unspecified short-distance physics (e.g., from integrating out heavy vector-like fermions). The relevant effective $X_\mu$ interactions are  given by
\begin{align}
  \label{eq:Lsimp}
   \L_{\rm eff}
    \supset &+ g_X\, (q_{V} + q_{A})\, \overline{\nu}_{\mu L}\, \slashed X\, \nu_{\mu L} +
    g_X\, \overline{\mu}\, \slashed X\, (q_{V} - q_{A}\, \gamma_5)\, \mu \nonumber\\
   &+ \left[ \overline{b}\, \slashed X\, (g_{L}^{bs}\,P_L + g_{R}^{bs}\, P_R)\, s \hc \right] \, ,
\end{align}
extending the effective Lagrangian in Eq.~\eqref{eq:lag-gm2}.
For $\U(1)_X$ models, the second line can arise from higher-dimension gauge invariant operators, for example from $(\phi^\dagger D_\mu \phi) (\overline{Q}_2 \gamma^\mu Q_3), \ldots$, after the breaking of $\U(1)_X$ by the $\phi$ VEV. The flavor-diagonal couplings to muons and muon neutrinos, on the other hand, occur in $\U(1)_X$ models directly from charging them under the $ \U(1)_X $ group.

The EFT in Eq.~\eqref{eq:Lsimp} now allows us to address the central question of this subsection: Can the anomalies in both $(g-2)_\mu$ and rare $B$ decays be explained by one-loop and tree-level exchanges of $X_\mu$, respectively, or is an additional short-distance contribution needed?

\begin{figure}[t]
	\centering
	\includegraphics[width=\columnwidth]{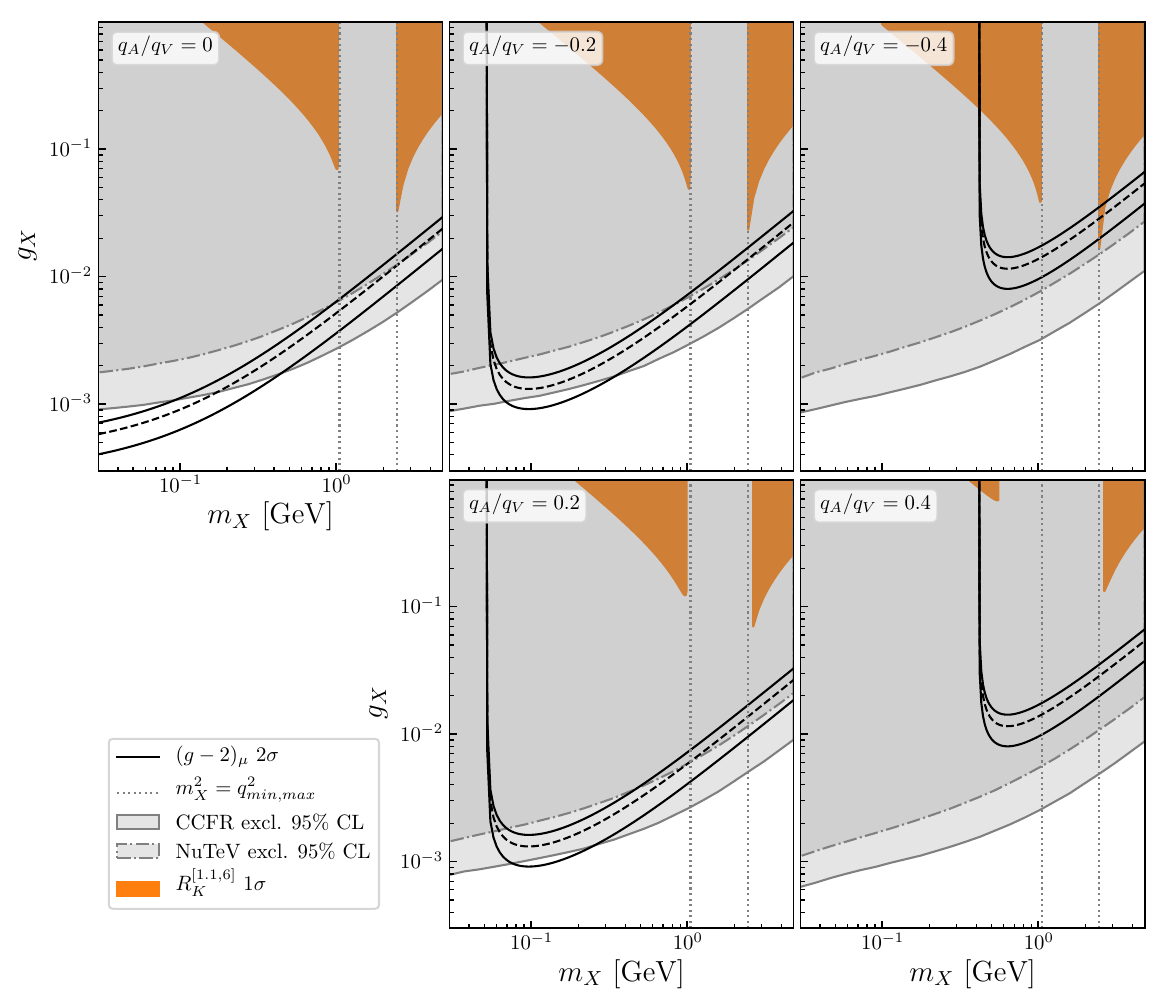}
	\caption{Parameter space for the light $ X_\mu $ solution to the $ (g-2)_\mu $ anomaly with the $2
	\sigma$  region and the central values in solid and dashed black, respectively.
	The gray region is excluded by the neutrino trident production (see Section~\ref{sec:single_mediator}). In the orange region, the LHCb measurement of $R_K^{[1.1,6]}$ can be explained at the $1 \sigma$ level while satisfying the constraint on $q_{L,R}^{bs}$ from $\text{BR}(B\to K\nu\nu)$. } \label{fig:qAqV}
\end{figure}

The NP effects that could explain \RK and the $b\to s\mu\mu$ anomalies can be expressed in terms of the modified Wilson coefficients $C_9$ and $C_{10}$ that we define as in Ref.~\cite{Altmannshofer:2021qrr}.
The tree-level $X_\mu$ exchanges contribute as (cf.~\cite{Sala:2017ihs})
\begin{align}\label{eq:WCs_q2}
    C_{9,10}^{(\prime)}
=  \frac{q_{V,A}}{N} \,\frac{g_X\, g_{L(R)}^{bs}}{q^2 - m_X^2 + i m_X\Gamma_X}\,,
\end{align}
where $N=\frac{G_F\,\alpha}{\sqrt{2}\,\pi}V_{tb}V_{ts}^*$ is a normalization factor.
The $Xbs$ couplings $g_{L(R)}^{bs}$ are stringently constrained, e.g.\ by $\text{BR}(B\to K\nu\nu)$ if $X_\mu$ is lighter than the $B$ meson or by the $B_s$ neutral meson mixing.
Since right-handed quark couplings are not needed to explain the anomalies in rare $B$ decays (cf.\ e.g.~\cite{Altmannshofer:2021qrr}), in the following we set $g_{R}^{bs}=0$.
Here we focus on light $X_\mu$ with $m_X<m_B-m_K$, such that the most important constraint on $g_{L}^{bs}$ comes from $\text{BR}(B\to K\nu\nu)$~\cite{Belle-II:2021rof,Belle:2017oht,BaBar:2013npw,Belle:2013tnz}.
For $\text{BR}(X\to\nu\nu)\simeq1$, this bound is given by~\cite{Sala:2017ihs}
\begin{equation}\label{eq:gbs_bound}
 g_{L}^{bs} \lesssim  0.7 \times 10^{-8}\, \frac{m_X}{\text{GeV}}\,.
\end{equation}
Consequently, an explanation of \RK and the $b\to s\mu\mu$ anomalies requires sizable $X \mu\mu$ couplings.

Let us now consider the one-loop contributions to $(g-2)_\mu$  from $X_\mu$ and either $e$, $\mu$ or $\tau$ running in the loop (forgetting for the moment about UV completions). For $e$ and $\tau$ running in the loop, the $X_\mu$ couplings are flavor violating, a possibility suggested in~\cite{Altmannshofer:2016brv}. The $(g-2)_\mu$ excess can, in principle, be explained by the flavor-violating $X\mu\tau$ coupling, with $m_X>1.7\,\GeV$.
In contrast, the $Xe \mu$ coupling leads to a negative contributions to $\Delta a_\mu$. For the flavor-conserving option, with $X_\mu$ coupling to muons, inducing the $(g-2)_\mu$ in the loop is viable, as long as vector couplings are larger than axial ones, see the discussion in Section~\ref{sec:g-2}.
If we want to explain $(g-2)_\mu$ and the anomalies in rare $B$ decays at the same time, the requirement of a sizable $X \mu\mu$ coupling to explain the latter precludes an explanation of $(g-2)_\mu$ predominantly through flavor-violating $X \mu\tau$ couplings. The presence of both $X\mu\tau$ and $X\mu\mu$ couplings would lead to a too large $\text{BR}(\tau \to 3 \mu)$ in conflict with experimental bounds~\cite{Hayasaka:2010np}.
This leaves us with the possibility that both the $(g-2)_\mu$ and rare $B$ decay anomalies could be due to flavor diagonal $X\mu\mu$ couplings.

In Fig.~\ref{fig:qAqV} we show the regions in the $m_X$ -- $g_X$ plane, in which a light $X_\mu$ with couplings given by Eq.~\eqref{eq:Lsimp} can explain the measured values of $(g-2)_\mu$ and $R_K$ for different choices of $q_A/ q_V$.\footnote{In Fig.~\ref{fig:qAqV} the $X e e$ couplings were set to zero. However, even if these were allowed to vary, the conclusions would not have changed. The $X \mu\mu$ coupling, needed to explain the $(g-2)_\mu$ anomaly, is expected to be much larger than the size of $Xee$ coupling allowed by  the bounds from the light resonance searches, cf. Section~\ref{sec:lightX}.}
The region where the $(g-2)_\mu$ measurement can be explained at the $2\sigma$ level is represented by solid black lines (dashed black lines give the central values).
The viable region becomes smaller and moves to larger values of $g_X$ as $|q_A/ q_V|$ increases. For $|q_A/ q_V|>1/\sqrt{5}\approx0.45$, no explanation is possible at all (cf. Section~\ref{sec:g-2}).
The orange shaded area shows the $m_X$ and $g_X$ values that can explain the LHCb measurement of $R_K$~\cite{LHCb:2021trn} at the $1\sigma$ level, while $g_L^{bs}$ satisfies Eq.~\eqref{eq:gbs_bound}.
Since $R_K$ is measured in a bin of the invariant dilepton mass squared, $q^2\in[1.1,6]\,\text{GeV}^2$, an $X_\mu$ with $m_X^2\in [1.1,6]\,\text{GeV}^2$ can only enhance $R_K$ if it couples to muons (this is further discussed in Appendix~\ref{app:RKX}). The boundaries of this region are shown in Fig.~\ref{fig:qAqV}, as gray dotted lines. In conclusion, the measured value of $R_K$ cannot be explained for $m_X^2\in [1.1,6]\,\text{GeV}^2$. Outside this mass window, as expected from fits to $R_K$ in the $C_9^\mu$ -- $C_{10}^\mu$ plane (see e.g.~\cite{Altmannshofer:2021qrr}), we find that $R_K$ can be explained for vanishing and negative $q_A/ q_V$ and that sizable positive $q_A/ q_V$ can preclude an explanation.

The most important constraint on the parameter space shown in Fig.~\ref{fig:qAqV} is due to neutrino trident production (cf.~Section~\ref{sec:trident}).
The gray shaded area is excluded at 95\% C.L. by the CCFR measurement (light gray region). A weaker bound from the NuTeV experiment mentioned in Section~\ref{sec:trident} is also shown in darker gray. When the CCFR constraint is satisfied, it is only possible to explain $(g-2)_\mu$ for $|q_A/q_V|\lesssim0.2$ and $m_X\lesssim 400\,\text{MeV}$.
An explanation of $R_K$ for $m_X<m_B-m_K$ is even completely excluded by the CCFR bound.\footnote{
Since the stringent bound on $g_L^{bs}$ from $\text{BR}(B\to K\nu\nu)$ does not apply for $m_X>m_B-m_K$, it is possible to explain $R_K$ in this higher mass region.
}
This, in particular, applies to the small window at $m_X\simeq2.5\,\text{GeV}$ and $q_A/ q_V\simeq-0.4$, where a combined explanation of $(g-2)_\mu$ and $R_K$ would be possible otherwise~\cite{Sala:2017ihs}.
We conclude that a single mediator $m_X$ cannot explain both the $(g-2)_\mu$ and rare $B$ decay anomalies while satisfying all constraints as long as the couplings of left-handed muon and muon neutrino are related by $\SU(2)_L$ gauge invariance.
The lack of a combined solution to the $ b\to s \mu \mu $ and \gmu from a light vector agrees with the findings of Refs.~\cite{Darme:2020hpo,Darme:2021qzw}. These analyses also pointed out that $ \mathrm{BR}(B_s \to \mu\mu) $ effectively rules out $ m_X \lesssim \SI{1.4}{GeV} $ as an explanation of \RK unless the $ Xbs$ coupling is due to an effective dipole operator.

\subsection{Benchmarks}
\label{sec:benchmarks}
For the benchmarks we focus on the extended gauge sector from gauging the $ \U(1)_X $ groups discussed in Section~\ref{sec:models}. After EWSB, the relevant part of the Lagrangian reads
\begin{equation}
\label{eq:L:benchmark}
    \mathcal{L} \supset - \tfrac{1}{4} F_{\mu\nu}^2 - \tfrac{1}{4} X_{\mu\nu}^2 + \tfrac{1}{2} \varepsilon \, F_{\mu\nu} X^{\mu\nu} +  \tfrac{1}{2} m_X^2 X_\mu^2 + e A_\mu J_\sscript{EM}^\mu + g_X X_\mu J_X^\mu,
\end{equation}
where $ J^\mu_{\sscript{EM}, X} $ are the EM current and the current associated with $ \U(1)_X $, respectively. This ignores a small $ X_\mu$--$ Z_\mu $ mixing proportional to the kinetic-mixing parameter $ \varepsilon $, which we will assume to be no larger than $ \mathcal{O}(g_X) $. We relegate further details about the mixing of EW and $ X_\mu$ gauge bosons to Appendix~\ref{app:gauge_mass_basis}.

The $ X_\mu $ couplings to the SM fermions are determined by the form of the current $ J^\mu_{X} $ and by the size of the kinetic mixing parameter $ \varepsilon $.
The $J_X^\mu$ is given by the $\U(1)_X$ SM fermion charges $X_f$,
\beq
J_X^\mu= \sum_f X_f \overline{f} \gamma^\mu f,
\eeq
 and is specific to each model we consider. The value of $\varepsilon$ we treat as a free parameter, with the exception of the $L_\mu-L_\tau$ model, where we follow the literature
and assume that the kinetic mixing is generated exclusively by the mass difference between the tau and the muon running in the loop, and is therefore predicted to be  $ \varepsilon \sim {\mathcal O}( e g_X /16\pi^2)$, see Section \ref{sec:bm3Lmu} for details.
This relies on the assumption that $ \varepsilon $ vanishes at the high scale (as required for gauge coupling unification). Furthermore, for $\U(1)_{L_\mu-L_\tau}$ the kinetic mixing parameter $\varepsilon$ has vanishing $\beta$ function at 1-loop above the EW scale , while the 2-loop contribution is $ (m_\tau^2 - m_\mu^2)/ v^2 $ suppressed. It is thus appropriate to make the approximation that in the $\U(1)_{L_\mu-L_\tau}$ model the kinetic mixing only receives the contributions from muons and taus.
The other benchmark models we consider have a nonzero $\beta$-function for the kinetic mixing parameter $ \varepsilon $. For these models we therefore do not expect to have $ \varepsilon =0 $, except for special tuning points. Typically, it is reasonable to assume $ \varepsilon \lesssim g_X $, which is what we use in the sample plots. The 1-loop running of the $ \varepsilon $ parameter is further analyzed in Appendix~\ref{app:epsilon_running}.

All the models allow for the $S_3$ muoquark to be part of the spectrum. We assume that this is the case, which means that both $(g-2)_\mu$ and the anomalies in rare $B$ decays can be explained simultaneously. In the rest of this section we focus on the part of the phenomenology that is relevant for explaining the $(g-2)_\mu$ anomaly, i.e., on the constraints on the couplings of the light gauge boson, $X_\mu$.

\subsubsection{Gauged $L_\mu-L_\tau$ }
\label{sec:LmuLtau}

\begin{figure}[t]
	\centering
	\includegraphics[width=0.8\columnwidth]{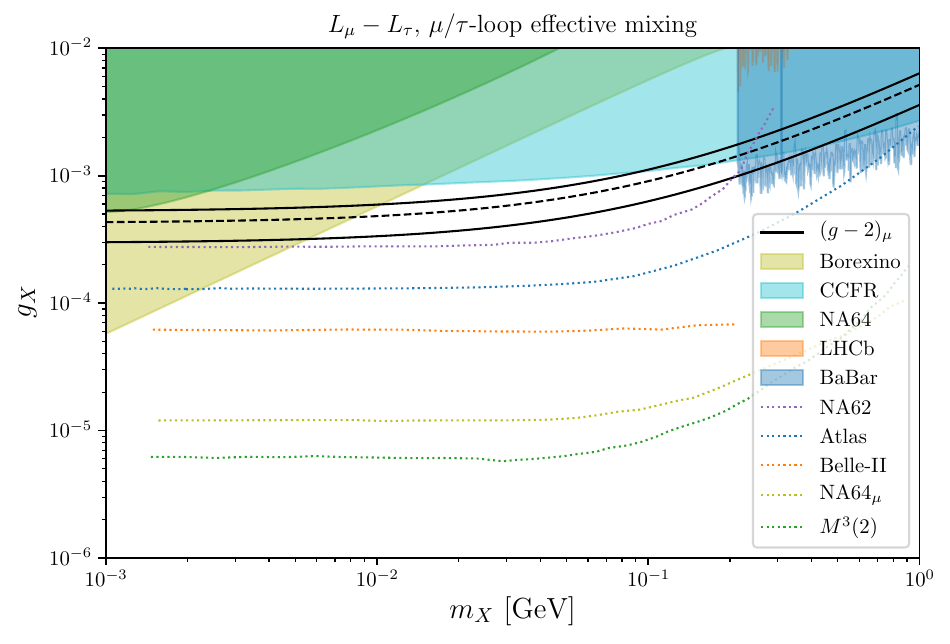}
	\caption{Parameter space for the light $ X_\mu$ solution of the $ (g-2)_\mu $ anomaly in the $ \U(1)_{L_\mu - L_\tau}$ model. The shaded regions are excluded by various experiments, as denoted in the legend, with dotted lines giving the future projections for the exclusions, while the $2\sigma$ region between the black solid lines is preferred by $ (g-2)_\mu $ (dashed line for central values).
	} \label{fig:Lmu-Ltau}
\end{figure}

For completeness we start the list of benchmark models with the
$L_\mu-L_\tau$ gauge symmetry, which has a long history as an explanation for the \gmu anomaly~\cite{Baek:2001kca,Ma:2001md,Altmannshofer:2014pba}.
The 2nd and 3rd generation SM  leptons carry opposite charges, and form vector-like representations under $ \U(1)_{L_\mu-L_\tau} $:
\begin{equation}
\label{eq:LmuLtau:charges}
    X_e=X_{N_1}=0, \qquad X_\mu= -X_\tau  =1, \qquad \qquad X_{N_2}=-X_{N_3}.
\end{equation}
The right-handed neutrinos $N_{2,3}$ can carry vector-like charges, $X_{N_2}=-X_{N_3}$, that are independent of the charges of the SM leptons. However, as long as $X_{N_2,N_3}$ are not excessively large (or if the right-handed neutrinos are heavy) their exact values are expected not to change appreciably the phenomenology of the gauged $L_\mu-L_\tau$ model.

In Fig.~\ref{fig:Lmu-Ltau} we show with black lines the $1\sigma$ parameter band (dashed black for central values), for which the $L_\mu-L_\tau$ model explains the observed \gmu. The model is not expected to have any appreciable kinetic mixing parameter and thus the UV value of $\varepsilon$ is set to zero. At one loop the muon and tau kinetic mixing diagrams lead to an effective momentum-dependent mixing, already taken into account in the \textsc{DarkCast} model~\cite{Amaral:2020tga,Altmannshofer:2019zhy}, which gives an effective coupling of $X_\mu$ to the electrons at the percent level. The resulting constraints (future projections)  in Fig.~\ref{fig:Lmu-Ltau} are thus taken directly from \textsc{DarkCast} and are shown as shaded colored regions (dashed lines) with color coding denoted in the legend.

\subsubsection{Gauged $ B - 3 L_\mu $}
\label{sec:bm3Lmu}

\begin{figure}[t]
	\centering
	\includegraphics[width=0.8\columnwidth]{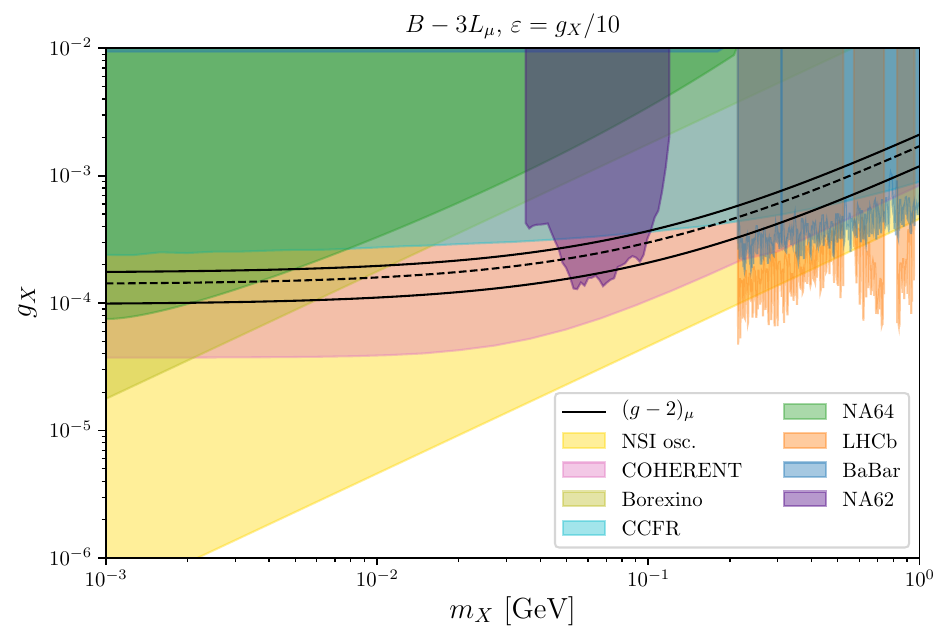}
	\caption{Parameter space for the light $ X_\mu$ solution of the $ (g-2)_\mu $ anomaly in the $ B-3L_\mu $ model. The shaded regions are excluded by various experiments, as denoted in the legend, while the $2\sigma$ region between the black solid lines is preferred by $ (g-2)_\mu $ (dashed line for central values).} \label{fig:B-3Lmu}
\end{figure}

The $ B-3 L_\mu $ gauge group has charges
    \begin{equation}
    X_\mu = X_{N_2}= -3, \qquad X_{e,\tau} = X_{N_1, N_2} = 0,  \qquad X_{Q_i, U_i, D_i} = \tfrac{1}{3}.
    \end{equation}
This is another example of the vector category of $\U(1)_X$ charge assignments. The $\U(1)_{B-3L_\mu}$ was the first gauge group used in a muoquark construction~\cite{Greljo:2021xmg}. This group is particularly suitable for the task at hand because it allows for a phenomenologically viable type-I seesaw, with only one symmetry-breaking scalar, while still forbidding $\U(1)_B$-violating dimension-5 operators that could otherwise be induced by $S_3$ leptoquark exchanges.
In contrast to the $ L_\mu - L_\tau $ model, a sizable kinetic mixing parameter is generated by the RG running. For this reason we use  the benchmark numerical value $ \varepsilon= 0.1\, g_X$.

In Fig.~\ref{fig:B-3Lmu} we report the bounds on the $ B-3 L_\mu $ model in the sub-GeV mass region, i.e., in the mass region where $X_\mu$ could potentially explain the \gmu anomaly ($1\sigma$ range is denoted with solid black lines, the central values with dashed black lines). Since $X_\mu$ couples to quarks, a significant part of the relevant parameters space is excluded by the di-muon resonance searches at the LHCb (orange region), and by the bounds on $\pi^0\to \gamma X_\mu$ decays from NA62 (purple region). Even more stringent constraints on the model are imposed by the bounds on the NSI interactions from the global fits to the neutrino oscillation data (yellow region) and from the measurements of the coherent neutrino--nucleus scattering by COHERENT (light magenta). These bounds are sufficiently strong to completely exclude the model.

The bounds on the $ B-3 L_\mu $ model,  Fig.~\ref{fig:B-3Lmu}, are representative of the bounds that would be obtained for a generic $U(1)_X$ model with nonzero $ B$ charges. In order for such models to lead to a viable explanation of the $(g-2)_\mu$ anomaly, one would need to increase the $ L_\mu $ charge assignment well over the  $ B $ charges. This can be achieved, for instance, by taking the linear combination  $c_1(B-3 L_\mu)+c_2(L_\tau-L_\mu)$ for the gauge group, where $c_2\gg c_1$.

\subsubsection{Gauged $L_e - L_\mu$} \label{sec:LmuLe}
The gauged $L_e-L_\mu $ model,
\begin{equation}
    X_\tau=X_{N_3}=0, \qquad X_e= -X_\mu  =1, \qquad \qquad X_{N_2}=-X_{N_1},
\end{equation}
is obtained from the $L_\mu-L_\tau$ model, Eq. \eqref{eq:LmuLtau:charges}, through a simple permutation of the charge assignments, but leads to a drastically changed phenomenology.
The large electron charge greatly enhances the $ X_\mu $ interactions with electrons. Since such couplings are targeted by  many of the light-sector searches one may expect that these exclude the $\U(1)_{L_e-L_\mu} $ model as an explanation of the \gmu anomaly.
\begin{figure}[t]
	\centering
	\includegraphics[width=\columnwidth]{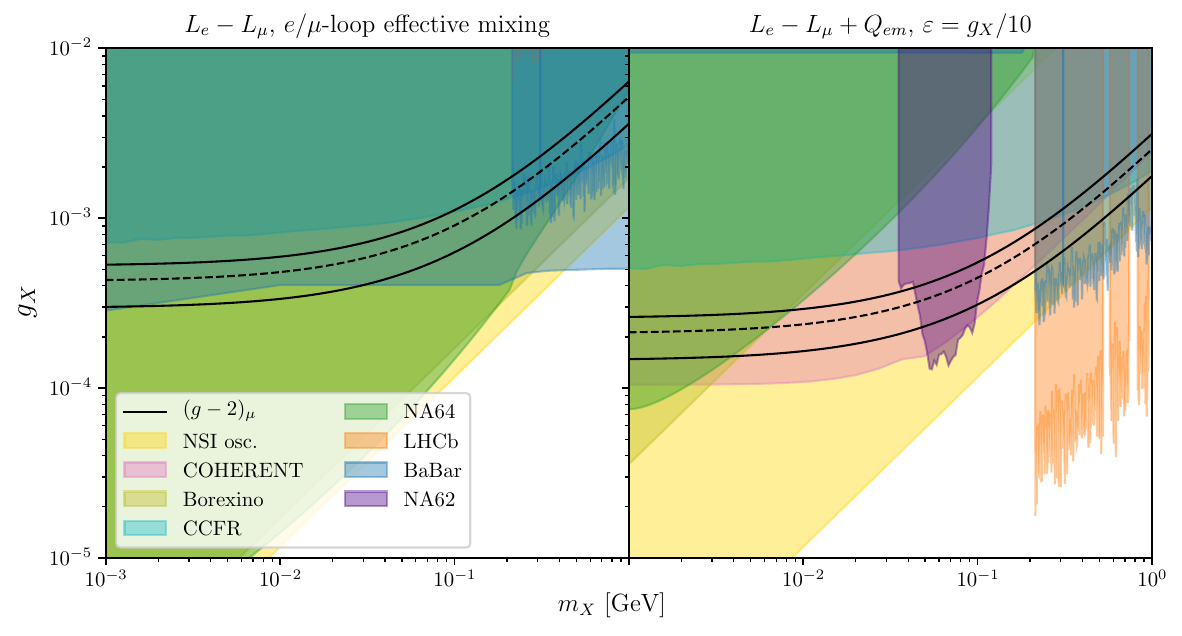}
	\caption{Parameter space for the light $ X $ solution to the $ (g-2)_\mu $ anomaly in $ \U(1)_{L_e - L_\mu}$ model (left plot) and $ \U(1)_{L_e - L_\mu - Q_\mathrm{em}}$ (right plot). The shaded regions are excluded by various experiments, as denoted in the legend, while the $2\sigma$ region between the black lines is preferred by $ (g-2)_\mu $ (dashed lines denote central values).} \label{fig:LeLmu}
\end{figure}

In Fig.~\ref{fig:LeLmu} (left) we show the constraints on the $ \U(1)_{L_e - L_\mu}$ model assuming vanishing kinetic mixing in the UV. The parameter space that would explain the  \gmu anomaly is excluded by a number of different experiments: both by NA64, Borexino, and constrains on NSI from neutrino oscillation fits. A different choice of the kinetic-mixing parameter can change the effective $X_\mu$ couplings to electrons, reducing the importance of some of the experimental constraints.
Taking for example $ \varepsilon= - g_X/e $ leads to effective $X_\mu$ charges that are equivalent to the charges in the $ L_e - L_\mu + Q_{\mathrm{em}} $ model, see App.~\ref{app:reparam}.
In this case, the electron charge vanishes.
However, this then introduces effective $X_\mu$ charges to quarks proportional to their electric charges $Q_{q, {\rm em}}$.
We illustrate this scenario in Fig.~\ref{fig:LeLmu} (right).
The NA64 and Borexino bounds become less stringent, the COHERENT bound becomes more stringent, while the choice of kinetic mixing parameter (i.e. shifting all $X_\mu$ couplings proportional to $Q_{\mathrm{em}}$) has no effect on the NSI constraints from neutrino oscillations. We can therefore conclude that it is not possible to explain the \gmu anomaly using just the $ L_e - L_\mu $ model or its $ L_e - L_\mu + Q_{\mathrm{em}} $ variant.

Both the $L_e - L_\mu$ model and the $B - 3 L_\mu$ model illustrate that if the $ X_\mu$ couplings to valence quarks and/or electrons are comparable to its couplings to muons, the   constraints from bounds on NSI  due to neutrino oscillations fits are going to be very stringent. A possible charge assignment that may circumvent the NSI bounds from atmospheric and long baseline neutrino oscillation experiments is $B - 2 L_e-  L_\mu$. For this charge assignment the ordinary earth matter is almost invisible to $X_\mu$. (The atoms with isospin symmetric nuclei are $X_\mu$ neutral, while for the average element composition of earth there are  slightly more  neutrons than protons, with approximate ratio $ \simeq 1.05:1.00$,  see, e.g.,  Eq.~(2.17) in Ref.~\cite{Esteban:2018ppq}). For $B - 2 L_e-  L_\mu$ with vanishing kinetic mixing in the UV, the $X_\mu$ couplings to electrons are large enough that NA64 and Borexino bounds  exclude such an explanation of \gmu. However, this may no longer be the case for judicial choices of kinetic mixing. Changing the kinetic mixing parameter $\varepsilon$ does not impact the NSI bounds from neutrino oscillations, while the value $ \varepsilon \simeq -2 g_X/ e $ effectively eliminates the electron charge, relaxing the NA64 and Borexino bounds (and simultaneously relaxing the COHERENT bounds). Such a value of $\varepsilon$ may at first appear to be tuned. However,  this is nothing but a reparameterization of a gauge group with nontrivial Higgs charge as described in Sec.~\ref{sec:qfu}. The choice $ \varepsilon = -2 g_X/ e $ is equivalent to the $ \U(1)_X $ group with $ X_H = -1 $ and $ X_{d_i} = - X_{u_i} = 1 $. A dedicated global fit of neutrino oscillations in this model is missing and is beyond the scope of this work.

\subsubsection{Gauged $ 9 B_3 - 8L_\mu - L_\tau $}
\label{sec:9B3}

\begin{figure}[t]
	\centering
	\includegraphics[width=.8\columnwidth]{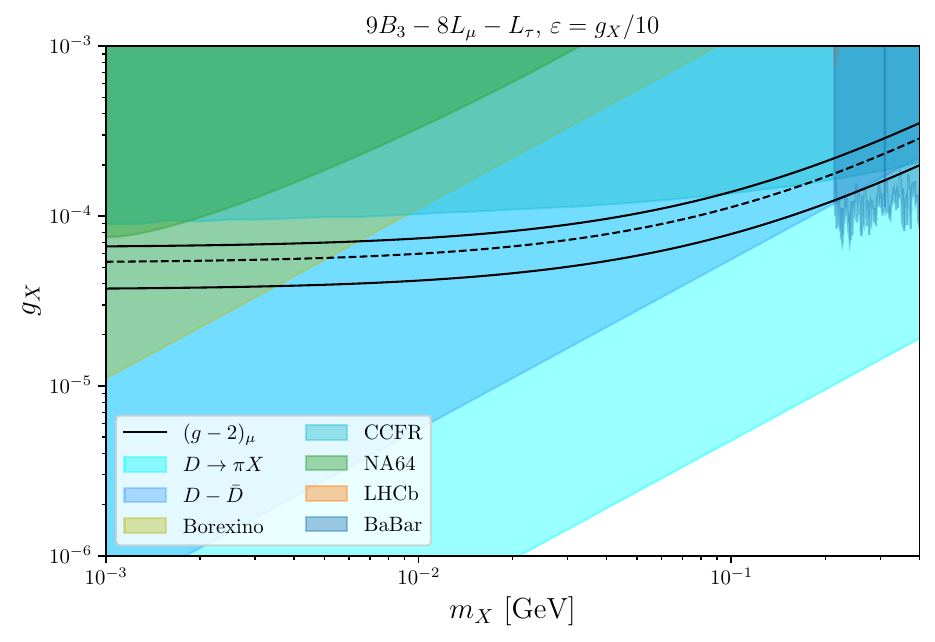}
	\caption{Parameter space for the light $ X $ solution to the $ (g-2)_\mu $ anomaly in the third-family quark model $ \U(1)_{9B_3 - 8L_\mu - L_\tau}$ model. The shaded regions are excluded by various experiments, as denoted in the legend, while the $2\sigma$ region between the black lines is preferred by $ (g-2)_\mu $ (dashed lines denote central values).} \label{fig:3B38LmuLtau}
\end{figure}

As the benchmark third-family-quark model we investigate the $ \U(1)_{9B_3-8L_\mu-L_\tau}$ model,  Eq.~\eqref{eq:3rd_family_charges}. As discussed in Sec.~\ref{sec:3rdFam}, a \gmu solution of this type should be down-aligned to avoid strong constraints from $ B $ decays. However, this still leaves strong bounds from $ D-\overline{D}$ mixing and $ D \to \pi X $ decays.

For the former, we reuse the results of Ref. \cite{Smolkovic:2019jow} where the operator product expansion, treating charm quark as heavy, was used to obtain the meson mixing bounds on flavored light  vector mediators. The $D-\bar D$ mixing measurements translate to a bound $m_X\gtrsim 60\, {\rm MeV} (g_X X_{q_{3}}/10^{-4})$, if the NP couplings are real, and $m_X\gtrsim 360\, {\rm MeV} (g_X X_{q_{3}}/10^{-4})$ for a NP contribution with the maximal weak phase.
To be conservative, we assume that the NP couplings are real. In particular, the off-diagonal couplings are only in the $X_\mu$ couplings to left-handed up quarks, and these are taken to be equal to $X_{q_3} g_X |V_{ub}^* V_{cb}|$. We show this bound in Fig.~\ref{fig:3B38LmuLtau} as the dark blue shaded region.

Potentially even more sensitive is the $D\to \pi X$ decay, where $X$ decays to neutrinos, resulting in the $D\to \pi +E_{\rm miss}$ signature. Unfortunately, there is no dedicated  experimental search for this signature yet. To indicate the potential experimental reach we can use the recast of the CLEO $D \to \mu \nu$ search \cite{CLEO:2008ffk} that was reinterpreted  in Ref. \cite{MartinCamalich:2020dfe} as the bound on $D\to \pi a$  decays for a massless invisible pseudoscalar $a$,  giving $\mathrm{Br}(D^+\to \pi^+ a )<8 \cdot 10^{-6}$.  We expect this recast to be valid also for $X_\mu$ masses of up to a few hundred MeV, while for heavier masses the recast should be repeated, which we do not attempt here. Using the expressions for decay branching ratios in \cite{Smolkovic:2019jow} we then obtain the bound  $m_X\gtrsim 70 {\rm MeV} (g_X X_{q_3}/10^{-5})$, assuming 100\% branching ratio to neutrinos, which is valid in our benchmark up to the muon threshold, $m_X<2m_\mu$. In Fig.~\ref{fig:3B38LmuLtau} the bound is indicated with a light blue dashed line. To obtain the complete sensitivity of rare meson decay bounds other decay channels such as $D\to \pi \mu^+\mu^-$, $D\to \rho \mu^+\mu^-$  should be considered (as well as $D\to \rho E_{\rm miss}$ once experimental searches are performed),   which goes beyond the scope of the present exploratory study.

Even so, Fig.~\ref{fig:3B38LmuLtau} clearly demonstrates that the $ D $ meson bounds are strong enough to completely rule out the light $ X_\mu$ solution to \gmu in this model. The only viable models of this kind would be the ones with $ |X_{L_2,E_2}| \gg |X_{q_3}| $, which would suppress the NP couplings to the quark sector relative to the coupling to the SM leptons. This could be achieved by modifying the present third quark family benchmark and replacing it with a linear combination of these third-family charge assignments and the $ \U(1)_{\mu-\tau}$ charge assignments (weighting more heavily toward the latter). Clearly, one can approximate arbitrarily well the physics of $ \U(1)_{\mu-\tau}$  by allowing for sufficiently large charges of the latter group.

\subsubsection{Chiral model: Gauged $\tilde L_{\mu-\tau}$}
\label{sec:skewed}

\begin{figure}[t]
	\centering
	\includegraphics[width=0.8\columnwidth]{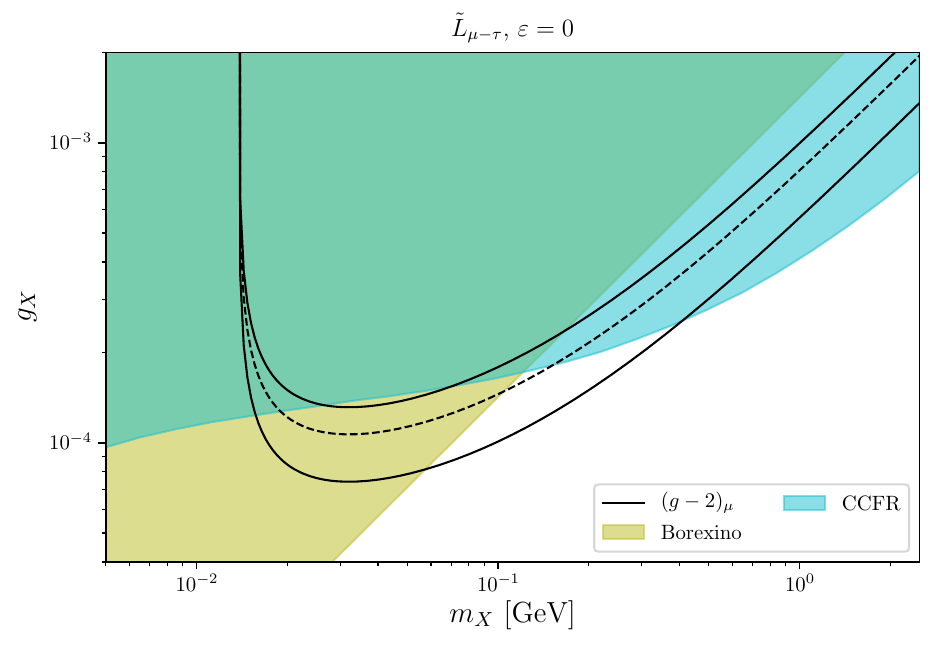}
	\caption{Parameter space for the light $ X $ solution to the $ (g-2)_\mu $ anomaly in the $\tilde{L}_{\mu-\tau}$ model. The shaded regions are excluded by various experiments, while the region between the black lines is preferred by $ (g-2)_\mu $.} \label{fig:Lt-3B}
\end{figure}

We now entertain a more exotic option for the $ \U(1)_X $ group from among the models of Sec.~\ref{sec:not:all} with chiral charge assignments. \\
    {\bfseries \boldmath $\tilde L_{\mu-\tau}$ model     :}
\begin{align}
    (X_{L_1},\,X_{L_2},\,X_{L_3}) &= (-1, 7, -6), & (X_{E_1},\,X_{E_2},\,X_{E_3}) &= (1, 6, -7), \nonumber \\ (X_{N_1},\,X_{N_2},\,X_{N_3}) &= (-7, -2, 9), & X_{Q_i,D_i,U_i} &= 0.
\end{align}
The model has a large vectorial coupling to muons and only a small axial component. This maximizes the NP contributions to the $ (g-2)_\mu $ with the right sign to explain the anomaly. The model also maintains a large ratio between the muon and electron charges, without which there is little hope of evading the Borexino bound (that the Borexino bound is relevant even for small induced couplings to electrons we saw already in the case of the $ L_{\mu-\tau} $ model). Finally, the model has purely axial couplings to electrons and no couplings to quarks such that NSI bounds are completely avoided. Fig.~\ref{fig:Lt-3B} shows that the Borexino and CCFR bounds are still strong enough to exclude most of the parameter space relevant for \gmu except for a small window around 100~MeV, assuming that the kinetic mixing vanishes in the UV. The bounds do not change significantly for other reasonable values of $\varepsilon$, for instance even for  values as large as $\varepsilon = \pm g_X / 10$. In Fig.~\ref{fig:Lt-3B} we do not show the collider bounds. We expect these to be qualitatively similar to the collider bounds for the $L_\mu - L_\tau$ model, cf. Fig. \ref{fig:Lmu-Ltau}. However, the \textsc{DarkCast} code, which we used to derive the collider bounds, only has vector couplings implemented at the moment. We therefore defer the complete phenomenological study of  the $\tilde L_{\mu-\tau}$ model to future work.

Finally, it would be interesting to relax the chiral model search criteria beyond $ |X_F|\leq 10$. We anticipate that this would generate additional feasible models with axial currents that are further suppressed and also have a larger muon-to-electron charge ratio, further relaxing the experimental bounds. Of course, this is just another way to approach the limit of the $L_\mu - L_\tau$ model.

\section{Muon mass and $(g-2)_\mu$ at one loop}
\label{sec:radiative}

\begin{figure}[t]
	\centering
	\includegraphics[width=0.32\columnwidth]{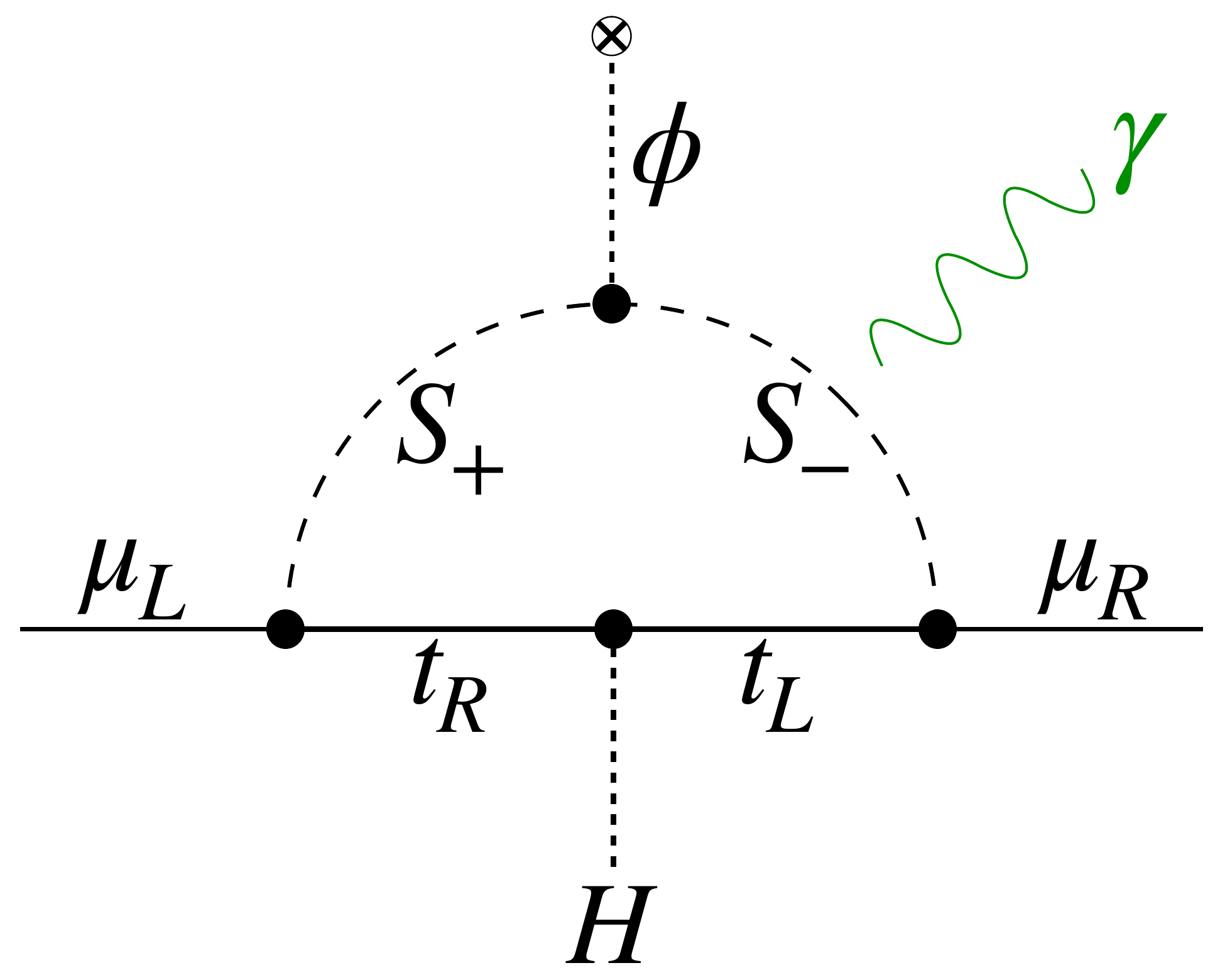}
	\caption{A muoquark model for the radiative muon mass generation that also contributes to the anomalous magnetic moment through the above diagram. See main text in Sec.~\ref{sec:radiative} for further details.} \label{fig:radiativeDiag}
\end{figure}

The observed smallness of Yukawa couplings can be explained in models in which fermion masses are generated from radiative corrections~\cite{Weinberg:2020zba,Baker:2020vkh}. Here we focus on radiatively generated muon mass.  Since both the muon mass and the muon anomalous magnetic moment are chirality flipping operators, the TeV-scale NP that at one-loop generates  the muon mass then generically  also  gives correlated one-loop NP contributions to $(g-2)_\mu$~\cite{Baker:2021yli}.

\subsection*{Model example}

Let us consider a scenario in which the SM is extended by two scalar leptoquarks, $S_+$ and $S_-$, in the $(\rep{3},\, \rep{2},\, 7/6)$ representation of the SM gauge group. This leptoquark representation is usually called $R_2$ as in Ref.~\cite{Dorsner:2016wpm}, however, for clarity we use a simpler notation in this section. We assume that the leptoquarks are coupled to the third generation quarks, $q_\LL^3, t_\RR$, and the second generation leptons, $\ell_\LL^2, \mu_\RR$. The model is assumed to have a $Z_2$ parity symmetry under which $ S_{-}$ and $\mu_R$ are odd, while all the other fields are even,
\begin{equation}
    \mathcal{L} \supset  \eta_\LL \, \overline{t}_\RR \ell^2_\LL \, i\sigma_2 S_+ - \eta_\RR \, \overline{q}^{3}_\LL \, \mu_\RR \, S_-  \hc .\label{eq:radiative}
\end{equation}
The global phase rotations can be used to make the couplings $\eta_{\LL,\RR}$ real without loss of generality. We assume that the left-handed quark doublet is defined in the down-quark mass eigenstate basis and take $V_{tb} =1$.

The $Z_2$ symmetry forbids the direct muon Yukawa coupling, $ \overline{\ell}^2_\LL \tilde H \mu_R$. This is then generated only radiatively due to the presence of a $Z_2$ soft breaking term,
\begin{equation}
    \mathcal{L}_{\rm{break}} \supset - \, \widetilde m^2_S S^\dagger_+ S_-  \hc ,
\end{equation}
which induces a finite one-loop contribution to the muon mass as well as to the anomalous magnetic moment.  For simplicity, we take the $Z_2$ symmetric mass terms,
\begin{equation}
    \mathcal{L} \supset - \, m^2_{S_-} S^\dagger_- S_- - m^2_{S_+} S^\dagger_+ S_+  ,
\end{equation}
to be degenerate, $ m^2_S \equiv m^2_{S_+} = m^2_{S_-} $,  with  $0 < \widetilde m^2_S \ll m^2_S$.  This induces maximal mixing, $\theta = \pi/4$, into the mass eigenstates  $S_{1,2}$,  with the corresponding physical masses given by $m^2_{S_{1(2)}} = m^2  \mp \widetilde m^2$.  Furthermore, we assume that $m_t \ll m_S$ as suggested by the direct searches for leptoquarks at the LHC~\cite{Aad:2020iuy,ATLAS:2020qoc}.

Expanding the muon mass and the anomalous magnetic moment to the leading order in $m_t / m_S $ and $\widetilde m_S / m_S$ (see~\cite{Dorsner:2019itg,Baker:2020vkh} for full expressions),
\begin{equation}
\begin{split}
    m_\mu &\approx  ~ \frac{3}{16 \pi^2} \,m_t \eta_\LL \eta_\RR \,  \frac{\widetilde m^2_S}{m^2_S} ~, \\
    \Delta a_\mu &\approx ~ \frac{3}{16 \pi^2} \, \eta_\LL \eta_\RR ~ \frac{m_\mu m_t \widetilde m^2_S}{m^4_S} \left ( 5 - 4 Q_S + 2 (1 - Q_S) \log \frac{m_t^2}{m^2_S} \right) \,.
\end{split}
\end{equation}
Assuming that $m_\mu$ is entirely generated by the above one loop radiative correction, the ratio $\Delta a_\mu / m_\mu$ depends only on two unknowns, $m_S$ and $Q_S$,
\beq
\label{eq:Delta:a:mu:approx}
\Delta a_\mu \approx \frac{m_\mu^2}{m^2_S}\left ( 5 - 4 Q_S + 2 (Q_S-1) \log \frac{m_S^2}{m_t^2} \right).
\eeq
Since $\Delta a_\mu$ needs to be positive, and $m_S\gg m_t$ experimentally, this puts a constraint on possible values of leptoquark charge, $Q_S\gtrsim 1$.  In our example, the electric charge of the scalars running in the loop is $Q_S = 5/3$. Consequently, $\Delta a_\mu{}{} = (251 \pm 59)\times 10^{-11}$ points to $m_S \in [5 - 7]$~TeV. For $\eta_\LL \eta_\RR \approx 1$, the soft breaking mass needed to match the muon mass is then $\widetilde m_S \approx 1 $~TeV.

\begin{figure}[t]
	\centering
	\includegraphics[width=0.4\columnwidth]{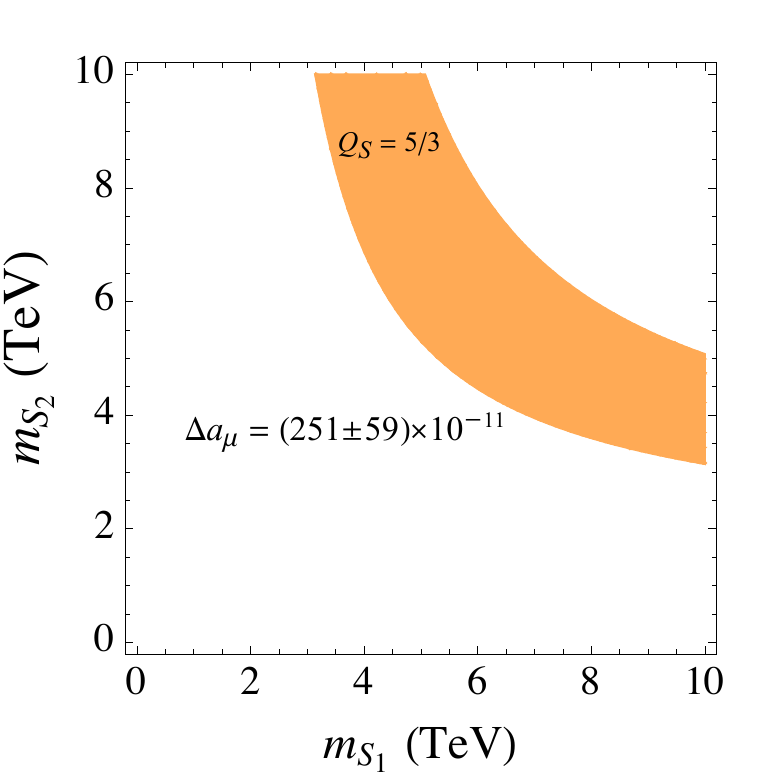}
	\caption{Preferred parameter space in the muoquark model for the radiative muon mass generation as implied by the $(g-2)_\mu$ anomaly. See the main text in Sec.~\ref{sec:radiative} for further details.} \label{fig:radiative}
\end{figure}

A wider parameter space opens up, if we move away from the limit $m_{S_+} \neq m_{S_-}$. Assuming as before that the muon mass is entirely due to the one loop radiative correction, Eq. \eqref{eq:Delta:a:mu:approx} generalizes to~\cite{Baker:2021yli}
\begin{equation}
    \Delta a_\mu = \frac{m^2_\mu}{m^2_t} \, \tilde F\left( \frac{m^2_{S_1}}{m^2_t}, \frac{m^2_{S_2}}{m^2_t} \right)~,
\end{equation}
where
\begin{equation}
\begin{split}
   \tilde F(x_1,x_2) =&
\left(\frac{x_1 \log x_1}{1 - x_1} - \frac{
   x_2 \log x_2}{1 - x_2} \right)^{-1}  \biggr[ \frac{3 x_1 - 1}{(1 - x_1)^2} - \frac{
     3 x_2 - 1}{(1 - x_2)^2} + \frac{2 x_1^2 \log x_1}{(1 - x_1)^3}
    \\
   &  - \frac{ 2 x_2^2 \log x_2}{(1 - x_2)^3}   +
   2 Q_S \left( \frac{1}{1 - x_1}- \frac{1}{1 - x_2}+ \frac{x_1 \log x_1 }{(1 - x_1)^2}- \frac{x_2 \log x_2}{(1 - x_2)^2} \right) \biggr]~.
\end{split}
\end{equation}
The preferred region in the $S_1, S_2$ mass plane, taking  $Q_S = 5 /3 $,  that explains the observed deviations in $\Delta a_\mu$ is shown in Fig.~\ref{fig:radiative} as the brown shaded band.\footnote{We note in passing  that requiring a positive contribution to $\Delta a_\mu$ even in this more general case still remains quite restrictive regarding the viable  choices for the leptoquark gauge representations. In particular, the $(\repbar{3},\, \rep{1},\, 1/3)$ scalar leptoquark is not a phenomenologically viable possibility.}

\subsection*{$\U(1)_X$ completion}

The above scenario can be elegantly UV completed in our setup. The scan over the anomaly free charge assignments in Section~\ref{sec:models} reveals a family of solutions for which the dimension-4 muon Yukawa is forbidden. This occurs when the $\U(1)_{X_\mu}$ charge of the left-handed muon is different from the charge of the right-handed muon. We assume that in addition to the SM there are three scalars, the $S_\pm$ in the $(\rep{3},\, \rep{2},\, 7/6)$ representation of the SM gauge group and the SM singlet $\phi$. The extra scalars carry the following charges under the $U(1)_{X_\mu}$ gauge symmetry
\begin{align}
    X_{S_+}  &= - X_{L_2} + X_{U_i} = - X_\mu + X_q~,\\
    X_{S_-}  &= - X_{E_2} + X_{Q_i} = - X_\mu - b_{E_2} + X_q~,\\
    X_{\phi} &= - X_{S_-} + X_{S_+} = b_{E_2}~.
\end{align}
where $X_q = ( - X_e - X_\mu - X_\tau) / 9$ (see Section~\ref{sec:not:all}).
The leptoquarks $S_+ (S_-)$ have allowed couplings to the left-(right-)handed muons, respectively, i.e., they are the muoquarks.
An explicit mass mixing between $S_+$ and $S_-$ is forbidden by the $\U(1)_{X_\mu}$ gauge symmetry. However, there is a gauge invariant trilinear scalar coupling
\begin{equation}
    \mathcal{L} \supset - A \phi S_+^\dagger S_- \hc,
\end{equation}
that gives rise to the $U(1)_{X_\mu}$ mass mixing term, $\widetilde m_S^2 = A v_\phi$, once the SM singlet $\phi$ gets a VEV, $\langle \phi \rangle =v_\phi$ and breaks $U(1)_{X_\mu}$ spontaneously.
The radiative generation of the muon mass requires a single insertion of $v_\phi$, a mechanism that is illustrated by the Feynman diagram in Fig.~\ref{fig:radiativeDiag}. An alternative charge assignment for $\phi$ can be $X_{\phi} = b_{E_2} / 2$, in which case the quartic interaction  $ \mathcal{L} \supset -\lambda \phi^2 S_+^\dagger  S_-$ is gauge invariant, giving $\widetilde m_S^2 = \lambda v_\phi^2$ and the radiative generation of the muon mass is induced by two insertions of the $\phi$ VEV.

In the above set-up the $X_\mu$ gauge boson is no longer needed to resolve $(g-2)_\mu$ and can be decoupled from the phenomenological discussion by taking the small gauge coupling limit. In addition, the corrections to the $h \to \mu^+ \mu^-$ decays start at $\mathcal{O}(m_t^2 / m_S^2)$ and are well within the present experimental bounds on the Higgs couplings to muons.

\section{A vector muoquark model}
\label{sec:4321}

For a leptoquark to become a muoquark, it has to be charged under a $\U(1)_X$ gauge symmetry discussed in Section~\ref{sec:models}.
While this is straightforward for a scalar leptoquark, charging a vector leptoquark in a UV complete model is more challenging.
In this section, we demonstrate that it is possible to construct UV-complete gauge models in which a vector leptoquark carries a $\U(1)_X$ charge.
We build a model that contains a $U_1$ muoquark with the SM quantum numbers $(\bm{3},\bm{1},2/3)$ as one of the leptoquarks charged under the gauged $\U(1)_{\mu-\tau}$ as an example.

\subsection*{Gauge group and fermion embedding}
We consider the gauge group
\begin{equation}
 G_{5321} =  \SU(5)\times \SU(3)' \times \SU(2)_L \times \U(1)'
\end{equation}
and embed the third generation of quarks together with the second and third lepton generations into fundamental $\bm{5}$ multiplets of $\SU(5)$:
\begin{equation}
 \psi_L \sim (q_L^3\; \ell_L^3\; \ell_L^2)^\intercal\,,
 \qquad
 \psi_R^+ \sim (u_R^3\; \nu_R^3\; \nu_R^2)^\intercal\,,
 \qquad
 \psi_R^- \sim (d_R^3\; e_R^3\; e_R^2)^\intercal\,.
\end{equation}
The quantum numbers of all SM-like fermions are collected in Table~\ref{tab:5321_fermions}.
\begin{table}[t]
\begin{center}
\begin{tabular}{ccccc}
\hline\hline
Field & $\SU(5)$ & $\SU(3)'$ & $\SU(2)_L$ & $\U(1)'$ \\
\hline
$\ell^1_L$ & $\mathbf{1}$ & $\mathbf{1}$ & $\mathbf{2}$ & $-1/2$ \\
$e^1_R$ & $\mathbf{1}$ & $\mathbf{1}$ & $\mathbf{1}$ & $-1$ \\
$q^i_L$ & $\mathbf{1}$ & $\mathbf{3}$ & $\mathbf{2}$ & $1/6$ \\
$u^i_R$ & $\mathbf{1}$ & $\mathbf{3}$ & $\mathbf{1}$ & $2/3$  \\
$d^i_R$ & $\mathbf{1}$ & $\mathbf{3}$ & $\mathbf{1}$ & $-1/3$  \\
$\psi_L$ & $\mathbf{5}$ & $\mathbf{1}$ & $\mathbf{2}$ & $-1/10$ \\
$\psi_R^+$ & $\mathbf{5}$ & $\mathbf{1}$ & $\mathbf{1}$ & $2/5$ \\
$\psi_R^-$ & $\mathbf{5}$ & $\mathbf{1}$ & $\mathbf{1}$ & $-3/5$ \\
\hline\hline
\end{tabular}
\end{center}
\caption{Quantum numbers of SM-like fermions under the $G_{5321}$ gauge symmetry; $i\in\{1,2\}$.}
\label{tab:5321_fermions}
\end{table}
Even though the $\SU(5)$ multiplets unify fields from different fermion generations, all gauge anomalies cancel.
The $\SU(5)$ group contains the subgroup $\SU(3)_5 \times \U(1)_5 \times \U(1)_{\mu-\tau} \subset \SU(5)$ under which the $\bm{5}$ decomposes as
\begin{equation}\label{eq:5_fermion_branching}
\bm{5}\to(\bm{3},4/15,0)\oplus(\bm{1},-2/5,-1)\oplus(\bm{1},-2/5,1)\,,
\end{equation}
i.e.\, there is one $SU(3)_5$ triplet that is uncharged under $\U(1)_{\mu-\tau}$ and is identified with a third-generation quark, as well as two $\SU(3)_5$ singlets charged under $\U(1)_{\mu-\tau}$ that are identified with third and second generation leptons.
The QCD and hypercharge gauge groups are given by the diagonal subgroups of $\SU(3)_5\times \SU(3)'$ and $\U(1)_5\times \U(1)'$, respectively,
\begin{equation}
 \SU(3)_c = [\SU(3)_5 \times \SU(3)']_{\rm diag}\,,
 \qquad
 \U(1)_Y = [\U(1)_5 \times \U(1)']_{\rm diag}\,.
\end{equation}
It can be easily verified that the sums of $U(1)'$ charges in Table~\ref{tab:5321_fermions} and $\U(1)_5$ charges in Eq.~\eqref{eq:5_fermion_branching} yield the correct hypercharges, see also Appendix~\ref{app:su5}.

\subsection*{Gauge sector and vector bosons}\label{sec:5321_gauge_sector}
There are different ways to break $G_{5321}\to G_{\rm SM}\times U(1)_{\mu-\tau}$ that result in different vector boson spectra.
In particular, there are two possible intermediate gauge groups:
\begin{itemize}
 \item $G_{\rm SM} \times \SU(2)_{\mu\tau}$

 In this case, the breaking $G_{5321}\to G_{\rm SM}\times U(1)_{\mu-\tau}$ is done in the following two steps.
 \begin{enumerate}
  \item $G_{5321}\to G_{\rm SM} \times \SU(2)_{\mu\tau}$

  This first step proceeds via the $\SU(5)$ subgroup $\SU(3)_5 \times \SU(2)_{\mu\tau} \times \U(1)_5 \subset \SU(5)$, under which the fundamental $\bm{5}$ and the adjoint $\bm{24}$ representations decompose as
  \begin{equation}
  \begin{aligned}
   \bm{5}&\to (\bm{3},\bm{1},4/15) \oplus (\bm{1},\bm{2}, \eminus 2/5)\,,
   \\
   \bm{24}&\to (\bm{1},\bm{1},0) \oplus(\bm{8},\bm{1},0) \oplus (\bm{1},\bm{3},0) \oplus (\bm{3},\bm{2},2/3) \oplus (\bar{\bm{3}},\bm{2},-2/3)\,,
  \end{aligned}
  \end{equation}
  such that the third and second lepton generations are contained in doublets of $\SU(2)_{\mu\tau}$.
  After this step, the spectrum of massless gauge bosons consists of the SM gauge bosons and an additional triplet of the unbroken $\SU(2)_{\mu\tau}$.
  The vector bosons that become massive by this breaking and their quantum numbers under $\SU(3)_c  \times \SU(2)_L  \times \U(1)_Y  \times \SU(2)_{\mu\tau}$ are
  \begin{itemize}
   \item $G'_\mu \sim (\bm{8},\bm{1},0;\bm{1})$, a heavy gluon-like vector boson also known as \emph{coloron},
   \item $B'_\mu \sim (\bm{1},\bm{1},0;\bm{1})$, a heavy $B$-boson-like neutral vector boson,
   \item $U_\mu \sim (\bm{3},\bm{1}, 2/3;\bm{2})$ (and its complex conjugate), an $\SU(2)_L$ singlet vector leptoquark known as $U_1$, coming in a doublet of $\SU(2)_{\mu\tau}$.
  \end{itemize}
  \item $\SU(2)_{\mu\tau}\to \U(1)_{\mu-\tau}$

  In this second step, the $\bm{2}$ of $SU(2)_{\mu\tau}$ decompose under $\U(1)_{\mu-\tau}$ as
  \begin{equation}
   \bm{2}\to -1 \oplus 1~,
  \end{equation}
  such that the $\tau$ and $\mu$ leptons inside the $\SU(2)_{\mu\tau}$ doublets receive $\U(1)_{\mu-\tau}$ charges $-1$ and $+1$, respectively.
  Similarly, the $\SU(2)_{\mu\tau}$ doublet of $U_1$ leptoquarks splits into two leptoquarks with $\U(1)_{\mu-\tau}$ charges $-1$ and $+1$.
  Furthermore, the vector bosons in the $\SU(2)_{\mu\tau}$ triplet that mediate $\tau$-$\mu$ transitions become massive, while the one related to the diagonal $\SU(2)_{\mu\tau}$ generator becomes the gauge boson of $\U(1)_{\mu-\tau}$.
 \end{enumerate}

 \item $\SU(4)_5\times \U(1)_{5'} \times \SU(3)' \times \SU(2)_L \times \U(1)'$

 The breaking $G_{5321}\to G_{\rm SM}\times \U(1)_{\mu-\tau}$ is again done in two steps:
 \begin{enumerate}
  \item $G_{5321}\to \SU(4)_5\times \U(1)_{5'} \times \SU(3)' \times \SU(2)_L \times \U(1)'$

  This breaking proceeds via the $\SU(5)$ subgroup $\SU(4)_5\times \U(1)_{5'}\subset \SU(5)$, under which the fundamental $\bm{5}$ and the adjoint $\bm{24}$ representations decompose as\footnote{%
  The generator $X_{5'}$ of $\U(1)_{5'}$ is given in terms of the $\SU(5)$ generators $T_5$ as $X_{5'}=\frac{2}{\sqrt{10}}\, T_5^{24}$.
  }
  \begin{equation}
  \begin{aligned}
   \bm{5}&\to (\bm{4},1/10) \otimes (\bm{1},-2/5)\,,
   \\
   \bm{24}&\to (\bm{1},0) \oplus(\bm{15},0) \oplus (\bm{4},1/2) \oplus (\bar{\bm{4}},-1/2)\,,
  \end{aligned}
  \end{equation}
  such that the third generation leptons stay unified with the third generation quarks in the $\bm{4}$ of $\SU(4)_5$, while the second generation leptons are singlets of $\SU(4)_5$.
  The vector bosons that become massive in this step and their quantum numbers under $\SU(4)_5\times \U(1)_{5'} \times \SU(3)' \times \SU(2)_L \times \U(1)'$ are
  \begin{itemize}
   \item $V_\mu \sim (\bm{4}, 1/2; \bm{1}, \bm{1},0)$ (and its complex conjugate), which contains the $U_1$ leptoquark that couples second generation leptons to third generation quarks as well as the SM neutral vector bosons that mediate $\tau$-$\mu$ transitions.
  \end{itemize}

 \item $\SU(4)_5\times \U(1)_{5'} \times \SU(3)' \times \U(1)' \to \SU(3)_c \times \U(1)_Y \times \U(1)_{\mu-\tau}$

 This breaking proceeds via the $\SU(4)_5$ subgroup $\SU(3)_4 \times \U(1)_4\subset \SU(4)_5$, under which the $\bm{4}$ decomposes as\footnote{%
  The generator $X_{4}$ of $\U(1)_{4}$ is given in terms of the $\SU(4)_5$ generators $T_4$ as $X_{4}=\frac{2}{\sqrt{6}}\, T_4^{15}$.
  }
 \begin{equation}
  \bm{4}\to (\bm{3}, 1/6)\oplus (\bm{1},-1/2).
 \end{equation}
 The $U(1)_{\mu-\tau}$ generator $X_{\mu-\tau}$ and the hypercharge generator $Y$ are given by the linear combinations of the generators $X_{5'}$, $X_{4}$, and $X_1$ of $\U(1)_{5'}$, $\U(1)_4$, and $\U(1)'$ as
 \begin{equation}
  X_{\mu-\tau} = \frac{3}{2}\, X_{4} - \frac{5}{2}\, X_{5'},
  \qquad
  Y =  X_{4} + X_{5'} + X_{1}\,.
 \end{equation}
  The vector bosons that become massive in this step and their quantum numbers under $\SU(3)_c  \times \SU(2)_L  \times \U(1)_Y  \times \U(1)_{\mu-\tau}$ are:
  \begin{itemize}
   \item $G'_\mu \sim (\bm{8},\bm{1},0,0)$, a heavy gluon-like ``coloron'',
   \item $B'_\mu \sim (\bm{1},\bm{1},0,0)$, a heavy $B$-boson-like neutral vector boson,
   \item $U_\mu^{(\tau)} \sim (\bm{3},\bm{1}, 2/3,1)$ (and its complex conjugate), a $U_1$ leptoquark coupling third generation leptons to third generation quarks.
  \end{itemize}
  In addition, the heavy vector boson four-plet $V_\mu$ splits into $(\bm{3},\bm{1},2/3,-1)\oplus(\bm{1},\bm{1},0,2)$, which is the $U_1$ leptoquark coupling second generation leptons to third generation quarks and the SM neutral vector that mediates $\tau$-$\mu$ transitions.

 \end{enumerate}

\end{itemize}

\section{Conclusions}
\label{sec:conc}

Lepton-flavored $\U(1)_X$ gauge extensions of the SM --- the focus of this work ---
provide a framework in which the first experimental deviations are expected to be seen in the lepton flavor universality violating~(LFUV) observables and not, as is more common in many other new physics models,  in the
lepton flavor-violating~(LFV) observables.
The $\U(1)_X$  extensions of the SM are especially interesting in view of the recent flavor anomalies, showing hints of new physics in LFUV ratios \RK, while the LFV transitions such as the $\tau \to 3\mu$ decay are absent.
In the $\U(1)_X$ models  the LFUV is hardcoded in the $X_\mu$ gauge boson interactions through generation-dependent charges (allowing, e.g., for the muonic force). In contrast, the LFV is forbidden in the infrared-relevant operators.
The LFV effects arise only  in the form of higher-dimensional operators, whose effects
are power suppressed by high-energy scales.

The bulk of the present manuscript dealt with different SM$+3 \nu_R$ anomaly-free charge assignments and their impact on the model building applications for the flavor physics anomalies involving muons: $(g-2)_\mu$ and anomalies in rare $B$-meson decays ($R_{K^{(*)}}$ and $b \to s \mu^+ \mu^-$ angular distributions and branching ratios). Our main finding is that the quark flavor universal (or third-family-quark), but lepton flavor non-universal, $\U(1)_X$ gauge symmetry extensions provide a natural framework for the introduction of a leptoquark coupled to a single lepton flavor. Our attention was focused on the TeV-scale \textit{muoquarks} --- leptoquarks that due to their $\U(1)_X$ charges are allowed to couple to muons but not to electrons and taus.  We required the cancellation of the chiral anomalies within the field content present at low-energies, i.e., already within the SM$+3 \nu_R$. This guarantees  the theory to be consistent without introducing  additional $\U(1)_X$ charged fields in the UV.

Our work builds on and extends the simple scenario for flavor anomalies that was first proposed in Ref.~\cite{Greljo:2021xmg}. Extending the SM gauge group by an additional $\U(1)_X$ gauge symmetry was found to have two desirable effects. Firstly, it guaranteed that the $S_3$ scalar muoquark, needed to explain the anomalies in rare $B$ decays through tree-level contributions, does not have LFV couplings. Secondly, the existence of the gauge boson $X_\mu$ gives a natural explanation of the $(g-2)_\mu$ anomaly through a one loop contribution.
In this manuscript we identified 273 inequivalent quark-flavor-universal models supporting the muoquark when demanding that the ratio of maximal to minimal $\U(1)_X$ charges is at most equal to ten. As shown in Section~\ref{sec:models} one can classify the $\U(1)_X$ models into two broad categories based on whether or not for all  three generations the dimension-4 lepton Yukawa interaction are allowed by the $\U(1)_X$ charges. The third-family-quark charge assignments are easily derived from the quark-flavor-universal solutions, but the muoquark conditions are slightly more involved (Section~\ref{sec:3rdFam}). In this case, $X_\mu$ boson interactions with quarks are flavor non-universal in the weak basis and flavor violating in the mass basis. This may introduce potential problems with FCNC constraints. However, the theoretical advantage of the third-family-quark class is that it can partially explain the approximate $SU(2)^3$ flavour symmetry observed in the SM quark sector~\cite{Barbieri:2011ci,Kagan:2009bn}.

In the rest of the manuscript, we showed three different uses for the $\U(1)_X$ model classifications.
The first question we addressed is how restricted are the new physics models that use
a light  $X_\mu$ vector boson to explain the $(g-2)_\mu$ via one loop contributions (while at the same time explaining the anomalies in rare $B$ decays through tree level leptoquark exchanges). We found that the constraints on $X_\mu$ couplings to quarks and leptons are so severe that only very few examples, such as the  widely popular $L_\mu - L_\tau$ model, remain viable. An interesting new example which was not yet discussed in the literature is the chiral model $\tilde L_{\mu-\tau}$ introduced in Section~\ref{sec:skewed}.
(See Section~\ref{sec:pheno} for further details and several benchmark models.)

Another application of the lepton-flavored $\U(1)_X$ is in the models of radiative muon mass generation that can simultaneously lead to the explanation of the $(g-2)_\mu$ anomaly.
The models with two TeV-scale leptoquarks exhibiting a parity symmetry were already known to realize such a scenario.  We point out that the parity symmetry can be an automatic consequence of the $\U(1)_X$ gauge symmetry.
The mixing between the scalars is then generated once the $\U(1)_X$ becomes spontaneously broken.
Further details on such scenarios are given in Section~\ref{sec:radiative}.

Our final application concerns the conundrum that whereas scalar leptoquarks are easily charged under the $\U(1)_X$ the vector leptoquarks are not.
In Section~\ref{sec:4321} we settle the question whether or not it is even possible to have a vector muoquark in a perturbative UV complete framework.
For this purpose we built a proof-of-principle UV complete gauge model for the Pati-Salam vector leptoquark, carrying a nonzero muon number.

In conclusion, the lepton-flavored $\U(1)_X$ extensions of the SM provide a powerful framework to address the current flavor anomalies.
There is a variety of mediators to be considered, such as light or heavy gauge bosons $X_\mu$, heavy scalar-, or vector-leptoquarks.
The richness of the phenomenology such extensions entail is apparent from the large number of possible chiral-anomaly-free charge assignments.
Within the present work we probably only scratched the surface of possibilities. One can imagine several different interesting future research directions. As shown in Ref.~\cite{Greljo:2021xmg}, the minimal realization of the neutrino masses imposes nontrivial requirements on the model building. Having a more complete study of implications from $\U(1)_X$ gauge symmetry on the neutrino sector would be highly desirable. Another open question is what would be the result of relaxing assumption about the cancellations of anomalies in the IR, allowing instead for chiral fermions that obtain the mass from $\U(1)_X$ breaking vacuum expectation values. We anticipate that many, if not all, of the possible models can be successfully probed experimentally, since the parameter space of interest for explaining the $(g-2)_\mu$ anomaly is mostly within reach of the next generation experiments.

\section*{Acknowledgments}
We thank Philip Ilten for updating the \textsc{DarkCast} in time for the updated version to be used in this work.
We also thank Pilar Coloma, Joe Davighi, Javier Fuentes-Martin, Kohsaku Tobioka, Werner Rodejohann, Enrico Sessolo and Ben Stefanek for useful discussions.
The work of AG and AET has received funding from the Swiss National Science Foundation (SNF) through the Eccellenza Professorial Fellowship ``Flavor Physics at the High Energy Frontier'' project number 186866.
The work of AG is also partially supported by the European Research Council (ERC) under the European Union’s Horizon 2020 research and innovation programme, grant agreement 833280 (FLAY).
The work of PS is supported by the SNF grant 200020175449/1.
The work of YS is supported by grants from the United States-Israel Binational Science Foundation (BSF) (NSF-BSF program grant No. 2018683), by the Israel Science Foundation (grant No. 482/20) and by the Azrieli foundation. YS is Taub fellow (supported by the Taub Family Foundation). JZ acknowledges support in part by the DOE grant de-sc0011784.

\appendix

\section{Kinetic mixing}
\label{app:kin:mix}

\subsection{Equivalence of charge assignments}
\label{app:reparam}
For a  gauge group with multiple Abelian factors, there is a continuum of physically equivalent choices for the products of $\U(1)$ subgroups.
The freedom in defining the Abelian factors then translates into a continuum of physically equivalent charge assignments for the matter field representations.

To demonstrate the equivalence of the charge assignments, let us consider a toy model consisting of a $ \U(1)^n=\U(1)_1\times \cdots\times \U(1)_n $ gauge group, with associated gauge fields $ A^\mu_i $, $i=1,\ldots, n$, and a set of matter fields $\{ f\}$ (fermions and/or scalars) carrying charges $ q^f_i $ under group $\U(1)_i$.
It is useful to absorb
the gauge couplings
into the definition of the gauge fields so that the kinetic term and the covariant derivative are given respectively by (see, e.g., Ref.~\cite{Poole:2019kcm})
    \begin{equation}
    \mathcal{L} \supset -\tfrac{1}{4} F_{i\mu\nu} h^{\eminus 1}_{ij} F_j^{\mu\nu}, \qquad D^f_\mu = \partial_\mu - i q_i^f A_{i\mu}.
    \end{equation}
All the information on the gauge couplings and kinetic mixing parameters is contained in the symmetric, positive-definite matrix $ h $.

Consider now the effect of a linear field transformation $ A_i^\mu \to L_{ij} A_j^\mu $, where $ L \in \mathbb{R}^{n\times n} $ is invertible.  The kinetic term and the covariant derivative expressed in terms of the transformed fields are given by
    \begin{equation}
    \mathcal{L} \supset -\tfrac{1}{4} F_{i\mu\nu} \tilde{h}^{\eminus 1}_{ij} F_j^{\mu\nu}, \qquad D^f_\mu = \partial_\mu - i \tilde{q}_i^f A_{i\mu},
    \end{equation}
where the new coupling matrix $ \tilde{h} = L^{\eminus 1} h (L^{\eminus 1})^ \intercal$ is still positive-definite due to invertibility of $ L $, ensuring the validity of the transformed theory. The matter charges are also linearly transformed, $ \tilde{q}^f = L^\intercal q^f $.
Since the field redefinitions give physically equivalent theories, the linear transformations $L_{ij}$ give a family of physically equivalent choices for the definitions of $\U(1)$ factors, with appropriately transformed matter field charges. We limit ourselves to rational charges, thus $ L \in \mathbb{Q}^{n\times n} $.

\subsection{Mass basis of the gauge sector} \label{app:gauge_mass_basis}
We next turn to the example at hand, the SM supplemented by the $\U(1)_X$ gauge group. To determine the mass eigenstates in the neutral gauge boson sector, it suffices to focus on the $\SU(2)_L\times \U(1)_Y\times \U(1)_X$ subgroup.
We assume that the $ \U(1)_X $ is spontaneously broken by a VEV of a SM singlet, resulting in a mass term, $\tfrac{1}{2} m_X^2 X_\mu^2$, but can otherwise remain agnostic about the specifics of the $\U(1)_X$ breaking sector.
That is, the part of the Lagrangian describing the EW and $ X_\mu $ gauge interactions is, after $\U(1)_X$ breaking, given by
	\begin{equation} \label{eq:lag_EWX}
	\begin{split}
	\L \supset &- \tfrac{1}{4} B_{\mu\nu}^2 - \tfrac{1}{4} X^2_{\mu\nu} + \tfrac{1}{2} \varepsilon \, B_{\mu\nu} X^{\mu\nu} - \tfrac{1}{4} (W_{\mu\nu}^a)^2 + |D_\mu H|^2 +  \tfrac{1}{2} m_X^2 X_\mu^2\\
	&+ g_1 B_\mu J_Y^\mu + g_2 W^a_\mu J_W^{a\mu} + g_X X_\mu J_X^\mu,
	\end{split}
	\end{equation}
where $ J^\mu_{X,Y,W} $ are the respective fermion currents.
After EW symmetry breaking due to the SM Higgs VEV the above Lagrangian is,
	\begin{equation}
	\begin{split}
	\L \supset &- \tfrac{1}{4} \big(A_{\mu\nu}^2 + X^2_{\mu\nu}+ Z_{\mu\nu}^2\big) + \tfrac{1}{2} \varepsilon \, \big(c_w A_{\mu\nu} - s_w Z_{\mu\nu} \big) X^{\mu\nu} + \tfrac{1}{2} M_Z^2 Z_\mu^2 + \tfrac{1}{2} m_X^2 X_\mu^2\\
	&+ e\, A_\mu J_\sscript{EM}^\mu + g_Z Z_\mu J_Z^{\mu} + g_X X_\mu J_X^\mu,
	\end{split}
	\end{equation}
where $Z_\mu, A_\mu$ are the would be mass eigenstates had we had only the SM gauge group and are given by $ B_\mu= c_w Z_\mu - s_w A_\mu $ and $ W_\mu^3 = c_w Z_\mu + s_w A_\mu $, where $c_w\equiv \cos \theta_w$, $s_w \equiv \sin \theta_w$ with $\theta_w$ the weak mixing angle. Also, $ g_Z = \sqrt{g_1^2 + g_2^2}$ is the usual coupling constant of the $ Z_\mu$.
A non-unitary transformation of the fields,
	\begin{equation}
	A\longrightarrow A + \dfrac{c_w \varepsilon}{r_\varepsilon} X, \qquad Z \longrightarrow Z - \dfrac{s_w \varepsilon}{r_\varepsilon} X, \andeq X\longrightarrow \dfrac{1}{r_\varepsilon} X, \qquad r_\varepsilon = \sqrt{1- \varepsilon^2},
	\end{equation}
eliminates the kinetic-mixing terms:
	\begin{equation}
	\begin{split}
	\L \supset &- \tfrac{1}{4} \big(A_{\mu\nu}^2 + X^2_{\mu\nu}+ Z_{\mu\nu}^2\big) + \dfrac{M_Z^2}{2} \left(Z_\mu - \dfrac{s_w \varepsilon}{r_\varepsilon} X_\mu \right)^{\!\! 2} + \dfrac{m_X^2}{2 r_\varepsilon^2} X_\mu^2\\
	&+ e\, A_\mu J_\sscript{EM}^\mu + g_Z Z_\mu J_Z^{\mu} + \dfrac{1}{r_\varepsilon} X_\mu \big(g_X J_X^\mu + c_w\varepsilon\, e J_A^\mu - s_w \varepsilon \, g_Z J_Z^\mu \big).
	\end{split}
	\end{equation}
For $ \varepsilon \ll 1 $, a rotation by $ s_w \varepsilon $ between $ Z $ and $ X $ suffices to mass-diagonalize the Lagrangian up to subleading terms, giving (in a slightly abused notation with $Z_\mu$, $X_\mu$ now denoting the rotated fields)
	\begin{equation}
	\begin{split}
	\L \supset &- \tfrac{1}{4} \big(A_{\mu\nu}^2 + X^2_{\mu\nu}+ Z_{\mu\nu}^2\big) + \tfrac{1}{2} M_Z^2 Z_\mu^2 + \tfrac{1}{2 }m_X^2 X_\mu^2\\
	&+ e\, A_\mu J_\sscript{EM}^\mu + g_Z Z_\mu J_Z^{\mu} + X_\mu \big(g_X J_X^\mu + c_w\varepsilon\, e J_A^\mu \big) \\
	&+ \mathcal{O}\big(g_X \varepsilon,\, \varepsilon^2,\, \varepsilon m_X^2/M_Z^2 \big).
	\end{split}
	\end{equation}
The result is that the effective $ \U(1)_X $ charge of the matter fields is
	\begin{equation}
	q_X^{\sscript{eff}} = q_X +\dfrac{c_w e\, \varepsilon}{g_X} q_\sscript{EM},
	\end{equation}
where $ q_\sscript{EM} $ is the ordinary EM charge.

\subsection{The RG running of the kinetic mixing parameter} \label{app:epsilon_running}
The 1-loop running of the kinetic mixing parameter, as in Eq.~\eqref{eq:lag_EWX}, is given by~\cite{Poole:2019kcm}
    \begin{equation}
    \label{eq:RG:varepsilon}
    \dfrac{\dd \varepsilon}{\dd t} = \dfrac{g_1 g_X }{48\pi^2} \left[2 s_{BX} + s_{BB} \varepsilon \dfrac{g_1}{g_X} + s_{XX} \varepsilon \dfrac{g_X}{g_1} \right], \qquad t = \ln \frac{\mu}{\mu_0},
    \end{equation}
    with $\mu$ the renormalization scale, and $\mu_0$ the starting scale for RG running, to be taken $\mu_0=3$~TeV in the numerics below.
The coefficients $ s_{xy} $ are determined by the charges of the matter fields of the model. We have
    \begin{equation}
    s_{xy} = 2 \Tr_F\big[Q_x Q_y \big] + \Tr_S\big[Q_x Q_y \big],
    \end{equation}
with the two traces running over the Weyl-spinor and complex-scalar degrees of freedom, respectively.
For SM+$ 3\nu_\RR $ with an $S_3$ muoquark the coefficients in the RG equation \eqref{eq:RG:varepsilon} are
    \begin{equation}
    \begin{split}
    s_{BB} = \frac{43}{2}, \qquad s_{BX} = 12 X_q - 3(X_q + X_{L_2})  - 2 \sum_i(X_{L_i} + X_{E_2}), \\
    s_{XX} = 72 X_q^2 + 9 (X_q + X_{L_2})^2  + 2 \sum_i\big[ 2X_{L_i}^2 + X_{E_i}^2 + X_{N_i}^2 \big].
    \end{split}
    \end{equation}
In the models we consider the contribution from $ s_{XX} $ is numerically  negligible.

The 1-loop RGE can be integrated analytically in the $ g_X, \varepsilon \ll g_B $ limit. The $ g_X $ gauge coupling can be treated as a constant to a very good approximation, since
    \begin{equation}
    \dfrac{\dd g_X}{\dd t} = \mathcal{O}\big(g_X \varepsilon^2, \, g_X^2 \varepsilon,\, g_X^3\big).
    \end{equation}
The running of the hypercharge coupling is given by
    \begin{equation}
    \dfrac{\dd g_1}{\dd t} = \dfrac{s_{BB}}{48\pi^2} g_1^3 + \mathcal{O} \big( g_X \varepsilon \big) \implies g_1(t) = \dfrac{\sqrt{24\pi^2/ s_{BB} } }{\sqrt{T- t}}, \qquad T= \dfrac{24 \pi^2 }{s_{BB}\, g^2_{1}(0) },
    \end{equation}
with $g_{1}(0)$ is the value of hypercharge coupling constant at scale $\mu_0$. The running of $ \varepsilon $ is driven by the hypercharge and integrates to
    \begin{equation}
    \varepsilon = \dfrac{\sqrt{T} \varepsilon_0 + b t}{ \sqrt{T-t}}, \qquad \text{where} \quad b= \dfrac{g_1 s_{BX}}{\sqrt{24 \pi^2 s_{BB}}}.
    \end{equation}
For RG running from the leptoquark mass scale, $ \mu_0 = \SI{3}{TeV} $, to the Planck mass, $M_{\rm Pl}\simeq 1.2 \cdot 10^{19}\SI{}{GeV}$, we have $ T\simeq 85 $, while $ t_\mathrm{Pl} = 36$. To a reasonable approximation, we have
    \begin{equation}
    \varepsilon(M_\mathrm{Pl}) - \varepsilon(\mu_0) \simeq \num{0.32} \varepsilon(\mu_0) + \num{0.072} s_{BX}\, g_X.
    \end{equation}

\section{The $X$ boson phenomenology}

\subsection{Decay channels}
\label{app:branchingratios}

\begin{figure}[t]
    \centering
    \includegraphics[width=\textwidth]{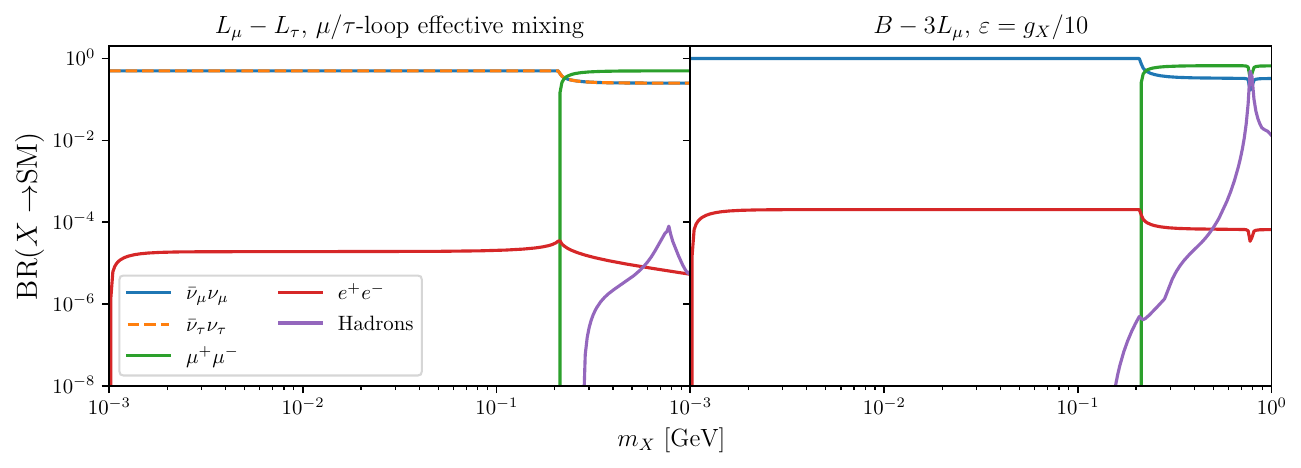}
    \caption{The $X$ boson decay branching ratios in $L_\mu - L_\tau$ model (left) and $B - 3 L_\mu$ model (right).}
    \label{fig:BR}
\end{figure}

In Fig.~\ref{fig:BR} we plot for clarity the branching ratios of the $X_\mu$ boson for several final states as a function of the $X_\mu$ mass $m_X$ derived with \textsc{DarkCast}. The two benchmark models are presented in Section~\ref{sec:LmuLtau} and Section~\ref{sec:bm3Lmu}.

\vspace{10pt}

\subsection{\RK from a light $X_\mu$ vector boson}
\label{app:RKX}

Here we extend the discussion given in Section~\ref{sec:single_mediator} regarding the predictions for \RK in the presence of a light vector boson $X_\mu$ .
The \RK observables are measured in bins of the invariant dilepton mass squared $q^2$, and
 are given in terms of the $q^2$-differential branching ratios by
\begin{equation}\label{eq:RK_def}
 R_{K^{(*)}}^{[q^2_\text{min},\,q^2_\text{max}]}
 =
 \frac{
 \int_{q^2_\text{min}}^{q^2_\text{max}} dq^2\, \frac{d\text{BR}(B\to K^{(*)}\mu^+\mu^-)}{dq^2}
 }{
 \int_{q^2_\text{min}}^{q^2_\text{max}} dq^2\, \frac{d\text{BR}(B\to K^{(*)}e^+e^-)}{dq^2}
 }\,.
\end{equation}
In the SM and for heavy NP particles, the differential branching ratios can be expressed as
\begin{equation}
 \frac{d\text{BR}(B\to K^{(*)}\ell^+\ell^-)}{dq^2} \simeq {\rm Re} \sum_{i,j} f_{ij}(q^2)\, C_i\, C_j^*\,,
\end{equation}
where $f_{ij}(q^2)$ are $q^2$-dependent functions that depend on the parameters such as meson masses and hadronic form factors, while $C_i$ denote the $q^2$-independent Wilson coefficients of effective operators in the weak Hamiltonian.
In the presence of a light NP mediator, its tree level effect can be modeled by introducing $q^2$-dependent Wilson coefficients.
In particular, the Wilson coefficients most relevant in the presence of the Lagrangian Eq.~\eqref{eq:Lsimp} are given in Eq.~\eqref{eq:WCs_q2} and repeated here for convenience,
\begin{align}\label{eq:WCs_q2_app}
    C_{9,10}^{(\prime)}(q^2)
=  \frac{q_{V,A}}{N} \,\frac{g_X\, g_{L(R)}^{bs}}{q^2 - m_X^2 + i m_X\Gamma_X}\,.
\end{align}
It is evident that due to these Wilson coefficients, the $q^2$ dependence of the differential branching ratios, and thus the value of the integrals in Eq.~\eqref{eq:RK_def}, strongly depends on the mass $m_X$.
\begin{figure}[t]
    \centering
    \includegraphics[width=0.8\textwidth]{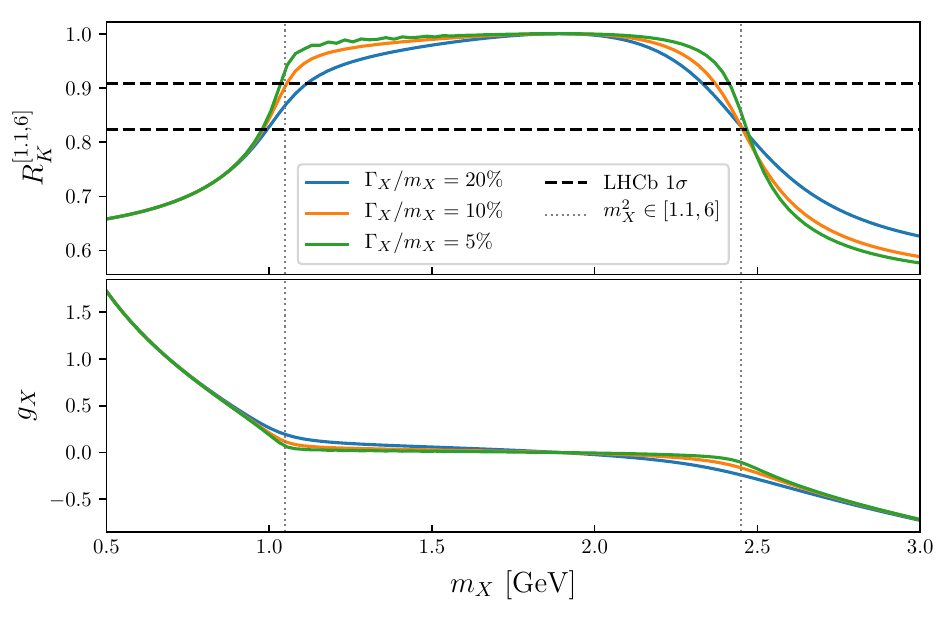}
    \caption{\emph{Upper panel:} Minimal values of $R_K^{[1.1,6]}$ from $q^2$-dependent Wilson coefficients in Eq.~\eqref{eq:WCs_q2_app} with fixed $q_{V}=1$, $q_{A}=0$, $g_L^{bs}=0.7 \times 10^{-8}\, \frac{m_X}{\text{GeV}}$ for different values of $m_X$ and $\Gamma_X$.
    Blue, orange, and green lines correspond to the values obtained for a relative width $\Gamma_X/m_X$ equal to 20\%, 10\%, and 5\%, respectively.
    Dashed black lines correspond to the $1\sigma$ region preferred by the LHCb measurement~\cite{LHCb:2021trn}. Dotted gray lines mark the points where $m_X$ coincides with the boundaries of the bin $q^2\in[1.1,6]\,\text{GeV}^2$.
    \emph{Lower panel:} Values of $g_X$ corresponding to the minimal values of $R_K^{[1.1,6]}$ shown in the upper panel.}
    \label{fig:RKX}
\end{figure}
To demonstrate this, we show in the upper panel of Fig.~\ref{fig:RKX}  the minimal value of $R_K^{[1.1,6]}$ that can be obtained using the Wilson coefficients in Eq.~\ref{eq:WCs_q2_app} for different masses $m_X$ and different widths $\Gamma_X$. We have fixed $q_{V}=1$, $q_{A}=0$, $g_R^{bs}=0$, $g_L^{bs}=0.7 \times 10^{-8}\, \frac{m_X}{\text{GeV}}$ (cf.~Eq.~\eqref{eq:gbs_bound}), but kept $g_X$ as a free parameter.
The lower panel in Fig.~\ref{fig:RKX} shows the $g_X$ values that correspond to the minimal values of $R_K^{[1.1,6]}$ shown in the upper panel.
The black dashed lines represent the $1\sigma$ region of the LHCb measurement~\cite{LHCb:2021trn}, while the gray dotted lines show the boundaries of the $q^2$ bin where $m_X^2=q^2_\text{min}=1.1\,\GeV^2$ and $m_X^2=q^2_\text{max}=6\,\GeV^2$.
One can clearly see that for $m_X$ inside the bin, the minimal value of $R_K$ is always close to the SM value $R_K^{[1.1,6]}\simeq1$, while the corresponding value of $g_X$ drops to zero.

To understand this behavior, it is convenient to consider the pure NP contribution proportional to $\left|C_9(q^2)\right|^2$,
\begin{equation}
 \left|C_{9}(q^2)\right|^2
=  \frac{q_V^2\,g_X^2\, (g_{L}^{bs})^2}{|N|^2} \,\frac{1}{(q^2 - m_X^2)^2 + m_X^2\Gamma_X^2}\,.
\end{equation}
For $\Gamma_X^2\ll m_X^2$, one can use the narrow width approximation (NWA),
\begin{equation}
 \frac{1}{(q^2 - m_X^2)^2 + m_X^2\Gamma_X^2}\ \xrightarrow{\Gamma_X^2/ m_X^2\to 0}\ \frac{\pi}{\Gamma_X\,m_X}\,\delta(q^2-m_X^2)\,,
\end{equation}
such that we get
\begin{equation}
 \left|C_{9}(q^2)\right|^2\Big|_{\Gamma_X^2\ll m_X^2} \approx
 \frac{\pi\,q_V^2\,g_X^2\, (g_{L}^{bs})^2}{|N|^2\, \Gamma_X\,m_X}\,\delta(q^2-m_X^2)\,.
\end{equation}
In the NWA, the $\delta$ function in the pure NP contribution dominates the integral in Eq.~\eqref{eq:RK_def} if $m_X^2$ is inside the interval of integration. This contribution is always \emph{positive}, so if $X_\mu$ couples to muons, the numerator of \RK is always enhanced in this case. Since no suppression is possible, the minimal value of $R_K^{[1.1,6]}$ is just the SM value. At the same time, $g_X$ has to vanish in order not to enhance $R_K^{[1.1,6]}$ even further from the experimental result.
On the other hand, if $m_X^2$ is outside the interval of integration, the pure NP contribution vanishes in the NWA and the interference with the SM contribution can lead to a suppression of the numerator in $R_K^{[1.1,6]}$.

The effect described above is most pronounced for the narrow $X_\mu$ width and is reduced for a broad $X_\mu$, see  Fig.~\ref{fig:RKX}.
However, even for rather wide resonances (i.e. $\Gamma_X / m_X = 20\%$), it is only possible to achieve the $R_K^{[1.1,6]}$ value preferred by the LHCb results for $m_X^2$ either very close to the boundaries of the $[1.1,6]\,\text{GeV}^2$ bin or for $m_X^2$ outside this interval.

\section{The generators of $\U(1)$ embeddings in $\SU(5)$}
\label{app:su5}

The diagonal generators of $\SU(5)$ that commute with all the generators of the conventional QCD gauge group embedding in the $\SU(5)$, $\SU(3)_5\subset \SU(5)$ can be chosen as
\begin{equation}
 T_5^{15} = \frac{1}{2\sqrt{6}}\diag(1,1,1,-3,0)\,,
 \qquad
 T_5^{24} = \frac{1}{2\sqrt{10}}\diag(1,1,1,1,-4)\,.
\end{equation}
The generators of $X_4$ and $X_{5'}$ of $\U(1)_4$ and $\U(1)_{5'}$ used in section~\ref{sec:5321_gauge_sector} are proportional to these $\SU(5)$ generators:
\begin{equation}
 X_{4}=\frac{2}{\sqrt{6}}\, T_5^{15} = \frac{1}{6}\diag(1,1,1,-3,0)\,,
\end{equation}
and
\begin{equation}
 X_{5'}=\frac{2}{\sqrt{10}}\, T_5^{24} = \frac{1}{10}\diag(1,1,1,1,-4)\,.
\end{equation}
The generators $X_5$ of $\U(1)_5$ and $X_{\mu-\tau}$  of $\U(1)_{\mu-\tau}$ are given by the two linear combinations
\begin{equation}
 X_5
 = X_{4} + X_{5'}
 = \frac{2}{15}\diag\left(2,2,2,-3,-3\right),
\end{equation}
and
\begin{equation}
 X_{\mu-\tau}
 = \frac{3}{2}\, X_{4} - \frac{5}{2}\, X_{5'}
 = \diag(0,0,0,-1,1)\,.
\end{equation}

\bibliographystyle{JHEP}
\bibliography{g2biblio}

\providecommand{\href}[2]{#2}\begingroup\raggedright\begin{thebibliography}{100}

\bibitem{Greljo:2021xmg}
A.~Greljo, P.~Stangl and A.~E. Thomsen, \emph{{A Model of Muon Anomalies}},
  \href{https://arxiv.org/abs/2103.13991}{{\ttfamily 2103.13991}}.

\bibitem{Bennett:2006fi}
{\scshape Muon g-2} collaboration, G.~W. Bennett et~al., \emph{{Final Report of
  the Muon E821 Anomalous Magnetic Moment Measurement at BNL}},
  \href{https://doi.org/10.1103/PhysRevD.73.072003}{\emph{Phys. Rev. D}
  {\bfseries 73} (2006) 072003},
  [\href{https://arxiv.org/abs/hep-ex/0602035}{{\ttfamily hep-ex/0602035}}].

\bibitem{Aoyama:2020ynm}
T.~Aoyama et~al., \emph{{The anomalous magnetic moment of the muon in the
  Standard Model}},
  \href{https://doi.org/10.1016/j.physrep.2020.07.006}{\emph{Phys. Rept.}
  {\bfseries 887} (2020) 1--166},
  [\href{https://arxiv.org/abs/2006.04822}{{\ttfamily 2006.04822}}].

\bibitem{Colangelo:2020lcg}
G.~Colangelo, M.~Hoferichter and P.~Stoffer, \emph{{Constraints on the two-pion
  contribution to hadronic vacuum polarization}},
  \href{https://doi.org/10.1016/j.physletb.2021.136073}{\emph{Phys. Lett. B}
  {\bfseries 814} (2021) 136073},
  [\href{https://arxiv.org/abs/2010.07943}{{\ttfamily 2010.07943}}].

\bibitem{aoyama:2012wk}
T.~Aoyama, M.~Hayakawa, T.~Kinoshita and M.~Nio, \emph{{Complete Tenth-Order
  QED Contribution to the Muon $g-2$}},
  \href{https://doi.org/10.1103/PhysRevLett.109.111808}{\emph{Phys. Rev. Lett.}
  {\bfseries 109} (2012) 111808},
  [\href{https://arxiv.org/abs/1205.5370}{{\ttfamily 1205.5370}}].

\bibitem{Aoyama:2019ryr}
T.~Aoyama, T.~Kinoshita and M.~Nio, \emph{{Theory of the Anomalous Magnetic
  Moment of the Electron}},
  \href{https://doi.org/10.3390/atoms7010028}{\emph{Atoms} {\bfseries 7} (2019)
  28}.

\bibitem{czarnecki:2002nt}
A.~Czarnecki, W.~J. Marciano and A.~Vainshtein, \emph{{Refinements in
  electroweak contributions to the muon anomalous magnetic moment}},
  \href{https://doi.org/10.1103/PhysRevD.67.073006}{\emph{Phys. Rev.}
  {\bfseries D67} (2003) 073006},
  [\href{https://arxiv.org/abs/hep-ph/0212229}{{\ttfamily hep-ph/0212229}}].

\bibitem{gnendiger:2013pva}
C.~Gnendiger, D.~St{\"o}ckinger and H.~St{\"o}ckinger-Kim, \emph{{The
  electroweak contributions to $(g-2)_\mu$ after the Higgs boson mass
  measurement}}, \href{https://doi.org/10.1103/PhysRevD.88.053005}{\emph{Phys.
  Rev.} {\bfseries D88} (2013) 053005},
  [\href{https://arxiv.org/abs/1306.5546}{{\ttfamily 1306.5546}}].

\bibitem{davier:2017zfy}
M.~Davier, A.~Hoecker, B.~Malaescu and Z.~Zhang, \emph{{Reevaluation of the
  hadronic vacuum polarisation contributions to the Standard Model predictions
  of the muon $g-2$ and ${\alpha (m_Z^2)}$ using newest hadronic cross-section
  data}}, \href{https://doi.org/10.1140/epjc/s10052-017-5161-6}{\emph{Eur.
  Phys. J.} {\bfseries C77} (2017) 827},
  [\href{https://arxiv.org/abs/1706.09436}{{\ttfamily 1706.09436}}].

\bibitem{keshavarzi:2018mgv}
A.~Keshavarzi, D.~Nomura and T.~Teubner, \emph{{Muon $g-2$ and $\alpha(M_Z^2)$:
  a new data-based analysis}},
  \href{https://doi.org/10.1103/PhysRevD.97.114025}{\emph{Phys. Rev.}
  {\bfseries D97} (2018) 114025},
  [\href{https://arxiv.org/abs/1802.02995}{{\ttfamily 1802.02995}}].

\bibitem{colangelo:2018mtw}
G.~Colangelo, M.~Hoferichter and P.~Stoffer, \emph{{Two-pion contribution to
  hadronic vacuum polarization}},
  \href{https://doi.org/10.1007/JHEP02(2019)006}{\emph{JHEP} {\bfseries 02}
  (2019) 006}, [\href{https://arxiv.org/abs/1810.00007}{{\ttfamily
  1810.00007}}].

\bibitem{hoferichter:2019gzf}
M.~Hoferichter, B.-L. Hoid and B.~Kubis, \emph{{Three-pion contribution to
  hadronic vacuum polarization}},
  \href{https://doi.org/10.1007/JHEP08(2019)137}{\emph{JHEP} {\bfseries 08}
  (2019) 137}, [\href{https://arxiv.org/abs/1907.01556}{{\ttfamily
  1907.01556}}].

\bibitem{davier:2019can}
M.~Davier, A.~Hoecker, B.~Malaescu and Z.~Zhang, \emph{{A new evaluation of the
  hadronic vacuum polarisation contributions to the muon anomalous magnetic
  moment and to $\mathbf{\boldsymbol\alpha(m_Z^2)}$}},
  \href{https://doi.org/10.1140/epjc/s10052-020-7792-2}{\emph{Eur. Phys. J.}
  {\bfseries C80} (2020) 241},
  [\href{https://arxiv.org/abs/1908.00921}{{\ttfamily 1908.00921}}].

\bibitem{keshavarzi:2019abf}
A.~Keshavarzi, D.~Nomura and T.~Teubner, \emph{{The $g-2$ of charged leptons,
  $\alpha(M_Z^2)$ and the hyperfine splitting of muonium}},
  \href{https://doi.org/10.1103/PhysRevD.101.014029}{\emph{Phys. Rev.}
  {\bfseries D101} (2020) 014029},
  [\href{https://arxiv.org/abs/1911.00367}{{\ttfamily 1911.00367}}].

\bibitem{kurz:2014wya}
A.~Kurz, T.~Liu, P.~Marquard and M.~Steinhauser, \emph{{Hadronic contribution
  to the muon anomalous magnetic moment to next-to-next-to-leading order}},
  \href{https://doi.org/10.1016/j.physletb.2014.05.043}{\emph{Phys. Lett.}
  {\bfseries B734} (2014) 144--147},
  [\href{https://arxiv.org/abs/1403.6400}{{\ttfamily 1403.6400}}].

\bibitem{melnikov:2003xd}
K.~Melnikov and A.~Vainshtein, \emph{{Hadronic light-by-light scattering
  contribution to the muon anomalous magnetic moment revisited}},
  \href{https://doi.org/10.1103/PhysRevD.70.113006}{\emph{Phys. Rev.}
  {\bfseries D70} (2004) 113006},
  [\href{https://arxiv.org/abs/hep-ph/0312226}{{\ttfamily hep-ph/0312226}}].

\bibitem{masjuan:2017tvw}
P.~Masjuan and P.~S{\'a}nchez-Puertas, \emph{{Pseudoscalar-pole contribution to
  the $(g_{\mu}-2)$: a rational approach}},
  \href{https://doi.org/10.1103/PhysRevD.95.054026}{\emph{Phys. Rev.}
  {\bfseries D95} (2017) 054026},
  [\href{https://arxiv.org/abs/1701.05829}{{\ttfamily 1701.05829}}].

\bibitem{Colangelo:2017fiz}
G.~Colangelo, M.~Hoferichter, M.~Procura and P.~Stoffer, \emph{{Dispersion
  relation for hadronic light-by-light scattering: two-pion contributions}},
  \href{https://doi.org/10.1007/JHEP04(2017)161}{\emph{JHEP} {\bfseries 04}
  (2017) 161}, [\href{https://arxiv.org/abs/1702.07347}{{\ttfamily
  1702.07347}}].

\bibitem{hoferichter:2018kwz}
M.~Hoferichter, B.-L. Hoid, B.~Kubis, S.~Leupold and S.~P. Schneider,
  \emph{{Dispersion relation for hadronic light-by-light scattering: pion
  pole}}, \href{https://doi.org/10.1007/JHEP10(2018)141}{\emph{JHEP} {\bfseries
  10} (2018) 141}, [\href{https://arxiv.org/abs/1808.04823}{{\ttfamily
  1808.04823}}].

\bibitem{gerardin:2019vio}
A.~G{\'e}rardin, H.~B. Meyer and A.~Nyffeler, \emph{{Lattice calculation of the
  pion transition form factor with $N_f=2+1$ Wilson quarks}},
  \href{https://doi.org/10.1103/PhysRevD.100.034520}{\emph{Phys. Rev.}
  {\bfseries D100} (2019) 034520},
  [\href{https://arxiv.org/abs/1903.09471}{{\ttfamily 1903.09471}}].

\bibitem{bijnens:2019ghy}
J.~Bijnens, N.~Hermansson-Truedsson and A.~Rodr{\'i}guez-S{\'a}nchez,
  \emph{{Short-distance constraints for the HLbL contribution to the muon
  anomalous magnetic moment}},
  \href{https://doi.org/10.1016/j.physletb.2019.134994}{\emph{Phys. Lett.}
  {\bfseries B798} (2019) 134994},
  [\href{https://arxiv.org/abs/1908.03331}{{\ttfamily 1908.03331}}].

\bibitem{colangelo:2019uex}
G.~Colangelo, F.~Hagelstein, M.~Hoferichter, L.~Laub and P.~Stoffer,
  \emph{{Longitudinal short-distance constraints for the hadronic
  light-by-light contribution to $(g-2)_\mu$ with large-$N_c$ Regge models}},
  \href{https://doi.org/10.1007/JHEP03(2020)101}{\emph{JHEP} {\bfseries 03}
  (2020) 101}, [\href{https://arxiv.org/abs/1910.13432}{{\ttfamily
  1910.13432}}].

\bibitem{Blum:2019ugy}
T.~Blum, N.~Christ, M.~Hayakawa, T.~Izubuchi, L.~Jin, C.~Jung et~al.,
  \emph{{The hadronic light-by-light scattering contribution to the muon
  anomalous magnetic moment from lattice QCD}},
  \href{https://doi.org/10.1103/PhysRevLett.124.132002}{\emph{Phys. Rev. Lett.}
  {\bfseries 124} (2020) 132002},
  [\href{https://arxiv.org/abs/1911.08123}{{\ttfamily 1911.08123}}].

\bibitem{colangelo:2014qya}
G.~Colangelo, M.~Hoferichter, A.~Nyffeler, M.~Passera and P.~Stoffer,
  \emph{{Remarks on higher-order hadronic corrections to the muon $g-2$}},
  \href{https://doi.org/10.1016/j.physletb.2014.06.012}{\emph{Phys. Lett.}
  {\bfseries B735} (2014) 90--91},
  [\href{https://arxiv.org/abs/1403.7512}{{\ttfamily 1403.7512}}].

\bibitem{Borsanyi:2020mff}
S.~Borsanyi et~al., \emph{{Leading hadronic contribution to the muon 2 magnetic
  moment from lattice QCD}},
  \href{https://arxiv.org/abs/2002.12347}{{\ttfamily 2002.12347}}.

\bibitem{LHCb:2020lmf}
{\scshape LHCb} collaboration, R.~Aaij et~al., \emph{{Measurement of
  $CP$-Averaged Observables in the $B^{0}\rightarrow K^{*0}\mu^{+}\mu^{-}$
  Decay}}, \href{https://doi.org/10.1103/PhysRevLett.125.011802}{\emph{Phys.
  Rev. Lett.} {\bfseries 125} (2020) 011802},
  [\href{https://arxiv.org/abs/2003.04831}{{\ttfamily 2003.04831}}].

\bibitem{LHCb:2020gog}
{\scshape LHCb} collaboration, R.~Aaij et~al., \emph{{Angular Analysis of the
  $B^{+}\rightarrow K^{\ast+}\mu^{+}\mu^{-}$ Decay}},
  \href{https://doi.org/10.1103/PhysRevLett.126.161802}{\emph{Phys. Rev. Lett.}
  {\bfseries 126} (2021) 161802},
  [\href{https://arxiv.org/abs/2012.13241}{{\ttfamily 2012.13241}}].

\bibitem{LHCb:2020zud}
{\scshape LHCb} collaboration, \emph{{Combination of the ATLAS, CMS and LHCb
  results on the $B^0_{(s)} \to \mu^+ \mu^-$ decays}}, .

\bibitem{LHCb:2021awg}
{\scshape LHCb} collaboration, R.~Aaij et~al., \emph{{Measurement of the
  $B^0_s\to\mu^+\mu^-$ decay properties and search for the $B^0\to\mu^+\mu^-$
  and $B^0_s\to\mu^+\mu^-\gamma$ decays}},
  \href{https://arxiv.org/abs/2108.09283}{{\ttfamily 2108.09283}}.

\bibitem{LHCb:2021vsc}
{\scshape LHCb} collaboration, R.~Aaij et~al., \emph{{Analysis of neutral
  $B$-meson decays into two muons}},
  \href{https://arxiv.org/abs/2108.09284}{{\ttfamily 2108.09284}}.

\bibitem{LHCb:2014cxe}
{\scshape LHCb} collaboration, R.~Aaij et~al., \emph{{Differential branching
  fractions and isospin asymmetries of $B \to K^{(*)} \mu^+ \mu^-$ decays}},
  \href{https://doi.org/10.1007/JHEP06(2014)133}{\emph{JHEP} {\bfseries 06}
  (2014) 133}, [\href{https://arxiv.org/abs/1403.8044}{{\ttfamily 1403.8044}}].

\bibitem{LHCb:2015wdu}
{\scshape LHCb} collaboration, R.~Aaij et~al., \emph{{Angular analysis and
  differential branching fraction of the decay $B^0_s\to\phi\mu^+\mu^-$}},
  \href{https://doi.org/10.1007/JHEP09(2015)179}{\emph{JHEP} {\bfseries 09}
  (2015) 179}, [\href{https://arxiv.org/abs/1506.08777}{{\ttfamily
  1506.08777}}].

\bibitem{LHCb:2016ykl}
{\scshape LHCb} collaboration, R.~Aaij et~al., \emph{{Measurements of the
  S-wave fraction in $B^{0}\rightarrow K^{+}\pi^{-}\mu^{+}\mu^{-}$ decays and
  the $B^{0}\rightarrow K^{\ast}(892)^{0}\mu^{+}\mu^{-}$ differential branching
  fraction}}, \href{https://doi.org/10.1007/JHEP11(2016)047}{\emph{JHEP}
  {\bfseries 11} (2016) 047},
  [\href{https://arxiv.org/abs/1606.04731}{{\ttfamily 1606.04731}}].

\bibitem{LHCb:2021zwz}
{\scshape LHCb} collaboration, R.~Aaij et~al., \emph{{Branching fraction
  measurements of the rare $B^0_s\rightarrow\phi\mu^+\mu^-$ and
  $B^0_s\rightarrow f_2^\prime(1525)\mu^+\mu^-$ decays}},
  \href{https://arxiv.org/abs/2105.14007}{{\ttfamily 2105.14007}}.

\bibitem{LHCb:2017avl}
{\scshape LHCb} collaboration, R.~Aaij et~al., \emph{{Test of lepton
  universality with $B^{0} \rightarrow K^{*0}\ell^{+}\ell^{-}$ decays}},
  \href{https://doi.org/10.1007/JHEP08(2017)055}{\emph{JHEP} {\bfseries 08}
  (2017) 055}, [\href{https://arxiv.org/abs/1705.05802}{{\ttfamily
  1705.05802}}].

\bibitem{LHCb:2021trn}
{\scshape LHCb} collaboration, R.~Aaij et~al., \emph{{Test of lepton
  universality in beauty-quark decays}},
  \href{https://arxiv.org/abs/2103.11769}{{\ttfamily 2103.11769}}.

\bibitem{Hiller:2003js}
G.~Hiller and F.~Kruger, \emph{{More model-independent analysis of $b \to s$
  processes}}, \href{https://doi.org/10.1103/PhysRevD.69.074020}{\emph{Phys.
  Rev. D} {\bfseries 69} (2004) 074020},
  [\href{https://arxiv.org/abs/hep-ph/0310219}{{\ttfamily hep-ph/0310219}}].

\bibitem{Bordone:2016gaq}
M.~Bordone, G.~Isidori and A.~Pattori, \emph{{On the Standard Model predictions
  for $R_K$ and $R_{K^*}$}},
  \href{https://doi.org/10.1140/epjc/s10052-016-4274-7}{\emph{Eur. Phys. J. C}
  {\bfseries 76} (2016) 440},
  [\href{https://arxiv.org/abs/1605.07633}{{\ttfamily 1605.07633}}].

\bibitem{Isidori:2020acz}
G.~Isidori, S.~Nabeebaccus and R.~Zwicky, \emph{{QED corrections in $
  \overline{B}\to \overline{K}{\mathrm{\ell}}^{+}{\mathrm{\ell}}^{-} $ at the
  double-differential level}},
  \href{https://doi.org/10.1007/JHEP12(2020)104}{\emph{JHEP} {\bfseries 12}
  (2020) 104}, [\href{https://arxiv.org/abs/2009.00929}{{\ttfamily
  2009.00929}}].

\bibitem{Altmannshofer:2021qrr}
W.~Altmannshofer and P.~Stangl, \emph{{New Physics in Rare B Decays after
  Moriond 2021}},  \href{https://arxiv.org/abs/2103.13370}{{\ttfamily
  2103.13370}}.

\bibitem{Geng:2021nhg}
L.-S. Geng, B.~Grinstein, S.~J\"ager, S.-Y. Li, J.~Martin~Camalich and R.-X.
  Shi, \emph{{Implications of new evidence for lepton-universality violation in
  $b\to s\ell^+\ell^-$ decays}},
  \href{https://arxiv.org/abs/2103.12738}{{\ttfamily 2103.12738}}.

\bibitem{Alguero:2021anc}
M.~Alguer\'o, B.~Capdevila, S.~Descotes-Genon, J.~Matias and M.~Novoa-Brunet,
  \emph{{$\boldsymbol{b\to s\ell\ell}$ global fits after Moriond 2021
  results}},  in \emph{{55th Rencontres de Moriond on QCD and High Energy
  Interactions}}, 4, 2021, \href{https://arxiv.org/abs/2104.08921}{{\ttfamily
  2104.08921}}.

\bibitem{Hurth:2021nsi}
T.~Hurth, F.~Mahmoudi, D.~M. Santos and S.~Neshatpour, \emph{{More Indications
  for Lepton Nonuniversality in $b \to s \ell^+ \ell^-$}},
  \href{https://arxiv.org/abs/2104.10058}{{\ttfamily 2104.10058}}.

\bibitem{Ciuchini:2020gvn}
M.~Ciuchini, M.~Fedele, E.~Franco, A.~Paul, L.~Silvestrini and M.~Valli,
  \emph{{Lessons from the $B^{0,+}\to K^{*0,+}\mu^+\mu^-$ angular analyses}},
  \href{https://doi.org/10.1103/PhysRevD.103.015030}{\emph{Phys. Rev. D}
  {\bfseries 103} (2021) 015030},
  [\href{https://arxiv.org/abs/2011.01212}{{\ttfamily 2011.01212}}].

\bibitem{TheMEG:2016wtm}
{\scshape MEG} collaboration, A.~M. Baldini et~al., \emph{{Search for the
  lepton flavour violating decay $\mu ^+ \rightarrow \mathrm {e}^+ \gamma $
  with the full dataset of the MEG experiment}},
  \href{https://doi.org/10.1140/epjc/s10052-016-4271-x}{\emph{Eur. Phys. J. C}
  {\bfseries 76} (2016) 434},
  [\href{https://arxiv.org/abs/1605.05081}{{\ttfamily 1605.05081}}].

\bibitem{BaBar:2009hkt}
{\scshape BaBar} collaboration, B.~Aubert et~al., \emph{{Searches for Lepton
  Flavor Violation in the Decays tau+- ---\ensuremath{>} e+- gamma and tau+-
  ---\ensuremath{>} mu+- gamma}},
  \href{https://doi.org/10.1103/PhysRevLett.104.021802}{\emph{Phys. Rev. Lett.}
  {\bfseries 104} (2010) 021802},
  [\href{https://arxiv.org/abs/0908.2381}{{\ttfamily 0908.2381}}].

\bibitem{Baek:2001kca}
S.~Baek, N.~G. Deshpande, X.~G. He and P.~Ko, \emph{{Muon anomalous g-2 and
  gauged L(muon) - L(tau) models}},
  \href{https://doi.org/10.1103/PhysRevD.64.055006}{\emph{Phys. Rev. D}
  {\bfseries 64} (2001) 055006},
  [\href{https://arxiv.org/abs/hep-ph/0104141}{{\ttfamily hep-ph/0104141}}].

\bibitem{Ma:2001md}
E.~Ma, D.~P. Roy and S.~Roy, \emph{{Gauged L(mu) - L(tau) with large muon
  anomalous magnetic moment and the bimaximal mixing of neutrinos}},
  \href{https://doi.org/10.1016/S0370-2693(01)01428-9}{\emph{Phys. Lett. B}
  {\bfseries 525} (2002) 101--106},
  [\href{https://arxiv.org/abs/hep-ph/0110146}{{\ttfamily hep-ph/0110146}}].

\bibitem{Harigaya:2013twa}
K.~Harigaya, T.~Igari, M.~M. Nojiri, M.~Takeuchi and K.~Tobe, \emph{{Muon g-2
  and LHC phenomenology in the $L_\mu-L_\tau$ gauge symmetric model}},
  \href{https://doi.org/10.1007/JHEP03(2014)105}{\emph{JHEP} {\bfseries 03}
  (2014) 105}, [\href{https://arxiv.org/abs/1311.0870}{{\ttfamily 1311.0870}}].

\bibitem{Altmannshofer:2014pba}
W.~Altmannshofer, S.~Gori, M.~Pospelov and I.~Yavin, \emph{{Neutrino Trident
  Production: A Powerful Probe of New Physics with Neutrino Beams}},
  \href{https://doi.org/10.1103/PhysRevLett.113.091801}{\emph{Phys. Rev. Lett.}
  {\bfseries 113} (2014) 091801},
  [\href{https://arxiv.org/abs/1406.2332}{{\ttfamily 1406.2332}}].

\bibitem{Altmannshofer:2019zhy}
W.~Altmannshofer, S.~Gori, J.~Mart\'\i{}n-Albo, A.~Sousa and M.~Wallbank,
  \emph{{Neutrino Tridents at DUNE}},
  \href{https://doi.org/10.1103/PhysRevD.100.115029}{\emph{Phys. Rev. D}
  {\bfseries 100} (2019) 115029},
  [\href{https://arxiv.org/abs/1902.06765}{{\ttfamily 1902.06765}}].

\bibitem{Crivellin:2016ejn}
A.~Crivellin, J.~Fuentes-Martin, A.~Greljo and G.~Isidori, \emph{{Lepton Flavor
  Non-Universality in B decays from Dynamical Yukawas}},
  \href{https://doi.org/10.1016/j.physletb.2016.12.057}{\emph{Phys. Lett. B}
  {\bfseries 766} (2017) 77--85},
  [\href{https://arxiv.org/abs/1611.02703}{{\ttfamily 1611.02703}}].

\bibitem{Crivellin:2015mga}
A.~Crivellin, G.~D'Ambrosio and J.~Heeck, \emph{{Explaining
  $h\to\mu^\pm\tau^\mp$, $B\to K^* \mu^+\mu^-$ and $B\to K \mu^+\mu^-/B\to K
  e^+e^-$ in a two-Higgs-doublet model with gauged $L_\mu-L_\tau$}},
  \href{https://doi.org/10.1103/PhysRevLett.114.151801}{\emph{Phys. Rev. Lett.}
  {\bfseries 114} (2015) 151801},
  [\href{https://arxiv.org/abs/1501.00993}{{\ttfamily 1501.00993}}].

\bibitem{Crivellin:2018qmi}
A.~Crivellin, M.~Hoferichter and P.~Schmidt-Wellenburg, \emph{{Combined
  explanations of $(g-2)_{\mu,e}$ and implications for a large muon EDM}},
  \href{https://doi.org/10.1103/PhysRevD.98.113002}{\emph{Phys. Rev. D}
  {\bfseries 98} (2018) 113002},
  [\href{https://arxiv.org/abs/1807.11484}{{\ttfamily 1807.11484}}].

\bibitem{Altmannshofer:2014cfa}
W.~Altmannshofer, S.~Gori, M.~Pospelov and I.~Yavin, \emph{{Quark flavor
  transitions in $L_\mu-L_\tau$ models}},
  \href{https://doi.org/10.1103/PhysRevD.89.095033}{\emph{Phys. Rev. D}
  {\bfseries 89} (2014) 095033},
  [\href{https://arxiv.org/abs/1403.1269}{{\ttfamily 1403.1269}}].

\bibitem{Altmannshofer:2015mqa}
W.~Altmannshofer and I.~Yavin, \emph{{Predictions for lepton flavor
  universality violation in rare B decays in models with gauged $L_\mu -
  L_\tau$}}, \href{https://doi.org/10.1103/PhysRevD.92.075022}{\emph{Phys. Rev.
  D} {\bfseries 92} (2015) 075022},
  [\href{https://arxiv.org/abs/1508.07009}{{\ttfamily 1508.07009}}].

\bibitem{He:1990pn}
X.~G. He, G.~C. Joshi, H.~Lew and R.~R. Volkas, \emph{{NEW Z-prime
  PHENOMENOLOGY}}, \href{https://doi.org/10.1103/PhysRevD.43.R22}{\emph{Phys.
  Rev. D} {\bfseries 43} (1991) 22--24}.

\bibitem{He:1991qd}
X.-G. He, G.~C. Joshi, H.~Lew and R.~R. Volkas, \emph{{Simplest Z-prime
  model}}, \href{https://doi.org/10.1103/PhysRevD.44.2118}{\emph{Phys. Rev. D}
  {\bfseries 44} (1991) 2118--2132}.

\bibitem{Altmannshofer:2019xda}
W.~Altmannshofer, J.~Davighi and M.~Nardecchia, \emph{{Gauging the accidental
  symmetries of the standard model, and implications for the flavor
  anomalies}}, \href{https://doi.org/10.1103/PhysRevD.101.015004}{\emph{Phys.
  Rev. D} {\bfseries 101} (2020) 015004},
  [\href{https://arxiv.org/abs/1909.02021}{{\ttfamily 1909.02021}}].

\bibitem{Alonso:2017uky}
R.~Alonso, P.~Cox, C.~Han and T.~T. Yanagida, \emph{{Flavoured $B?L$ local
  symmetry and anomalous rare $B$ decays}},
  \href{https://doi.org/10.1016/j.physletb.2017.10.027}{\emph{Phys. Lett. B}
  {\bfseries 774} (2017) 643--648},
  [\href{https://arxiv.org/abs/1705.03858}{{\ttfamily 1705.03858}}].

\bibitem{Bonilla:2017lsq}
C.~Bonilla, T.~Modak, R.~Srivastava and J.~W.~F. Valle,
  \emph{{$U(1)_{B_3-3L_\mu}$ gauge symmetry as a simple description of $b\to s$
  anomalies}}, \href{https://doi.org/10.1103/PhysRevD.98.095002}{\emph{Phys.
  Rev. D} {\bfseries 98} (2018) 095002},
  [\href{https://arxiv.org/abs/1705.00915}{{\ttfamily 1705.00915}}].

\bibitem{Allanach:2020kss}
B.~C. Allanach, \emph{{$U(1)_{B_3-L_2}$ explanation of the neutral current
  $B$\ensuremath{-}anomalies}},
  \href{https://doi.org/10.1140/epjc/s10052-021-08855-w}{\emph{Eur. Phys. J. C}
  {\bfseries 81} (2021) 56},
  [\href{https://arxiv.org/abs/2009.02197}{{\ttfamily 2009.02197}}].

\bibitem{Allanach:2018lvl}
B.~C. Allanach and J.~Davighi, \emph{{Third family hypercharge model for $
  {R}_{K^{\left(\ast \right)}} $ and aspects of the fermion mass problem}},
  \href{https://doi.org/10.1007/JHEP12(2018)075}{\emph{JHEP} {\bfseries 12}
  (2018) 075}, [\href{https://arxiv.org/abs/1809.01158}{{\ttfamily
  1809.01158}}].

\bibitem{Allanach:2019iiy}
B.~C. Allanach and J.~Davighi, \emph{{Naturalising the third family hypercharge
  model for neutral current $B$-anomalies}},
  \href{https://doi.org/10.1140/epjc/s10052-019-7414-z}{\emph{Eur. Phys. J. C}
  {\bfseries 79} (2019) 908},
  [\href{https://arxiv.org/abs/1905.10327}{{\ttfamily 1905.10327}}].

\bibitem{Bhatia:2017tgo}
D.~Bhatia, S.~Chakraborty and A.~Dighe, \emph{{Neutrino mixing and $R_K$
  anomaly in U(1)$_X$ models: a bottom-up approach}},
  \href{https://doi.org/10.1007/JHEP03(2017)117}{\emph{JHEP} {\bfseries 03}
  (2017) 117}, [\href{https://arxiv.org/abs/1701.05825}{{\ttfamily
  1701.05825}}].

\bibitem{AristizabalSierra:2015vqb}
D.~Aristizabal~Sierra, F.~Staub and A.~Vicente, \emph{{Shedding light on the
  $b\to s$ anomalies with a dark sector}},
  \href{https://doi.org/10.1103/PhysRevD.92.015001}{\emph{Phys. Rev. D}
  {\bfseries 92} (2015) 015001},
  [\href{https://arxiv.org/abs/1503.06077}{{\ttfamily 1503.06077}}].

\bibitem{Celis:2015ara}
A.~Celis, J.~Fuentes-Martin, M.~Jung and H.~Serodio, \emph{{Family nonuniversal
  Z' models with protected flavor-changing interactions}},
  \href{https://doi.org/10.1103/PhysRevD.92.015007}{\emph{Phys. Rev. D}
  {\bfseries 92} (2015) 015007},
  [\href{https://arxiv.org/abs/1505.03079}{{\ttfamily 1505.03079}}].

\bibitem{Falkowski:2015zwa}
A.~Falkowski, M.~Nardecchia and R.~Ziegler, \emph{{Lepton Flavor
  Non-Universality in B-meson Decays from a U(2) Flavor Model}},
  \href{https://doi.org/10.1007/JHEP11(2015)173}{\emph{JHEP} {\bfseries 11}
  (2015) 173}, [\href{https://arxiv.org/abs/1509.01249}{{\ttfamily
  1509.01249}}].

\bibitem{Chiang:2016qov}
C.-W. Chiang, X.-G. He and G.~Valencia, \emph{{Z' model for
  b\textrightarrow{}s\ensuremath{\ell}$\overline{?}$ flavor anomalies}},
  \href{https://doi.org/10.1103/PhysRevD.93.074003}{\emph{Phys. Rev. D}
  {\bfseries 93} (2016) 074003},
  [\href{https://arxiv.org/abs/1601.07328}{{\ttfamily 1601.07328}}].

\bibitem{Boucenna:2016wpr}
S.~M. Boucenna, A.~Celis, J.~Fuentes-Martin, A.~Vicente and J.~Virto,
  \emph{{Non-abelian gauge extensions for B-decay anomalies}},
  \href{https://doi.org/10.1016/j.physletb.2016.06.067}{\emph{Phys. Lett. B}
  {\bfseries 760} (2016) 214--219},
  [\href{https://arxiv.org/abs/1604.03088}{{\ttfamily 1604.03088}}].

\bibitem{Boucenna:2016qad}
S.~M. Boucenna, A.~Celis, J.~Fuentes-Martin, A.~Vicente and J.~Virto,
  \emph{{Phenomenology of an $SU(2) \times SU(2) \times U(1)$ model with
  lepton-flavour non-universality}},
  \href{https://doi.org/10.1007/JHEP12(2016)059}{\emph{JHEP} {\bfseries 12}
  (2016) 059}, [\href{https://arxiv.org/abs/1608.01349}{{\ttfamily
  1608.01349}}].

\bibitem{Ko:2017lzd}
P.~Ko, Y.~Omura, Y.~Shigekami and C.~Yu, \emph{{LHCb anomaly and B physics in
  flavored Z' models with flavored Higgs doublets}},
  \href{https://doi.org/10.1103/PhysRevD.95.115040}{\emph{Phys. Rev. D}
  {\bfseries 95} (2017) 115040},
  [\href{https://arxiv.org/abs/1702.08666}{{\ttfamily 1702.08666}}].

\bibitem{Alonso:2017bff}
R.~Alonso, P.~Cox, C.~Han and T.~T. Yanagida, \emph{{Anomaly-free local
  horizontal symmetry and anomaly-full rare B-decays}},
  \href{https://doi.org/10.1103/PhysRevD.96.071701}{\emph{Phys. Rev. D}
  {\bfseries 96} (2017) 071701},
  [\href{https://arxiv.org/abs/1704.08158}{{\ttfamily 1704.08158}}].

\bibitem{Tang:2017gkz}
Y.~Tang and Y.-L. Wu, \emph{{Flavor non-universal gauge interactions and
  anomalies in B-meson decays}},
  \href{https://doi.org/10.1088/1674-1137/42/3/033104}{\emph{Chin. Phys. C}
  {\bfseries 42} (2018) 033104},
  [\href{https://arxiv.org/abs/1705.05643}{{\ttfamily 1705.05643}}].

\bibitem{Fuyuto:2017sys}
K.~Fuyuto, H.-L. Li and J.-H. Yu, \emph{{Implications of hidden gauged $U(1)$
  model for $B$ anomalies}},
  \href{https://doi.org/10.1103/PhysRevD.97.115003}{\emph{Phys. Rev. D}
  {\bfseries 97} (2018) 115003},
  [\href{https://arxiv.org/abs/1712.06736}{{\ttfamily 1712.06736}}].

\bibitem{Bian:2017xzg}
L.~Bian, H.~M. Lee and C.~B. Park, \emph{{$B$-meson anomalies and Higgs physics
  in flavored $U(1)'$ model}},
  \href{https://doi.org/10.1140/epjc/s10052-018-5777-1}{\emph{Eur. Phys. J. C}
  {\bfseries 78} (2018) 306},
  [\href{https://arxiv.org/abs/1711.08930}{{\ttfamily 1711.08930}}].

\bibitem{King:2018fcg}
S.~F. King, \emph{{$ {R}_{K^{\left(*\right)}} $ and the origin of Yukawa
  couplings}}, \href{https://doi.org/10.1007/JHEP09(2018)069}{\emph{JHEP}
  {\bfseries 09} (2018) 069},
  [\href{https://arxiv.org/abs/1806.06780}{{\ttfamily 1806.06780}}].

\bibitem{Duan:2018akc}
G.~H. Duan, X.~Fan, M.~Frank, C.~Han and J.~M. Yang, \emph{{A minimal
  $U(1)^\prime$ extension of MSSM in light of the B decay anomaly}},
  \href{https://doi.org/10.1016/j.physletb.2018.12.005}{\emph{Phys. Lett. B}
  {\bfseries 789} (2019) 54--58},
  [\href{https://arxiv.org/abs/1808.04116}{{\ttfamily 1808.04116}}].

\bibitem{Dorsner:2016wpm}
I.~Dor\v{s}ner, S.~Fajfer, A.~Greljo, J.~F. Kamenik and N.~Ko\v{s}nik,
  \emph{{Physics of leptoquarks in precision experiments and at particle
  colliders}}, \href{https://doi.org/10.1016/j.physrep.2016.06.001}{\emph{Phys.
  Rept.} {\bfseries 641} (2016) 1--68},
  [\href{https://arxiv.org/abs/1603.04993}{{\ttfamily 1603.04993}}].

\bibitem{Hambye:2017qix}
T.~Hambye and J.~Heeck, \emph{{Proton decay into charged leptons}},
  \href{https://doi.org/10.1103/PhysRevLett.120.171801}{\emph{Phys. Rev. Lett.}
  {\bfseries 120} (2018) 171801},
  [\href{https://arxiv.org/abs/1712.04871}{{\ttfamily 1712.04871}}].

\bibitem{Davighi:2020qqa}
J.~Davighi, M.~Kirk and M.~Nardecchia, \emph{{Anomalies and accidental
  symmetries: charging the scalar leptoquark under L$_{\mu}$ \ensuremath{-}
  L$_{\tau}$}}, \href{https://doi.org/10.1007/JHEP12(2020)111}{\emph{JHEP}
  {\bfseries 12} (2020) 111},
  [\href{https://arxiv.org/abs/2007.15016}{{\ttfamily 2007.15016}}].

\bibitem{Wess:1971yu}
J.~Wess and B.~Zumino, \emph{{Consequences of anomalous Ward identities}},
  \href{https://doi.org/10.1016/0370-2693(71)90582-X}{\emph{Phys. Lett. B}
  {\bfseries 37} (1971) 95--97}.

\bibitem{Davighi:2021oel}
J.~Davighi, \emph{{Anomalous $Z^\prime$ bosons for anomalous $B$ decays}},
  \href{https://arxiv.org/abs/2105.06918}{{\ttfamily 2105.06918}}.

\bibitem{Allanach:2018vjg}
B.~C. Allanach, J.~Davighi and S.~Melville, \emph{{An Anomaly-free Atlas:
  charting the space of flavour-dependent gauged $U(1)$ extensions of the
  Standard Model}}, \href{https://doi.org/10.1007/JHEP02(2019)082}{\emph{JHEP}
  {\bfseries 02} (2019) 082},
  [\href{https://arxiv.org/abs/1812.04602}{{\ttfamily 1812.04602}}].

\bibitem{Asai:2019ciz}
K.~Asai, \emph{{Predictions for the neutrino parameters in the minimal model
  extended by linear combination of U(1)$_{L_e-L_\mu}$, U(1)$_{L_\mu-L_\tau}$
  and U(1)$_{B-L}$ gauge symmetries}},
  \href{https://doi.org/10.1140/epjc/s10052-020-7622-6}{\emph{Eur. Phys. J. C}
  {\bfseries 80} (2020) 76},
  [\href{https://arxiv.org/abs/1907.04042}{{\ttfamily 1907.04042}}].

\bibitem{Araki:2019rmw}
T.~Araki, K.~Asai, J.~Sato and T.~Shimomura, \emph{{Low scale seesaw models for
  low scale $U(1)_{L_\mu-L_\tau}$ symmetry}},
  \href{https://doi.org/10.1103/PhysRevD.100.095012}{\emph{Phys. Rev. D}
  {\bfseries 100} (2019) 095012},
  [\href{https://arxiv.org/abs/1909.08827}{{\ttfamily 1909.08827}}].

\bibitem{Hiller:2014yaa}
G.~Hiller and M.~Schmaltz, \emph{{$R_K$ and future $b \to s \ell \ell$ physics
  beyond the standard model opportunities}},
  \href{https://doi.org/10.1103/PhysRevD.90.054014}{\emph{Phys. Rev. D}
  {\bfseries 90} (2014) 054014},
  [\href{https://arxiv.org/abs/1408.1627}{{\ttfamily 1408.1627}}].

\bibitem{Buttazzo:2017ixm}
D.~Buttazzo, A.~Greljo, G.~Isidori and D.~Marzocca, \emph{{B-physics anomalies:
  a guide to combined explanations}},
  \href{https://doi.org/10.1007/JHEP11(2017)044}{\emph{JHEP} {\bfseries 11}
  (2017) 044}, [\href{https://arxiv.org/abs/1706.07808}{{\ttfamily
  1706.07808}}].

\bibitem{Crivellin:2017zlb}
A.~Crivellin, D.~M\"uller and T.~Ota, \emph{{Simultaneous explanation of
  R(D$^{(?)}$) and b\textrightarrow{}s\ensuremath{\mu}$^{+}$
  \ensuremath{\mu}$^{?}$: the last scalar leptoquarks standing}},
  \href{https://doi.org/10.1007/JHEP09(2017)040}{\emph{JHEP} {\bfseries 09}
  (2017) 040}, [\href{https://arxiv.org/abs/1703.09226}{{\ttfamily
  1703.09226}}].

\bibitem{Hiller:2017bzc}
G.~Hiller and I.~Nisandzic, \emph{{$R_K$ and $R_{K^{\ast}}$ beyond the standard
  model}}, \href{https://doi.org/10.1103/PhysRevD.96.035003}{\emph{Phys. Rev.
  D} {\bfseries 96} (2017) 035003},
  [\href{https://arxiv.org/abs/1704.05444}{{\ttfamily 1704.05444}}].

\bibitem{Gherardi:2020qhc}
V.~Gherardi, D.~Marzocca and E.~Venturini, \emph{{Low-energy phenomenology of
  scalar leptoquarks at one-loop accuracy}},
  \href{https://doi.org/10.1007/JHEP01(2021)138}{\emph{JHEP} {\bfseries 01}
  (2021) 138}, [\href{https://arxiv.org/abs/2008.09548}{{\ttfamily
  2008.09548}}].

\bibitem{Angelescu:2021lln}
A.~Angelescu, D.~Be\v{c}irevi\'c, D.~A. Faroughy, F.~Jaffredo and O.~Sumensari,
  \emph{{On the single leptoquark solutions to the $B$-physics anomalies}},
  \href{https://arxiv.org/abs/2103.12504}{{\ttfamily 2103.12504}}.

\bibitem{Marzocca:2018wcf}
D.~Marzocca, \emph{{Addressing the B-physics anomalies in a fundamental
  Composite Higgs Model}},
  \href{https://doi.org/10.1007/JHEP07(2018)121}{\emph{JHEP} {\bfseries 07}
  (2018) 121}, [\href{https://arxiv.org/abs/1803.10972}{{\ttfamily
  1803.10972}}].

\bibitem{Dorsner:2017ufx}
I.~Dor\v{s}ner, S.~Fajfer, D.~A. Faroughy and N.~Ko\v{s}nik, \emph{{The role of
  the $S_3$ GUT leptoquark in flavor universality and collider searches}},
  \href{https://doi.org/10.1007/JHEP10(2017)188}{\emph{JHEP} {\bfseries 10}
  (2017) 188}, [\href{https://arxiv.org/abs/1706.07779}{{\ttfamily
  1706.07779}}].

\bibitem{Babu:2020hun}
K.~S. Babu, P.~S.~B. Dev, S.~Jana and A.~Thapa, \emph{{Unified framework for
  $B$-anomalies, muon $g-2$ and neutrino masses}},
  \href{https://doi.org/10.1007/JHEP03(2021)179}{\emph{JHEP} {\bfseries 03}
  (2021) 179}, [\href{https://arxiv.org/abs/2009.01771}{{\ttfamily
  2009.01771}}].

\bibitem{Bauer:2015knc}
M.~Bauer and M.~Neubert, \emph{{Minimal Leptoquark Explanation for the
  R$_{D^{(*)}}$ , R$_K$ , and $(g-2)_g$ Anomalies}},
  \href{https://doi.org/10.1103/PhysRevLett.116.141802}{\emph{Phys. Rev. Lett.}
  {\bfseries 116} (2016) 141802},
  [\href{https://arxiv.org/abs/1511.01900}{{\ttfamily 1511.01900}}].

\bibitem{Dorsner:2019itg}
I.~Dor\v{s}ner, S.~Fajfer and O.~Sumensari, \emph{{Muon $g-2$ and scalar
  leptoquark mixing}},
  \href{https://doi.org/10.1007/JHEP06(2020)089}{\emph{JHEP} {\bfseries 06}
  (2020) 089}, [\href{https://arxiv.org/abs/1910.03877}{{\ttfamily
  1910.03877}}].

\bibitem{Brdar:2020quo}
V.~Brdar, A.~Greljo, J.~Kopp and T.~Opferkuch, \emph{{The Neutrino Magnetic
  Moment Portal: Cosmology, Astrophysics, and Direct Detection}},
  \href{https://doi.org/10.1088/1475-7516/2021/01/039}{\emph{JCAP} {\bfseries
  01} (2021) 039}, [\href{https://arxiv.org/abs/2007.15563}{{\ttfamily
  2007.15563}}].

\bibitem{Queiroz:2014zfa}
F.~S. Queiroz and W.~Shepherd, \emph{{New Physics Contributions to the Muon
  Anomalous Magnetic Moment: A Numerical Code}},
  \href{https://doi.org/10.1103/PhysRevD.89.095024}{\emph{Phys. Rev. D}
  {\bfseries 89} (2014) 095024},
  [\href{https://arxiv.org/abs/1403.2309}{{\ttfamily 1403.2309}}].

\bibitem{Arnold:2013cva}
J.~M. Arnold, B.~Fornal and M.~B. Wise, \emph{{Phenomenology of scalar
  leptoquarks}}, \href{https://doi.org/10.1103/PhysRevD.88.035009}{\emph{Phys.
  Rev. D} {\bfseries 88} (2013) 035009},
  [\href{https://arxiv.org/abs/1304.6119}{{\ttfamily 1304.6119}}].

\bibitem{Assad:2017iib}
N.~Assad, B.~Fornal and B.~Grinstein, \emph{{Baryon Number and Lepton
  Universality Violation in Leptoquark and Diquark Models}},
  \href{https://doi.org/10.1016/j.physletb.2017.12.042}{\emph{Phys. Lett. B}
  {\bfseries 777} (2018) 324--331},
  [\href{https://arxiv.org/abs/1708.06350}{{\ttfamily 1708.06350}}].

\bibitem{Barbieri:2015yvd}
R.~Barbieri, G.~Isidori, A.~Pattori and F.~Senia, \emph{{Anomalies in
  $B$-decays and $U(2)$ flavour symmetry}},
  \href{https://doi.org/10.1140/epjc/s10052-016-3905-3}{\emph{Eur. Phys. J. C}
  {\bfseries 76} (2016) 67},
  [\href{https://arxiv.org/abs/1512.01560}{{\ttfamily 1512.01560}}].

\bibitem{DiLuzio:2017vat}
L.~Di~Luzio, A.~Greljo and M.~Nardecchia, \emph{{Gauge leptoquark as the origin
  of B-physics anomalies}},
  \href{https://doi.org/10.1103/PhysRevD.96.115011}{\emph{Phys. Rev. D}
  {\bfseries 96} (2017) 115011},
  [\href{https://arxiv.org/abs/1708.08450}{{\ttfamily 1708.08450}}].

\bibitem{Greljo:2018tuh}
A.~Greljo and B.~A. Stefanek, \emph{{Third family quark\textendash{}lepton
  unification at the TeV scale}},
  \href{https://doi.org/10.1016/j.physletb.2018.05.033}{\emph{Phys. Lett. B}
  {\bfseries 782} (2018) 131--138},
  [\href{https://arxiv.org/abs/1802.04274}{{\ttfamily 1802.04274}}].

\bibitem{Bordone:2017bld}
M.~Bordone, C.~Cornella, J.~Fuentes-Martin and G.~Isidori, \emph{{A three-site
  gauge model for flavor hierarchies and flavor anomalies}},
  \href{https://doi.org/10.1016/j.physletb.2018.02.011}{\emph{Phys. Lett. B}
  {\bfseries 779} (2018) 317--323},
  [\href{https://arxiv.org/abs/1712.01368}{{\ttfamily 1712.01368}}].

\bibitem{Bordone:2018nbg}
M.~Bordone, C.~Cornella, J.~Fuentes-Mart\'\i{}n and G.~Isidori,
  \emph{{Low-energy signatures of the $\mathrm{PS}^3$ model: from $B$-physics
  anomalies to LFV}},
  \href{https://doi.org/10.1007/JHEP10(2018)148}{\emph{JHEP} {\bfseries 10}
  (2018) 148}, [\href{https://arxiv.org/abs/1805.09328}{{\ttfamily
  1805.09328}}].

\bibitem{Cornella:2019hct}
C.~Cornella, J.~Fuentes-Martin and G.~Isidori, \emph{{Revisiting the vector
  leptoquark explanation of the B-physics anomalies}},
  \href{https://doi.org/10.1007/JHEP07(2019)168}{\emph{JHEP} {\bfseries 07}
  (2019) 168}, [\href{https://arxiv.org/abs/1903.11517}{{\ttfamily
  1903.11517}}].

\bibitem{Fornal:2018dqn}
B.~Fornal, S.~A. Gadam and B.~Grinstein, \emph{{Left-Right SU(4) Vector
  Leptoquark Model for Flavor Anomalies}},
  \href{https://doi.org/10.1103/PhysRevD.99.055025}{\emph{Phys. Rev. D}
  {\bfseries 99} (2019) 055025},
  [\href{https://arxiv.org/abs/1812.01603}{{\ttfamily 1812.01603}}].

\bibitem{Blanke:2018sro}
M.~Blanke and A.~Crivellin, \emph{{$B$ Meson Anomalies in a Pati-Salam Model
  within the Randall-Sundrum Background}},
  \href{https://doi.org/10.1103/PhysRevLett.121.011801}{\emph{Phys. Rev. Lett.}
  {\bfseries 121} (2018) 011801},
  [\href{https://arxiv.org/abs/1801.07256}{{\ttfamily 1801.07256}}].

\bibitem{Fuentes-Martin:2019ign}
J.~Fuentes-Mart\'\i{}n, G.~Isidori, M.~K\"onig and N.~Selimovi\'c,
  \emph{{Vector Leptoquarks Beyond Tree Level}},
  \href{https://doi.org/10.1103/PhysRevD.101.035024}{\emph{Phys. Rev. D}
  {\bfseries 101} (2020) 035024},
  [\href{https://arxiv.org/abs/1910.13474}{{\ttfamily 1910.13474}}].

\bibitem{Guadagnoli:2020tlx}
D.~Guadagnoli, M.~Reboud and P.~Stangl, \emph{{The Dark Side of 4321}},
  \href{https://doi.org/10.1007/JHEP10(2020)084}{\emph{JHEP} {\bfseries 10}
  (2020) 084}, [\href{https://arxiv.org/abs/2005.10117}{{\ttfamily
  2005.10117}}].

\bibitem{Heeck:2018ntp}
J.~Heeck and D.~Teresi, \emph{{Pati-Salam explanations of the B-meson
  anomalies}}, \href{https://doi.org/10.1007/JHEP12(2018)103}{\emph{JHEP}
  {\bfseries 12} (2018) 103},
  [\href{https://arxiv.org/abs/1808.07492}{{\ttfamily 1808.07492}}].

\bibitem{Fuentes-Martin:2020bnh}
J.~Fuentes-Mart\'\i{}n and P.~Stangl, \emph{{Third-family quark-lepton
  unification with a fundamental composite Higgs}},
  \href{https://doi.org/10.1016/j.physletb.2020.135953}{\emph{Phys. Lett. B}
  {\bfseries 811} (2020) 135953},
  [\href{https://arxiv.org/abs/2004.11376}{{\ttfamily 2004.11376}}].

\bibitem{Fuentes-Martin:2019bue}
J.~Fuentes-Mart\'\i{}n, M.~Reig and A.~Vicente, \emph{{Strong $CP$ problem with
  low-energy emergent QCD: The 4321 case}},
  \href{https://doi.org/10.1103/PhysRevD.100.115028}{\emph{Phys. Rev. D}
  {\bfseries 100} (2019) 115028},
  [\href{https://arxiv.org/abs/1907.02550}{{\ttfamily 1907.02550}}].

\bibitem{Fuentes-Martin:2020luw}
J.~Fuentes-Mart\'\i{}n, G.~Isidori, M.~K\"onig and N.~Selimovi\'c,
  \emph{{Vector leptoquarks beyond tree level. II. $\mathcal{O}(\alpha_s)$
  corrections and radial modes}},
  \href{https://doi.org/10.1103/PhysRevD.102.035021}{\emph{Phys. Rev. D}
  {\bfseries 102} (2020) 035021},
  [\href{https://arxiv.org/abs/2006.16250}{{\ttfamily 2006.16250}}].

\bibitem{Fuentes-Martin:2020hvc}
J.~Fuentes-Mart\'\i{}n, G.~Isidori, M.~K\"onig and N.~Selimovi\'c,
  \emph{{Vector Leptoquarks Beyond Tree Level III: Vector-like Fermions and
  Flavor-Changing Transitions}},
  \href{https://doi.org/10.1103/PhysRevD.102.115015}{\emph{Phys. Rev. D}
  {\bfseries 102} (2020) 115015},
  [\href{https://arxiv.org/abs/2009.11296}{{\ttfamily 2009.11296}}].

\bibitem{Dobrescu:2020evn}
B.~A. Dobrescu and P.~J. Fox, \emph{{Diophantine equations with sum of cubes
  and cube of sum}},  \href{https://arxiv.org/abs/2012.04139}{{\ttfamily
  2012.04139}}.

\bibitem{Costa:2019zzy}
D.~B. Costa, B.~A. Dobrescu and P.~J. Fox, \emph{{General Solution to the U(1)
  Anomaly Equations}},
  \href{https://doi.org/10.1103/PhysRevLett.123.151601}{\emph{Phys. Rev. Lett.}
  {\bfseries 123} (2019) 151601},
  [\href{https://arxiv.org/abs/1905.13729}{{\ttfamily 1905.13729}}].

\bibitem{Allanach:2020zna}
B.~C. Allanach, B.~Gripaios and J.~Tooby-Smith, \emph{{Anomaly cancellation
  with an extra gauge boson}},
  \href{https://doi.org/10.1103/PhysRevLett.125.161601}{\emph{Phys. Rev. Lett.}
  {\bfseries 125} (2020) 161601},
  [\href{https://arxiv.org/abs/2006.03588}{{\ttfamily 2006.03588}}].

\bibitem{Barbieri:2011ci}
R.~Barbieri, G.~Isidori, J.~Jones-Perez, P.~Lodone and D.~M. Straub,
  \emph{{$U(2)$ and Minimal Flavour Violation in Supersymmetry}},
  \href{https://doi.org/10.1140/epjc/s10052-011-1725-z}{\emph{Eur. Phys. J. C}
  {\bfseries 71} (2011) 1725},
  [\href{https://arxiv.org/abs/1105.2296}{{\ttfamily 1105.2296}}].

\bibitem{Kagan:2009bn}
A.~L. Kagan, G.~Perez, T.~Volansky and J.~Zupan, \emph{{General Minimal Flavor
  Violation}}, \href{https://doi.org/10.1103/PhysRevD.80.076002}{\emph{Phys.
  Rev. D} {\bfseries 80} (2009) 076002},
  [\href{https://arxiv.org/abs/0903.1794}{{\ttfamily 0903.1794}}].

\bibitem{Fuentes-Martin:2019mun}
J.~Fuentes-Mart\'\i{}n, G.~Isidori, J.~Pag\`es and K.~Yamamoto, \emph{{With or
  without U(2)? Probing non-standard flavor and helicity structures in
  semileptonic B decays}},
  \href{https://doi.org/10.1016/j.physletb.2019.135080}{\emph{Phys. Lett. B}
  {\bfseries 800} (2020) 135080},
  [\href{https://arxiv.org/abs/1909.02519}{{\ttfamily 1909.02519}}].

\bibitem{Jegerlehner:2009ry}
F.~Jegerlehner and A.~Nyffeler, \emph{{The Muon g-2}},
  \href{https://doi.org/10.1016/j.physrep.2009.04.003}{\emph{Phys. Rept.}
  {\bfseries 477} (2009) 1--110},
  [\href{https://arxiv.org/abs/0902.3360}{{\ttfamily 0902.3360}}].

\bibitem{Altmannshofer:2016brv}
W.~Altmannshofer, C.-Y. Chen, P.~S. Bhupal~Dev and A.~Soni, \emph{{Lepton
  flavor violating Z' explanation of the muon anomalous magnetic moment}},
  \href{https://doi.org/10.1016/j.physletb.2016.09.046}{\emph{Phys. Lett. B}
  {\bfseries 762} (2016) 389--398},
  [\href{https://arxiv.org/abs/1607.06832}{{\ttfamily 1607.06832}}].

\bibitem{Abi:2021gix}
B.~Abi et~al., \emph{{Measurement of the Positive Muon Anomalous Magnetic
  Moment to 0.46 ppm}},  \href{https://arxiv.org/abs/2104.03281}{{\ttfamily
  2104.03281}}.

\bibitem{Albahri:2021kmg}
{\scshape Muon g-2} collaboration, T.~Albahri et~al., \emph{{Magnetic-field
  measurement and analysis for the Muon $g-2$ Experiment at Fermilab}},
  \href{https://doi.org/10.1103/PhysRevA.103.042208}{\emph{Phys. Rev. A}
  {\bfseries 103} (2021) 042208},
  [\href{https://arxiv.org/abs/2104.03201}{{\ttfamily 2104.03201}}].

\bibitem{Albahri:2021ixb}
{\scshape Muon g-2} collaboration, T.~Albahri et~al., \emph{{Measurement of the
  anomalous precession frequency of the muon in the Fermilab Muon $g-2$
  Experiment}}, \href{https://doi.org/10.1103/PhysRevD.103.072002}{\emph{Phys.
  Rev. D} {\bfseries 103} (2021) 072002},
  [\href{https://arxiv.org/abs/2104.03247}{{\ttfamily 2104.03247}}].

\bibitem{Crivellin:2020zul}
A.~Crivellin, M.~Hoferichter, C.~A. Manzari and M.~Montull, \emph{{Hadronic
  Vacuum Polarization: $(g-2)_\mu$ versus Global Electroweak Fits}},
  \href{https://doi.org/10.1103/PhysRevLett.125.091801}{\emph{Phys. Rev. Lett.}
  {\bfseries 125} (2020) 091801},
  [\href{https://arxiv.org/abs/2003.04886}{{\ttfamily 2003.04886}}].

\bibitem{Keshavarzi:2020bfy}
A.~Keshavarzi, W.~J. Marciano, M.~Passera and A.~Sirlin, \emph{{Muon $g-2$ and
  $\Delta \alpha$ connection}},
  \href{https://doi.org/10.1103/PhysRevD.102.033002}{\emph{Phys. Rev. D}
  {\bfseries 102} (2020) 033002},
  [\href{https://arxiv.org/abs/2006.12666}{{\ttfamily 2006.12666}}].

\bibitem{Jackiw:1972jz}
R.~Jackiw and S.~Weinberg, \emph{{Weak interaction corrections to the muon
  magnetic moment and to muonic atom energy levels}},
  \href{https://doi.org/10.1103/PhysRevD.5.2396}{\emph{Phys. Rev. D} {\bfseries
  5} (1972) 2396--2398}.

\bibitem{Mishra:1991bv}
{\scshape CCFR} collaboration, S.~R. Mishra et~al., \emph{{Neutrino tridents
  and W Z interference}},
  \href{https://doi.org/10.1103/PhysRevLett.66.3117}{\emph{Phys. Rev. Lett.}
  {\bfseries 66} (1991) 3117--3120}.

\bibitem{NuTeV:1999wlw}
{\scshape NuTeV} collaboration, T.~Adams et~al., \emph{{Evidence for
  diffractive charm production in muon-neutrino Fe and anti-muon-neutrino Fe
  scattering at the Tevatron}},
  \href{https://doi.org/10.1103/PhysRevD.61.092001}{\emph{Phys. Rev. D}
  {\bfseries 61} (2000) 092001},
  [\href{https://arxiv.org/abs/hep-ex/9909041}{{\ttfamily hep-ex/9909041}}].

\bibitem{Ballett:2019xoj}
P.~Ballett, M.~Hostert, S.~Pascoli, Y.~F. Perez-Gonzalez, Z.~Tabrizi and
  R.~Zukanovich~Funchal, \emph{{$Z^\prime$s in neutrino scattering at DUNE}},
  \href{https://doi.org/10.1103/PhysRevD.100.055012}{\emph{Phys. Rev. D}
  {\bfseries 100} (2019) 055012},
  [\href{https://arxiv.org/abs/1902.08579}{{\ttfamily 1902.08579}}].

\bibitem{Ballett:2018uuc}
P.~Ballett, M.~Hostert, S.~Pascoli, Y.~F. Perez-Gonzalez, Z.~Tabrizi and
  R.~Zukanovich~Funchal, \emph{{Neutrino Trident Scattering at Near
  Detectors}}, \href{https://doi.org/10.1007/JHEP01(2019)119}{\emph{JHEP}
  {\bfseries 01} (2019) 119},
  [\href{https://arxiv.org/abs/1807.10973}{{\ttfamily 1807.10973}}].

\bibitem{Bellini:2011rx}
G.~Bellini et~al., \emph{{Precision measurement of the 7Be solar neutrino
  interaction rate in Borexino}},
  \href{https://doi.org/10.1103/PhysRevLett.107.141302}{\emph{Phys. Rev. Lett.}
  {\bfseries 107} (2011) 141302},
  [\href{https://arxiv.org/abs/1104.1816}{{\ttfamily 1104.1816}}].

\bibitem{Borexino:2017rsf}
{\scshape Borexino} collaboration, M.~Agostini et~al., \emph{{First
  Simultaneous Precision Spectroscopy of $pp$, $^7$Be, and $pep$ Solar
  Neutrinos with Borexino Phase-II}},
  \href{https://doi.org/10.1103/PhysRevD.100.082004}{\emph{Phys. Rev. D}
  {\bfseries 100} (2019) 082004},
  [\href{https://arxiv.org/abs/1707.09279}{{\ttfamily 1707.09279}}].

\bibitem{Ilten:2018crw}
P.~Ilten, Y.~Soreq, M.~Williams and W.~Xue, \emph{{Serendipity in dark photon
  searches}}, \href{https://doi.org/10.1007/JHEP06(2018)004}{\emph{JHEP}
  {\bfseries 06} (2018) 004},
  [\href{https://arxiv.org/abs/1801.04847}{{\ttfamily 1801.04847}}].

\bibitem{Bauer:2018onh}
M.~Bauer, P.~Foldenauer and J.~Jaeckel, \emph{{Hunting All the Hidden
  Photons}}, \href{https://doi.org/10.1007/JHEP07(2018)094}{\emph{JHEP}
  {\bfseries 07} (2018) 094},
  [\href{https://arxiv.org/abs/1803.05466}{{\ttfamily 1803.05466}}].

\bibitem{Krnjaic:2019rsv}
G.~Krnjaic, G.~Marques-Tavares, D.~Redigolo and K.~Tobioka, \emph{{Probing
  Muonphilic Force Carriers and Dark Matter at Kaon Factories}},
  \href{https://doi.org/10.1103/PhysRevLett.124.041802}{\emph{Phys. Rev. Lett.}
  {\bfseries 124} (2020) 041802},
  [\href{https://arxiv.org/abs/1902.07715}{{\ttfamily 1902.07715}}].

\bibitem{NA62:2021bji}
{\scshape NA62} collaboration, E.~Cortina~Gil et~al., \emph{{Search for $K^+$
  decays to a muon and invisible particles}},
  \href{https://doi.org/10.1016/j.physletb.2021.136259}{\emph{Phys. Lett. B}
  {\bfseries 816} (2021) 136259},
  [\href{https://arxiv.org/abs/2101.12304}{{\ttfamily 2101.12304}}].

\bibitem{BaBar:2016sci}
{\scshape BaBar} collaboration, J.~P. Lees et~al., \emph{{Search for a muonic
  dark force at BABAR}},
  \href{https://doi.org/10.1103/PhysRevD.94.011102}{\emph{Phys. Rev. D}
  {\bfseries 94} (2016) 011102},
  [\href{https://arxiv.org/abs/1606.03501}{{\ttfamily 1606.03501}}].

\bibitem{Ilten:2016tkc}
P.~Ilten, Y.~Soreq, J.~Thaler, M.~Williams and W.~Xue, \emph{{Proposed
  Inclusive Dark Photon Search at LHCb}},
  \href{https://doi.org/10.1103/PhysRevLett.116.251803}{\emph{Phys. Rev. Lett.}
  {\bfseries 116} (2016) 251803},
  [\href{https://arxiv.org/abs/1603.08926}{{\ttfamily 1603.08926}}].

\bibitem{LHCb:2017trq}
{\scshape LHCb} collaboration, R.~Aaij et~al., \emph{{Search for Dark Photons
  Produced in 13 TeV $pp$ Collisions}},
  \href{https://doi.org/10.1103/PhysRevLett.120.061801}{\emph{Phys. Rev. Lett.}
  {\bfseries 120} (2018) 061801},
  [\href{https://arxiv.org/abs/1710.02867}{{\ttfamily 1710.02867}}].

\bibitem{LHCb:2019vmc}
{\scshape LHCb} collaboration, R.~Aaij et~al., \emph{{Search for
  $A'\to\mu^+\mu^-$ Decays}},
  \href{https://doi.org/10.1103/PhysRevLett.124.041801}{\emph{Phys. Rev. Lett.}
  {\bfseries 124} (2020) 041801},
  [\href{https://arxiv.org/abs/1910.06926}{{\ttfamily 1910.06926}}].

\bibitem{Banerjee:2019pds}
D.~Banerjee et~al., \emph{{Dark matter search in missing energy events with
  NA64}}, \href{https://doi.org/10.1103/PhysRevLett.123.121801}{\emph{Phys.
  Rev. Lett.} {\bfseries 123} (2019) 121801},
  [\href{https://arxiv.org/abs/1906.00176}{{\ttfamily 1906.00176}}].

\bibitem{BaBar:2014zli}
{\scshape BaBar} collaboration, J.~P. Lees et~al., \emph{{Search for a Dark
  Photon in $e^+e^-$ Collisions at BaBar}},
  \href{https://doi.org/10.1103/PhysRevLett.113.201801}{\emph{Phys. Rev. Lett.}
  {\bfseries 113} (2014) 201801},
  [\href{https://arxiv.org/abs/1406.2980}{{\ttfamily 1406.2980}}].

\bibitem{NA62:2019meo}
{\scshape NA62} collaboration, E.~Cortina~Gil et~al., \emph{{Search for
  production of an invisible dark photon in $\pi^0$ decays}},
  \href{https://doi.org/10.1007/JHEP05(2019)182}{\emph{JHEP} {\bfseries 05}
  (2019) 182}, [\href{https://arxiv.org/abs/1903.08767}{{\ttfamily
  1903.08767}}].

\bibitem{Chen:2018vkr}
C.-Y. Chen, J.~Kozaczuk and Y.-M. Zhong, \emph{{Exploring leptophilic dark
  matter with NA64-$\mu$}},
  \href{https://doi.org/10.1007/JHEP10(2018)154}{\emph{JHEP} {\bfseries 10}
  (2018) 154}, [\href{https://arxiv.org/abs/1807.03790}{{\ttfamily
  1807.03790}}].

\bibitem{Batell:2016ove}
B.~Batell, N.~Lange, D.~McKeen, M.~Pospelov and A.~Ritz, \emph{{Muon anomalous
  magnetic moment through the leptonic Higgs portal}},
  \href{https://doi.org/10.1103/PhysRevD.95.075003}{\emph{Phys. Rev. D}
  {\bfseries 95} (2017) 075003},
  [\href{https://arxiv.org/abs/1606.04943}{{\ttfamily 1606.04943}}].

\bibitem{Gninenko:2014pea}
S.~N. Gninenko, N.~V. Krasnikov and V.~A. Matveev, \emph{{Muon g-2 and searches
  for a new leptophobic sub-GeV dark boson in a missing-energy experiment at
  CERN}}, \href{https://doi.org/10.1103/PhysRevD.91.095015}{\emph{Phys. Rev. D}
  {\bfseries 91} (2015) 095015},
  [\href{https://arxiv.org/abs/1412.1400}{{\ttfamily 1412.1400}}].

\bibitem{NA64:2016oww}
{\scshape NA64} collaboration, D.~Banerjee et~al., \emph{{Search for invisible
  decays of sub-GeV dark photons in missing-energy events at the CERN SPS}},
  \href{https://doi.org/10.1103/PhysRevLett.118.011802}{\emph{Phys. Rev. Lett.}
  {\bfseries 118} (2017) 011802},
  [\href{https://arxiv.org/abs/1610.02988}{{\ttfamily 1610.02988}}].

\bibitem{Kahn:2018cqs}
Y.~Kahn, G.~Krnjaic, N.~Tran and A.~Whitbeck, \emph{{M$^{3}$: a new muon
  missing momentum experiment to probe (g \ensuremath{-} 2)$_{\mu}$ and dark
  matter at Fermilab}},
  \href{https://doi.org/10.1007/JHEP09(2018)153}{\emph{JHEP} {\bfseries 09}
  (2018) 153}, [\href{https://arxiv.org/abs/1804.03144}{{\ttfamily
  1804.03144}}].

\bibitem{Jho:2019cxq}
Y.~Jho, Y.~Kwon, S.~C. Park and P.-Y. Tseng, \emph{{Search for muon-philic new
  light gauge boson at Belle II}},
  \href{https://doi.org/10.1007/JHEP10(2019)168}{\emph{JHEP} {\bfseries 10}
  (2019) 168}, [\href{https://arxiv.org/abs/1904.13053}{{\ttfamily
  1904.13053}}].

\bibitem{Galon:2019owl}
I.~Galon, E.~Kajamovitz, D.~Shih, Y.~Soreq and S.~Tarem, \emph{{Searching for
  muonic forces with the ATLAS detector}},
  \href{https://doi.org/10.1103/PhysRevD.101.011701}{\emph{Phys. Rev. D}
  {\bfseries 101} (2020) 011701},
  [\href{https://arxiv.org/abs/1906.09272}{{\ttfamily 1906.09272}}].

\bibitem{Kamada:2018zxi}
A.~Kamada, K.~Kaneta, K.~Yanagi and H.-B. Yu, \emph{{Self-interacting dark
  matter and muon $g-2$ in a gauged U$(1)_{L_{\mu} - L_{\tau}}$ model}},
  \href{https://doi.org/10.1007/JHEP06(2018)117}{\emph{JHEP} {\bfseries 06}
  (2018) 117}, [\href{https://arxiv.org/abs/1805.00651}{{\ttfamily
  1805.00651}}].

\bibitem{Croon:2020lrf}
D.~Croon, G.~Elor, R.~K. Leane and S.~D. McDermott, \emph{{Supernova Muons: New
  Constraints on $Z$' Bosons, Axions and ALPs}},
  \href{https://doi.org/10.1007/JHEP01(2021)107}{\emph{JHEP} {\bfseries 01}
  (2021) 107}, [\href{https://arxiv.org/abs/2006.13942}{{\ttfamily
  2006.13942}}].

\bibitem{Bar:2019ifz}
N.~Bar, K.~Blum and G.~D'Amico, \emph{{Is there a supernova bound on axions?}},
  \href{https://doi.org/10.1103/PhysRevD.101.123025}{\emph{Phys. Rev. D}
  {\bfseries 101} (2020) 123025},
  [\href{https://arxiv.org/abs/1907.05020}{{\ttfamily 1907.05020}}].

\bibitem{Wolfenstein:1977ue}
L.~Wolfenstein, \emph{{Neutrino Oscillations in Matter}},
  \href{https://doi.org/10.1103/PhysRevD.17.2369}{\emph{Phys. Rev. D}
  {\bfseries 17} (1978) 2369--2374}.

\bibitem{Mikheyev:1985zog}
S.~P. Mikheyev and A.~Y. Smirnov, \emph{{Resonance Amplification of
  Oscillations in Matter and Spectroscopy of Solar Neutrinos}}, {\emph{Sov. J.
  Nucl. Phys.} {\bfseries 42} (1985) 913--917}.

\bibitem{Esteban:2018ppq}
I.~Esteban, M.~C. Gonzalez-Garcia, M.~Maltoni, I.~Martinez-Soler and
  J.~Salvado, \emph{{Updated constraints on non-standard interactions from
  global analysis of oscillation data}},
  \href{https://doi.org/10.1007/JHEP08(2018)180}{\emph{JHEP} {\bfseries 08}
  (2018) 180}, [\href{https://arxiv.org/abs/1805.04530}{{\ttfamily
  1805.04530}}].

\bibitem{Heeck:2018nzc}
J.~Heeck, M.~Lindner, W.~Rodejohann and S.~Vogl, \emph{{Non-Standard Neutrino
  Interactions and Neutral Gauge Bosons}},
  \href{https://doi.org/10.21468/SciPostPhys.6.3.038}{\emph{SciPost Phys.}
  {\bfseries 6} (2019) 038},
  [\href{https://arxiv.org/abs/1812.04067}{{\ttfamily 1812.04067}}].

\bibitem{Coloma:2020gfv}
P.~Coloma, M.~C. Gonzalez-Garcia and M.~Maltoni, \emph{{Neutrino oscillation
  constraints on U(1)' models: from non-standard interactions to long-range
  forces}}, \href{https://doi.org/10.1007/JHEP01(2021)114}{\emph{JHEP}
  {\bfseries 01} (2021) 114},
  [\href{https://arxiv.org/abs/2009.14220}{{\ttfamily 2009.14220}}].

\bibitem{Freedman:1973yd}
D.~Z. Freedman, \emph{{Coherent Neutrino Nucleus Scattering as a Probe of the
  Weak Neutral Current}},
  \href{https://doi.org/10.1103/PhysRevD.9.1389}{\emph{Phys. Rev. D} {\bfseries
  9} (1974) 1389--1392}.

\bibitem{Drukier:1984vhf}
A.~Drukier and L.~Stodolsky, \emph{{Principles and Applications of a Neutral
  Current Detector for Neutrino Physics and Astronomy}},
  \href{https://doi.org/10.1103/PhysRevD.30.2295}{\emph{Phys. Rev. D}
  {\bfseries 30} (1984) 2295}.

\bibitem{COHERENT:2017ipa}
{\scshape COHERENT} collaboration, D.~Akimov et~al., \emph{{Observation of
  Coherent Elastic Neutrino-Nucleus Scattering}},
  \href{https://doi.org/10.1126/science.aao0990}{\emph{Science} {\bfseries 357}
  (2017) 1123--1126}, [\href{https://arxiv.org/abs/1708.01294}{{\ttfamily
  1708.01294}}].

\bibitem{Altmannshofer:2018xyo}
W.~Altmannshofer, M.~Tammaro and J.~Zupan, \emph{{Non-standard neutrino
  interactions and low energy experiments}},
  \href{https://arxiv.org/abs/1812.02778}{{\ttfamily 1812.02778}}.

\bibitem{Coloma:2019mbs}
P.~Coloma, I.~Esteban, M.~C. Gonzalez-Garcia and M.~Maltoni, \emph{{Improved
  global fit to Non-Standard neutrino Interactions using COHERENT energy and
  timing data}}, \href{https://doi.org/10.1007/JHEP02(2020)023}{\emph{JHEP}
  {\bfseries 02} (2020) 023},
  [\href{https://arxiv.org/abs/1911.09109}{{\ttfamily 1911.09109}}].

\bibitem{Denton:2020hop}
P.~B. Denton and J.~Gehrlein, \emph{{A Statistical Analysis of the COHERENT
  Data and Applications to New Physics}},
  \href{https://doi.org/10.1007/JHEP04(2021)266}{\emph{JHEP} {\bfseries 04}
  (2021) 266}, [\href{https://arxiv.org/abs/2008.06062}{{\ttfamily
  2008.06062}}].

\bibitem{COHERENT:2018imc}
{\scshape COHERENT} collaboration, D.~Akimov et~al., \emph{{COHERENT
  Collaboration data release from the first observation of coherent elastic
  neutrino-nucleus scattering}},
  \href{https://arxiv.org/abs/1804.09459}{{\ttfamily 1804.09459}}.

\bibitem{Sala:2017ihs}
F.~Sala and D.~M. Straub, \emph{{A New Light Particle in B Decays?}},
  \href{https://doi.org/10.1016/j.physletb.2017.09.072}{\emph{Phys. Lett. B}
  {\bfseries 774} (2017) 205--209},
  [\href{https://arxiv.org/abs/1704.06188}{{\ttfamily 1704.06188}}].

\bibitem{Belle-II:2021rof}
{\scshape Belle-II} collaboration, F.~Abudin\'en et~al., \emph{{Search for
  $B^{+}\to K^{+}\nu\bar{\nu}$ decays using an inclusive tagging method at
  Belle II}},  \href{https://arxiv.org/abs/2104.12624}{{\ttfamily 2104.12624}}.

\bibitem{Belle:2017oht}
{\scshape Belle} collaboration, J.~Grygier et~al., \emph{{Search for
  $\boldsymbol{B\to h\nu\bar{\nu}}$ decays with semileptonic tagging at
  Belle}}, \href{https://doi.org/10.1103/PhysRevD.96.091101}{\emph{Phys. Rev.
  D} {\bfseries 96} (2017) 091101},
  [\href{https://arxiv.org/abs/1702.03224}{{\ttfamily 1702.03224}}].

\bibitem{BaBar:2013npw}
{\scshape BaBar} collaboration, J.~P. Lees et~al., \emph{{Search for $B \to
  K^{(*)} \nu \overline \nu$ and invisible quarkonium decays}},
  \href{https://doi.org/10.1103/PhysRevD.87.112005}{\emph{Phys. Rev. D}
  {\bfseries 87} (2013) 112005},
  [\href{https://arxiv.org/abs/1303.7465}{{\ttfamily 1303.7465}}].

\bibitem{Belle:2013tnz}
{\scshape Belle} collaboration, O.~Lutz et~al., \emph{{Search for $B \to
  h^{(*)} \nu \bar{\nu}$ with the full Belle $\Upsilon(4S)$ data sample}},
  \href{https://doi.org/10.1103/PhysRevD.87.111103}{\emph{Phys. Rev. D}
  {\bfseries 87} (2013) 111103},
  [\href{https://arxiv.org/abs/1303.3719}{{\ttfamily 1303.3719}}].

\bibitem{Hayasaka:2010np}
K.~Hayasaka et~al., \emph{{Search for Lepton Flavor Violating Tau Decays into
  Three Leptons with 719 Million Produced Tau+Tau- Pairs}},
  \href{https://doi.org/10.1016/j.physletb.2010.03.037}{\emph{Phys. Lett. B}
  {\bfseries 687} (2010) 139--143},
  [\href{https://arxiv.org/abs/1001.3221}{{\ttfamily 1001.3221}}].

\bibitem{Darme:2020hpo}
L.~Darm\'e, M.~Fedele, K.~Kowalska and E.~M. Sessolo, \emph{{Flavour anomalies
  from a split dark sector}},
  \href{https://doi.org/10.1007/JHEP08(2020)148}{\emph{JHEP} {\bfseries 08}
  (2020) 148}, [\href{https://arxiv.org/abs/2002.11150}{{\ttfamily
  2002.11150}}].

\bibitem{Darme:2021qzw}
L.~Darm\'e, M.~Fedele, K.~Kowalska and E.~M. Sessolo, \emph{{Flavour anomalies
  and the muon $g-2$ from feebly interacting particles}},
  \href{https://arxiv.org/abs/2106.12582}{{\ttfamily 2106.12582}}.

\bibitem{Amaral:2020tga}
D.~W. P.~d. Amaral, D.~G. Cerdeno, P.~Foldenauer and E.~Reid, \emph{{Solar
  neutrino probes of the muon anomalous magnetic moment in the gauged $
  \mathrm{U}{(1)}_{L_{\mu }-{L}_{\tau }} $}},
  \href{https://doi.org/10.1007/JHEP12(2020)155}{\emph{JHEP} {\bfseries 12}
  (2020) 155}, [\href{https://arxiv.org/abs/2006.11225}{{\ttfamily
  2006.11225}}].

\bibitem{Smolkovic:2019jow}
A.~Smolkovi\v{c}, M.~Tammaro and J.~Zupan, \emph{{Anomaly free Froggatt-Nielsen
  models of flavor}},
  \href{https://doi.org/10.1007/JHEP10(2019)188}{\emph{JHEP} {\bfseries 10}
  (2019) 188}, [\href{https://arxiv.org/abs/1907.10063}{{\ttfamily
  1907.10063}}].

\bibitem{CLEO:2008ffk}
{\scshape CLEO} collaboration, B.~I. Eisenstein et~al., \emph{{Precision
  Measurement of B(D+ ---\ensuremath{>} mu+ nu) and the Pseudoscalar Decay
  Constant f(D+)}},
  \href{https://doi.org/10.1103/PhysRevD.78.052003}{\emph{Phys. Rev. D}
  {\bfseries 78} (2008) 052003},
  [\href{https://arxiv.org/abs/0806.2112}{{\ttfamily 0806.2112}}].

\bibitem{MartinCamalich:2020dfe}
J.~Martin~Camalich, M.~Pospelov, P.~N.~H. Vuong, R.~Ziegler and J.~Zupan,
  \emph{{Quark Flavor Phenomenology of the QCD Axion}},
  \href{https://doi.org/10.1103/PhysRevD.102.015023}{\emph{Phys. Rev. D}
  {\bfseries 102} (2020) 015023},
  [\href{https://arxiv.org/abs/2002.04623}{{\ttfamily 2002.04623}}].

\bibitem{Weinberg:2020zba}
S.~Weinberg, \emph{{Models of Lepton and Quark Masses}},
  \href{https://doi.org/10.1103/PhysRevD.101.035020}{\emph{Phys. Rev. D}
  {\bfseries 101} (2020) 035020},
  [\href{https://arxiv.org/abs/2001.06582}{{\ttfamily 2001.06582}}].

\bibitem{Baker:2020vkh}
M.~J. Baker, P.~Cox and R.~R. Volkas, \emph{{Has the Origin of the Third-Family
  Fermion Masses been Determined?}},
  \href{https://doi.org/10.1007/JHEP04(2021)151}{\emph{JHEP} {\bfseries 04}
  (2021) 151}, [\href{https://arxiv.org/abs/2012.10458}{{\ttfamily
  2012.10458}}].

\bibitem{Baker:2021yli}
M.~J. Baker, P.~Cox and R.~R. Volkas, \emph{{Radiative muon mass models and
  $(g-2)_\mu$}}, \href{https://doi.org/10.1007/JHEP05(2021)174}{\emph{JHEP}
  {\bfseries 05} (2021) 174},
  [\href{https://arxiv.org/abs/2103.13401}{{\ttfamily 2103.13401}}].

\bibitem{Aad:2020iuy}
{\scshape ATLAS} collaboration, G.~Aad et~al., \emph{{Search for pairs of
  scalar leptoquarks decaying into quarks and electrons or muons in $ \sqrt{s}
  $ = 13 TeV $pp$ collisions with the ATLAS detector}},
  \href{https://doi.org/10.1007/JHEP10(2020)112}{\emph{JHEP} {\bfseries 10}
  (2020) 112}, [\href{https://arxiv.org/abs/2006.05872}{{\ttfamily
  2006.05872}}].

\bibitem{ATLAS:2020qoc}
{\scshape ATLAS} collaboration, \emph{{Search for pair production of scalar
  leptoquarks decaying to first or second generation leptons and boosted
  hadronically decaying top quarks in $pp$ collisions at $\sqrt{s}=13 \
  \mathrm{TeV}$ with the ATLAS detector}}, .

\bibitem{Poole:2019kcm}
C.~Poole and A.~E. Thomsen, \emph{{Constraints on 3- and 4-loop
  $\beta$-functions in a general four-dimensional Quantum Field Theory}},
  \href{https://doi.org/10.1007/JHEP09(2019)055}{\emph{JHEP} {\bfseries 09}
  (2019) 055}, [\href{https://arxiv.org/abs/1906.04625}{{\ttfamily
  1906.04625}}].

\end{thebibliography}\endgroup

\end{document}